\documentclass[11pt, a4paper, copyright, gdm]{google}

\usepackage[numbers, sort&compress, square]{natbib}

\usepackage{mathtools}
\usepackage{bm}
\usepackage{multirow}
\usepackage{array}
\usepackage{tikz}
\usepackage{enumitem}
\usepackage{amsmath}
\numberwithin{equation}{section}
\usetikzlibrary{arrows.meta, positioning, calc, fit, backgrounds}

\newcommand{\cA}{\mathcal{A}}

\newcommand{\cH}{\mathcal{H}}
\newcommand{\cJ}{\mathcal{J}}
\newcommand{\cK}{\mathcal{K}}
\newcommand{\cL}{\mathcal{L}}
\newcommand{\cM}{\mathcal{M}}
\newcommand{\cN}{\mathcal{N}}
\newcommand{\cO}{\mathcal{O}}
\newcommand{\cP}{\mathcal{P}}
\newcommand{\cV}{\mathcal{V}}
\newcommand{\cW}{\mathcal{W}}

\newcommand{\PP}{\mathbb{P}}
\newcommand{\CC}{\mathbb{C}}
\newcommand{\RR}{\mathbb{R}}
\newcommand{\ZZ}{\mathbb{Z}}
\newcommand{\dd}{\mathrm{d}}
\newcommand{\ii}{\mathrm{i}}
\DeclareMathOperator{\rk}{rank}
\DeclareMathOperator{\Res}{Res}
\DeclareMathOperator{\Tr}{Tr}
\DeclareMathOperator{\im}{im}
\DeclareMathOperator{\coker}{coker}
\DeclareMathOperator{\End}{End}
\newcommand{\thead}[1]{\begin{tabular}{@{}c@{}}#1\end{tabular}}
\keywords{Heterotic String Theory, String Phenomenology, Calabi--Yau Manifolds, Yukawa Couplings, Quark Masses}

\uselogo{gdm}

\title{From Top to Bottom: Quark Mass Hierarchies in Heterotic String Theory}

\correspondingauthor{Yuri Chervonyi (cyuri@google.com), Fabian Ruehle (ruehlef@google.com)}

\author[1,*]{Yuri Chervonyi}
\author[1,2,*]{Fabian Ruehle}
\author[1]{Nenad Tomasev}

\affil[1]{Google DeepMind}
\affil[2]{Northeastern University}
\affil[*]{Equal contribution. Corresponding authors.\protect\\Work done at Google DeepMind.}

\begin{abstract}
Standard Model fermion masses exhibit large hierarchies, spanning more than five orders of magnitude in the quark sector alone. In this paper, we study whether these hierarchies can be generated perturbatively at tree level in smooth heterotic Calabi--Yau compactifications, where the supergravity approximation remains under computational control. In such backgrounds, physical Yukawa couplings combine holomorphic superpotential integrals with matter field K\"ahler metrics. The former can be computed algebraically, while the latter require solving a coupled system of partial differential equations for the Ricci-flat metric, Hermitian Yang--Mills connections, and harmonic bundle-valued forms. Within this framework, we identify two new tree-level mechanisms by which complex structure moduli can generate fermion mass hierarchies and show analytically and numerically that both mechanisms are obstructed for previously studied tetra-quadric models. Motivated by these obstructions, we scan the existing databases of tens of thousands of complete intersection Calabi--Yau line bundle models and find a unique smooth quotient admitting a rank-three up-quark Yukawa matrix at tree level. Evaluating the physical couplings on this manifold and evolving them under the renormalization group, we obtain realistic mass hierarchies for the top, charm, up, and bottom quarks in the smooth supergravity regime.
\end{abstract}

\begin{document}
\maketitle

\newpage
\begingroup
\setlength{\parskip}{0pt}
\hrule
\vspace{0.25em}
\tableofcontents
\vspace{1em}
\hrule
\endgroup
\newpage

\section{Introduction}
\label{sec:intro}

Computing the Standard Model fermion masses and explaining their pronounced hierarchies across the three generations from an ultraviolet-complete theory remains one of the central open problems in string phenomenology. In $E_8 \times E_8$ heterotic compactifications on smooth Calabi--Yau (CY) threefolds~\cite{candelas1985vacuum,green2012superstring}, four-dimensional physical Yukawa couplings arise at tree level from overlap integrals of internal zero-mode wavefunctions~\cite{Strominger:1985ks,Candelas:1987rx}. Whereas the chiral matter spectrum and gauge group are fixed by topological invariants, physical Yukawa matrices depend on two distinct geometric ingredients: holomorphic trilinear superpotential couplings, which vary with the complex structure moduli $C$ and bundle moduli $S_{ab}$ and can often be extracted algebraically~\cite{Candelas:1987rx,Blesneag_2016,Blesneag_2017}, and the matter field K\"ahler metric, which also depends on the K\"ahler moduli $t$ and requires solving the Ricci-flat Monge--Amp\`ere equation for the metric, the Hermitian Yang--Mills (HYM) equations~\cite{Donaldson:1985,Uhlenbeck:1986}, and the harmonic $(0,1)$-form equations on the internal manifold.

Heterotic line bundle models on free quotients $X/\Gamma$ of complete intersection Calabi--Yau (CICY) threefolds~\cite{Candelas:1987kf,Braun:2010} provide a tractable setting in which realistic low-energy spectra have been constructed systematically~\cite{Anderson:2011ns,Anderson:2012yf,Anderson:2013xka}. At a split locus in bundle moduli space, the visible-sector vector bundle on the covering space $X$ decomposes into a direct sum of five line bundles $V = \bigoplus_{a=1}^5 L_a$ with $c_1(V) = 0$, breaking $E_8$ to $SU(5) \times S(U(1)^5)$, while a discrete Wilson line on $X/\Gamma$ breaks $SU(5)$ to the Standard Model gauge group. Systematic scans over CICY quotients have produced large databases of three-generation $SU(5)$ and Minimal Supersymmetric Standard Model (MSSM) vacua free of exotic matter~\cite{Anderson:2012yf,Anderson:2013xka}, and Koszul-sequence techniques have uncovered rich selection rules and vanishing theorems for their holomorphic Yukawa couplings~\cite{Blesneag_2016,Blesneag_2017,Gray:2019,Anderson:2021}.

Until recently, computing the matter field K\"ahler metrics on compact CY threefolds was out of reach. Progress in neural network approximations of Ricci-flat CY metrics, HYM connections, and harmonic bundle-valued forms~\cite{Douglas:2020hpv,anderson2021modulidependentcalabiyausu3structuremetrics,Jejjala:2020wcc,larfors2022numericalmetricscompleteintersection,Berglund:2022gvm,constantin2024computationquarkmassesstring,berglund2024precision,Lust:2026warped} has made direct numerical evaluation of physical Yukawa couplings possible. In~\cite{constantin2024computationquarkmassesstring}, physical up-type quark masses were computed in a heterotic line bundle model on a $\Gamma = \ZZ_2 \times \ZZ_2$ quotient of the tetra-quadric threefold $X \subset (\PP^1)^4$ (CICY 7862). However, across the one-parameter Dwork family and at generic points in complex structure moduli space, the up-quark Yukawa matrix $Y^u$ had rank two ($m_u = 0$) and nearly degenerate non-zero singular values, with $m_t / m_c \approx 1.0$ on the Dwork slice and $m_t / m_c \approx 1.78$ at a generic complex structure point, well below the grand unified theory (GUT) scale ratio $m_t / m_c \sim 200\text{--}350$~\cite{Ross:2007az,Xing:2007fb,Antusch:2013jca}. Subsequent studies~\cite{Constantin:2024fn,Constantin:2025fn} therefore moved away from the split-bundle locus to implement heterotic Froggatt--Nielsen models, generating flavor hierarchies from postulated vacuum expectation values (VEVs) of $S(U(1)^5)$-charged bundle moduli singlets and worldsheet instantons.

These findings raise two natural questions. First, why do physical quark masses on the split-bundle tetra-quadric remain nearly degenerate across moduli space, and can complex structure deformations alone produce realistic hierarchies at tree level? Second, across the existing databases of CICY heterotic line bundle models~\cite{Anderson:2012yf,Anderson:2013xka}, are there smooth compactifications that support a rank-three quark Yukawa matrix and realistic mass hierarchies within the regime of validity of ten-dimensional supergravity?

In this paper, we address both questions analytically and numerically. We identify two general mechanisms by which complex structure moduli can generate tree-level hierarchies among physical Yukawa couplings. The first mechanism relies on destructive interference in the holomorphic trilinear coupling $W_{IJK}(C) \propto \int_{X} \Omega \wedge \nu_I \wedge \nu_J \wedge \nu_K$ near smooth zero-crossing loci in the interior of complex structure moduli space. The second mechanism relies on K\"ahler metric localization near factorized, completed-square polynomial degenerations, where the CY volume measure concentrates onto the vanishing locus of a matter wavefunction; for the resulting hierarchy to survive canonical normalization with the K\"ahler potential, the competing zero-mode wavefunctions must remain cohomologically non-exact on this pinched locus. To find promising loci in the complex structure moduli space, we extremize the holomorphic superpotential integrals numerically. However, we observe that the optimizer exploits numerical noise of the grid on which we evaluate the integrals, which necessitates developing an exact algebraic integration method for the holomorphic part of the Yukawa couplings.

Applying this analysis to the tetra-quadric model (CICY 7862), we show that both mechanisms we identified above are obstructed at tree level at the split-bundle locus. The first mechanism fails because the moduli-dependent defining polynomial drops out of the holomorphic Yukawa couplings~\cite{Blesneag_2016}, and an accidental ambient $U(1)$ symmetry fixes the holomorphic Yukawa texture across all $20$ complex structure moduli. The second mechanism fails for a more subtle reason related to the fact that on $\PP^1$ ambient space factors there is not enough freedom to accommodate the mechanism. Approaching a completed-square locus produces a large hierarchy $(m_t / m_c)_{\mathrm{ref}} \approx 220$ in the algebraic reference basis, but because squaring a bilinear section exhausts the $(2,2)$ degree on two of the model's ambient space $\PP^1$ factors, the competing Koszul $(0,1)$-form becomes $\bar{\partial}_X$-exact on this locus. Passing to the numerically computed harmonic (0,1)-form basis reduces its norm by more than three orders of magnitude and collapses the physical quark mass ratio back to $m_t / m_c \simeq 2.1$, while across the smooth interior of moduli space $m_t / m_c$ remains of order unity, $m_t / m_c \lesssim 2\text{--}3$, and $\rk Y^u = 2$ keeps $m_u = 0$ in these models.

Motivated by these obstructions, we classify the tree-level Yukawa rank across all $63{,}325$ upstairs $SU(5)$ models of~\cite{Anderson:2013xka} and all $2{,}112$ downstairs Standard Models of~\cite{Anderson:2012yf}. Achieving $\rk Y^u = 3$ at the split locus requires placing the three chiral $\mathbf{10}$ families into three distinct line bundles, keeping three light up-type Higgs sectors, and projecting left- and right-handed quarks onto distinct eigenspaces of a discrete hypercharge Wilson line so as to lift the singular-value bound $m_t \le m_c + m_u$ of complex symmetric matrices with zeros on the diagonal. This scan requires first working out the Wilson-line actions for these upstairs $SU(5)$ models~\cite{Anderson:2013xka}. An exhaustive search over both databases reveals a single smooth quotient which satisfies all topological, equivariant, and smoothness conditions: CICY 5273, Model 500 on $\cA = (\PP^1)^4 \times (\PP^2)^2$ with $|\Gamma| = 2$, which has $\rk Y^u = 3$ and $\rk Y^d = 1$ at tree level. Because its ambient space contains $\PP^2 \times \PP^2$ factors at codimension five, both hierarchy mechanisms operate unobstructed along the smooth complex structure locus where all three Higgs pairs become massless. Solving the coupled PDE system numerically and matching to two-loop renormalization group evolution, we reproduce the five-order-of-magnitude hierarchy across all four tree-level massive quarks $(t, c, u, b)$ well inside the smooth supergravity regime, while the strange and down quarks remain massless prior to turning on non-Abelian bundle deformations in this model. This also showcases the applicability of our numerical and analytic techniques at high codimension and in CICYs with $h^{1,1}(X) = 6$.

Methodologically, both the analytical obstructions and the CICY 5273 benchmarks were developed through an interactive human--AI discovery loop~\cite{Chervonyi:2026lehmer} powered by internal versions of Gemini~\cite{geminiteam2023gemini,geminiteam2024gemini15,geminiteam2025gemini25}, including Gemini~4 Argon~\cite{gemini4argon2026} and Carbon~\cite{gemini4carbon2026}, paired with the agentic research environment Google Antigravity~\cite{antigravity2026}. Following the multi-stage search paradigm of~\cite{Chervonyi:2026lehmer}, our numerical pipeline combines fast local algebraic and reference-metric optimization, distributed multi-worker searches augmented with global smoothness checks, and full multi-seed neural PDE solutions on Google Cloud CPU and GPU infrastructure. Within this loop, numerical anomalies directly guided our analytic derivations, including cases where the moduli-space optimizer exploited approximation errors. On the tetra-quadric, for instance, we found large quark mass hierarchies prior to neural network training that vanished once we included the learned solutions, leading us to discover that the competing $(0,1)$-forms become exact.

This paper is organized as follows. Section~\ref{sec:review} reviews heterotic line bundle models and the normalization and moduli dependence of physical Yukawa couplings (Sections~\ref{subsec:line_bundle_models}--\ref{subsec:moduli_dependence}), and then formulates the two tree-level hierarchy mechanisms and our novel algebraic Koszul--Dolbeault integration method (Sections~\ref{subsec:two_mechanisms}--\ref{subsec:exact_integration}). Section~\ref{sec:tetraquadric} illustrates this framework on the tetra-quadric model, deriving the obstructions to both hierarchy-tuning mechanisms and verifying them numerically. Section~\ref{sec:cicy_scan} establishes general Yukawa rank bounds and presents our scan of the CICY line bundle databases singling out CICY 5273, Model 500. Section~\ref{sec:cicy_5273} describes the geometry, K\"ahler cone, Higgs jumping locus, three-stage search workflow, and physical quark mass results for this model. We conclude in Section~\ref{sec:conclusions}. Appendix~\ref{app:koszul} collects technical details about the explicit Koszul $(0,1)$-forms and total-cocycle contractions, Appendix~\ref{app:benchmarks} records the defining polynomials, benchmark complex structure points, and renormalization group matching conventions that lead to realistic quark masses in the CICY 5273 model, and Appendix~\ref{app:ml_setup} details our neural network architectures, PDE loss functionals, and training diagnostics.

\section{Heterotic Line Bundle Models and Yukawa Hierarchies}
\label{sec:review}

Before turning to concrete Calabi--Yau geometries in Sections~\ref{sec:tetraquadric}--\ref{sec:cicy_5273}, we establish the general framework needed to compute physical Yukawa couplings and analyze tree-level mass hierarchies in heterotic line bundle models. Following~\cite{Anderson:2011ns,Anderson:2012yf,Anderson:2013xka,Blesneag_2016,Blesneag_2017,constantin2024computationquarkmassesstring}, in Sections~\ref{subsec:line_bundle_models}--\ref{subsec:moduli_dependence} we review four-dimensional $\cN=1$ heterotic compactifications with Abelian line bundle sums, fix the ten-dimensional supergravity normalization of physical Yukawa couplings, relate analytic reference quantities to the exact metric and harmonic-form corrections, and discuss their dependence on geometric and bundle moduli.
Building on this setup, Sections~\ref{subsec:two_mechanisms} and~\ref{subsec:exact_integration} present new results: we formulate two tree-level mechanisms for generating fermion mass hierarchies and develop an exact algebraic method to integrate holomorphic superpotential couplings via the Koszul--Dolbeault double complex.

\subsection{Line bundle models}
\label{subsec:line_bundle_models}

We consider ten-dimensional $\cN=1$ $E_8 \times E_8$ heterotic string theory compactified on a smooth CY threefold quotient $\hat{X} = X/\Gamma$. The covering space $X \subset \cA = \prod_{r=1}^m \PP^{n_r}$ is a simply connected CICY threefold of codimension $K = \sum_{r=1}^m n_r - 3$. It is defined as the common zero locus of $K$ homogeneous polynomials $P = (P_1, \dots, P_K)$ whose multidegrees $d_\alpha = (d_\alpha^1, \dots, d_\alpha^m)$ obey the CY condition $\sum_{\alpha=1}^K d_\alpha^r = n_r + 1$ for each ambient projective factor $r = 1, \dots, m$~\cite{Candelas:1987kf}. Here $\Gamma$ is a finite group acting freely and holomorphically on $X$~\cite{Braun:2010}.

To break one of the $E_8$ factors to the Standard Model gauge group, we equip $X$ with a holomorphic vector bundle $V \to X$. 
In heterotic line bundle models~\cite{Anderson:2011ns,Anderson:2012yf,Anderson:2013xka}, $V$ is chosen at a split locus in bundle moduli space to be a sum of five line bundles,
\begin{equation}
  V = \bigoplus_{a=1}^5 L_a\,, \qquad c_1(V) = \sum_{a=1}^5 c_1(L_a) = 0\,,
  \label{eq:rank5_line_bundle_sum}
\end{equation}
where on favorable embeddings each $L_a = \cO_X(q_a)$ is the restriction of an ambient line bundle $\cO_{\cA}(q_a^1, \dots, q_a^m)$. When each $L_a$ carries a $\Gamma$-equivariant structure, $V$ descends to a rank-five bundle on $\hat{X}$. To keep notation light, we denote the descended line bundles on $\hat{X}$ by the same symbols $L_a$. Its structure group $S(U(1)^5) \subset SU(5) \subset E_8$ breaks the visible $E_8$ gauge group to $SU(5) \times S(U(1)^5)$, and turning on a flat discrete Wilson line on $\hat{X}$ along the $U(1)$ hypercharge direction breaks $SU(5)$ further to $G_{\mathrm{SM}} = SU(3)_C \times SU(2)_L \times U(1)_Y$. Although the four independent Abelian factors in $S(U(1)^5)$ are generically Green--Schwarz anomalous and acquire St\"uckelberg masses near the compactification scale~\cite{Blumenhagen:2005ga,Anderson:2011ns}, they survive as global selection rules on the perturbative superpotential as long as all $S(U(1)^5)$-charged bundle moduli singlets have vanishing VEVs.

A consistent, supersymmetric compactification imposes both topological and K\"ahler-moduli constraints on $V$. Topologically, cancellation of heterotic five-brane and gauge anomalies without anti-five-branes requires $c_2(TX) - c_2(V)$ to be an effective curve class in the Mori cone of $X$. For the split bundle~\eqref{eq:rank5_line_bundle_sum}, this imposes $c_2(V) = -\frac{1}{2}\sum_{a=1}^5 c_1^2(L_a) \le c_2(TX)$. Preservation of four-dimensional $\cN=1$ supersymmetry further requires $V$ to be poly-stable with vanishing slope. By the Donaldson--Uhlenbeck--Yau theorem~\cite{Donaldson:1985,Uhlenbeck:1986}, $V$ admits a supersymmetric HYM connection if and only if all five line bundles have vanishing slope at a common K\"ahler class $J = \sum_{i=1}^{h^{1,1}} t^i J_i$ inside the K\"ahler cone,
\begin{equation}
  \mu_t(L_a) = \int_X c_1(L_a) \wedge J \wedge J = d_{ijk}\,c_1^i(L_a)\,t^j t^k = 0\,, \qquad a = 1, \dots, 5\,,
  \label{eq:duy_slope_zero}
\end{equation}
where $d_{ijk} = \int_X J_i \wedge J_j \wedge J_k$ are the triple intersection numbers on $X$.

\begin{table}[t]
\centering
\small
\setlength{\tabcolsep}{8pt}
\begin{tabular}{@{}llll@{}}
\toprule
$SU(5)$ sector & SM multiplets & $S(U(1)^5)$ charge & Cohomology on $\hat{X}$ \\
\midrule
$\mathbf{10}_a$ & $Q,\; u^c,\; e^c$ & $e_a$ & $H^1(\hat{X}, L_a \otimes \cW_R)$ \\
$\overline{\mathbf{5}}_{ab}$ ($a < b$) & $d^c,\; L,\; H_d$ & $e_a + e_b$ & $H^1(\hat{X}, L_a \otimes L_b \otimes \cW_R)$ \\
$\mathbf{5}^H_{ab}$ ($a < b$) & $H_u$ & $-e_a - e_b$ & $H^1(\hat{X}, L_a^* \otimes L_b^* \otimes \cW_R)$ \\
$\mathbf{1}_{ab}$ ($a \neq b$) & $S_{ab}$ & $e_a - e_b$ & $H^1(\hat{X}, L_a \otimes L_b^*)$ \\
\bottomrule
\end{tabular}
\caption{Zero-mode sectors arising from the branching of the adjoint $\mathbf{248}$ of $E_8$ under $SU(5) \times S(U(1)^5)$, where $e_a$ ($a = 1, \dots, 5$) denote the standard $S(U(1)^5)$ weights satisfying $\sum_{a=1}^5 e_a = 0$ and $\cW_R$ is the flat line bundle on $\hat{X}$ induced by the hypercharge Wilson line on the Standard Model representation $R$~\cite{Anderson:2012yf,Anderson:2013xka}.}
\label{tab:su5_branching}
\end{table}

Given a consistent background $(\hat{X}, V)$, the massless four-dimensional matter spectrum arises from the ten-dimensional $E_8$ gauge multiplet, which transforms in the adjoint representation $\mathbf{248}$. 
Decomposing the adjoint $\mathbf{248}$ of $E_8$ under $SU(5) \times S(U(1)^5)$ associates each low-energy multiplet to a cohomology group $H^1(\hat{X}, U \otimes \cW_R)$ on the quotient, or equivalently, to the $\Gamma$-equivariant subspace of $H^1(X, U)$ carrying the Wilson-line charge of $R$ upstairs, as summarized in Table~\ref{tab:su5_branching}. Whenever the Standard Model representation $R$ is clear from context, we leave the flat twist $\cW_R$ implicit and simply write $H^1(\hat{X}, U)$. Obtaining a realistic MSSM-like spectrum imposes conditions on both the chiral families and the Higgs sector. Getting three chiral generations without $\overline{\mathbf{10}}$ mirror families requires $\sum_{a=1}^5 \chi(X, L_a) = -3|\Gamma|$, $\chi(X, L_a) \equiv 0 \pmod{|\Gamma|}$, and $h^2(X, L_a) = 0$ for each $a$.\footnote{Any non-trivial line bundle of vanishing slope automatically satisfies $h^0(X, L_a) = h^3(X, L_a) = 0$, so $h^2(X, L_a) = 0$ sets $\chi(X, L_a) = -h^1(X, L_a)$ and excludes $\overline{\mathbf{10}}$ mirror generations.} In addition, the Wilson-line projection onto $\hat{X}$ must retain at least one pair of electroweak Higgs doublets $(H_u, H_d)$ while projecting out all color-triplet partners.

\subsection{Yukawa couplings}
\label{subsec:yukawa_couplings}

Dimensional reduction of the ten-dimensional $E_8 \times E_8$ heterotic theory on $\hat{X}$ yields a four-dimensional $\cN=1$ effective supergravity action for the massless chiral superfields $\Phi^I$~\cite{Strominger:1985ks,Lukas:1998tt,constantin2024computationquarkmassesstring}. Each superfield $\Phi^I$ in representation $R_I$ takes values in the associated line bundle 
\begin{equation}
    U_I \in \{L_a,\, L_a \otimes L_b,\, L_a^* \otimes L_b^*,\, L_a \otimes L_b^*\}
\end{equation}
of Table~\ref{tab:su5_branching} and corresponds to a harmonic bundle-valued $(0,1)$-form $\nu_I \in \cH^{0,1}(\hat{X}, U_I)$. Explicitly, $\nu_I$ satisfies the closedness and co-closedness conditions $\bar{\partial}_X \nu_I = \bar{\partial}_{X,H_I}^\dagger \nu_I = 0$ on $X$ with respect to the Ricci-flat K\"ahler metric $g_{\mathrm{CY}}$ and the slope-zero HYM bundle metric $H_I \in \{H_a,\, H_a H_b,\, (H_a H_b)^{-1},\, H_a H_b^{-1}\}$ induced from the five fundamental HYM metrics $H_a$ on $L_a$ with $\prod_{a=1}^5 H_a = 1$. For any gauge-invariant cubic superpotential coupling $W = \frac{1}{6} W_{IJK} \Phi^I \Phi^J \Phi^K$, where $U_I \otimes U_J \otimes U_K \cong \cO_X$ by $S(U(1)^5)$ gauge invariance, canonical normalization with the K\"ahler potential
\begin{equation}
  \cK = \cK_{\mathrm{mod}} + \cV_X^{-1/3}\,\cK_{I\bar{J}}\,\Phi^I \bar{\Phi}^{\bar{J}}\,, \qquad
  \cK_{\mathrm{mod}} = -\ln(S + \bar{S}) - \ln(8\cV_{\hat{X}}) - \ln\!\Big(\ii \int_{\hat{X}} \Omega \wedge \bar{\Omega}\Big)\,,
  \label{eq:kahler_potential}
\end{equation}
where $S$ is the axio-dilaton and $\cV_X=1/3!\int_X J^3$ is the CY volume, results in the physical Yukawa couplings~\cite{Cremmer:1982en,Dixon:1989fj,Candelas:1990pi,Kaplunovsky:1993rd}
\begin{equation}
  Y_{IJK} = e^{\cK_{\mathrm{mod}}/2}\,\cV_X^{1/2}\,\Big(\cK^{-1/2}\Big)_I^{I'} \Big(\cK^{-1/2}\Big)_J^{J'} \Big(\cK^{-1/2}\Big)_K^{K'} W_{I'J'K'}\,.
  \label{eq:4d_sugra_yukawa}
\end{equation}
Here $\cV_{\hat{X}} = \cV_X / |\Gamma|$ is the quotient volume, and the factor $\cV_X^{-1/3}$ is separated out in Eq.~\eqref{eq:kahler_potential} so that the matter field K\"ahler metric $\cK_{I\bar{J}}$ is dimensionless and invariant under overall volume rescalings~\cite{constantin2024computationquarkmassesstring}.

To express $W_{IJK}$ and $\cK_{I\bar{J}}$ as integrals over $X$ and fix all numerical prefactors, we relate the four-dimensional fields $\Phi^I$ to the underlying ten-dimensional heterotic theory~\cite{candelas1985vacuum,Strominger:1985ks,Lukas:1998tt,constantin2024computationquarkmassesstring}. In heterotic compactifications, both the visible-sector gauge bosons and all charged matter multiplets $\Phi^I$ originate from the ten-dimensional $E_8$ vector multiplet $(A_M, \Psi)$, governed by the super-Yang--Mills sector of the heterotic action,
\begin{equation}
  S_{\mathrm{YM,10D}} = -\frac{1}{4 g_{10}^2} \int_{M_4 \times \hat{X}} \dd^{10}x \sqrt{-G_{10}}\,\Big(\Tr F_{MN} F^{MN} + 2\bar{\Psi} \Gamma^M D_M \Psi\Big)\,.
  \label{eq:10d_sym_action}
\end{equation}
Matching the four-dimensional GUT gauge coupling $g_{\mathrm{GUT}}^{-2} = \cV_{\hat{X}} / g_{10}^2 = \operatorname{Re}S$, or equivalently $\alpha_{\mathrm{GUT}} = g_{\mathrm{GUT}}^2 / (4\pi)$, and comparing the Hodge-star inner product
\begin{equation}
  \nu_I \wedge \star_{g_{\mathrm{CY}}} \bar{\nu}_J = g_{\mathrm{CY}}^{a\bar{b}} \nu_{I,\bar{b}} \overline{\nu_{J,\bar{a}}}\, \dd\mathrm{Vol}_{\mathrm{CY}}
\end{equation}
of $(0,1)$-forms against the coordinate-volume convention used in numerical CY solvers~\cite{larfors2022numericalmetricscompleteintersection,constantin2024computationquarkmassesstring} fixes the overall normalization. Writing $\Omega = \Omega_{123}\,\dd z^1 \wedge \dd z^2 \wedge \dd z^3$ in local complex coordinates $z^a = x^a + \ii y^a$ and denoting by $\hat{g}_{\mathrm{CY}} = \cV_X^{-1/3} g_{\mathrm{CY}}$ the unit-volume Ricci-flat metric ($\int_X \dd\mathrm{Vol}_{\hat{g}_{\mathrm{CY}}} = 1$), we define
\begin{equation}
\begin{aligned}
  I_{\Omega} &:= 6 \int_X |\Omega_{123}|^2\,\dd^6 x = \frac{3\ii}{4} \int_X \Omega \wedge \bar{\Omega}\,, \\
  I_{IJK}(C) &:= 6 \int_X \Omega_{123} \det(\nu_I, \nu_J, \nu_K)\,\dd^6 x = \frac{3\ii}{4} \int_X \Omega \wedge \nu_I \wedge \nu_J \wedge \nu_K\,,
\end{aligned}
  \label{eq:raw_integral_conventions}
\end{equation}
where we indicate explicitly that $I_{IJK}$ depends on the complex structure moduli $C$. The dimensionless matter field K\"ahler metric~\cite{constantin2024computationquarkmassesstring} for zero modes $\nu_I, \nu_J$ valued in a common line bundle $U$ with HYM metric $H_U$ is
\begin{equation}
\begin{aligned}
  \cK_{I\bar{J}} &:= \frac{1}{4\,|\Gamma|\,\cV_X^{2/3}} \int_X H_U\,g_{\mathrm{CY}}^{a\bar{b}}\,\nu_{I,\bar{b}}\,\bar{\nu}_{\bar{J},a}\,\dd\mathrm{Vol}_{\mathrm{CY}} \\
  &\phantom{:}= \frac{1}{4\,|\Gamma|} \int_X H_U\,\hat{g}_{\mathrm{CY}}^{a\bar{b}}\,\nu_{I,\bar{b}}\,\bar{\nu}_{\bar{J},a}\,\dd\mathrm{Vol}_{\hat{g}_{\mathrm{CY}}}\,,
\end{aligned}
  \label{eq:dimensionless_matter_metric}
\end{equation}
where $\cK_{I\bar{J}} = 0$ when $U_I \neq U_J$. Dimensional reduction of the 10D super Yang--Mills action~\eqref{eq:10d_sym_action} fixes $W_{IJK} \propto I_{IJK}(C)$, and after canceling $\cV_X^{-1/2}$ from $e^{\cK_{\mathrm{mod}}/2}$ against $\cV_X^{1/2}$ in the physical Yukawa couplings $Y_{IJK}$~\eqref{eq:4d_sugra_yukawa}, we obtain
\begin{equation}
  Y_{IJK} = \frac{g_{\mathrm{GUT}}\,c_{IJK}}{8\,|\Gamma|^{3/2}\,\sqrt{6\,I_{\Omega}}}
  \Big(\cK^{-1/2}\Big)_I^{I'} \Big(\cK^{-1/2}\Big)_J^{J'} \Big(\cK^{-1/2}\Big)_K^{K'} I_{I'J'K'}(C)\,,\quad c_{IJK} = \sqrt{H_I H_J H_K}\,,
  \label{eq:exact_physical_yukawa}
\end{equation}
which simplifies to
\begin{equation}
  Y_{IJK} = \frac{g_{\mathrm{GUT}}\,c_{IJK}}{8\,|\Gamma|^{3/2}\,\sqrt{6\,I_{\Omega}}} \,
  \frac{I_{IJK}(C)}{\sqrt{\cK_{I\bar{I}}\,\cK_{J\bar{J}}\,\cK_{K\bar{K}}}}\,,
\end{equation}
when $\dim H^1(\hat{X}, U_I) = 1$. The factor of $1/\sqrt{6}$ in Eq.~\eqref{eq:exact_physical_yukawa} arises from the factor of $6$ relating $I_{\Omega}$ in Eq.~\eqref{eq:raw_integral_conventions} to the $(0,1)$-form inner product~\eqref{eq:dimensionless_matter_metric}. Moreover, the bundle-metric factor $c_{IJK}$ simplifies to unity under standard normalization: writing the slope-zero HYM metrics relative to the ambient Fubini--Study metric as $H_I = H_I^{\mathrm{FS}} e^{\beta_I}$ with $\Delta_{g_{\mathrm{CY}}} \beta_I \propto \star_{g_{\mathrm{CY}}} F_{I,\mathrm{FS}} \wedge J^2$, gauge invariance implies
\begin{equation}
  H_I^{\mathrm{FS}} H_J^{\mathrm{FS}} H_K^{\mathrm{FS}} = 1\qquad\text{and}\qquad \Delta_{g_{\mathrm{CY}}}(\beta_I + \beta_J + \beta_K) = 0\,.
\end{equation}
Thus, $c_{IJK}$ is constant on $X$, and by normalizing $\beta_I$ to zero mean on $X$, we get $c_{IJK} = 1$.

Note that since the exact data $(g_{\mathrm{CY}}, H_I, \nu_I^{\mathrm{harm}})$ admit no closed-form expressions on $X$, it is useful to relate Eq.~\eqref{eq:exact_physical_yukawa} to known analytic reference quantities $(g_{\mathrm{FS}}|_{TX}, H_I^{\mathrm{FS}}, \nu_I^{\mathrm{ref}})$ inherited from the ambient Fubini--Study geometry.  Specifically, since $\nu_I^{\mathrm{harm}}$ lies in the same cohomology class as the $\bar{\partial}_X$-closed reference form $\nu_I^{\mathrm{ref}} \in \Omega^{0,1}(X, U_I)$, any representative in this class takes the form $\nu_I(\sigma_I) = \nu_I^{\mathrm{ref}} + \bar{\partial}_X \sigma_I$ for some smooth section $\sigma_I \in \Omega^{0,0}(X, U_I)$. In the holomorphic numerator $I_{IJK}(C)$, this exact shift produces a total derivative: since $\Omega$, $\nu_J$, and $\nu_K$ are $\bar{\partial}_X$-closed, the identity
\begin{equation}
  \Omega \wedge (\bar{\partial}_X \sigma_I) \wedge \nu_J \wedge \nu_K = -\dd(\Omega \wedge \sigma_I \wedge \nu_J \wedge \nu_K)
\end{equation}
implies
\begin{equation}
  \int_X \Omega \wedge \nu_I(\sigma_I) \wedge \nu_J(\sigma_J) \wedge \nu_K(\sigma_K)
  = \int_X \Omega \wedge \nu_I^{\mathrm{ref}} \wedge \nu_J^{\mathrm{ref}} \wedge \nu_K^{\mathrm{ref}}\,,
  \label{eq:stokes_holomorphic_invariance}
\end{equation}
where we used that exact forms integrate to zero on compact spaces without boundary.
The holomorphic numerator $I_{IJK}(C)$ is therefore independent of $(g_{\mathrm{CY}}, H_I, \sigma_I)$. In the denominator, the harmonic representative $\nu_I^{\mathrm{harm}} = \nu_I^{\mathrm{ref}} + \bar{\partial}_X \sigma_I^*$ minimizes the $L^2(g_{\mathrm{CY}}, H_I)$ norm over $\sigma_I$ and is orthogonal to $\im(\bar{\partial}_X)$, thus
\begin{equation}
  \|\nu_I^{\mathrm{harm}}\|^2 = \|\nu_I^{\mathrm{ref}}\|^2 - \|\bar{\partial}_X \sigma_I^*\|^2 \le \|\nu_I^{\mathrm{ref}}\|^2\,.
  \label{eq:pythagorean_monotonicity}
\end{equation}
This means that the exact harmonic representatives decrease $\cK_{I\bar{I}}$ and hence increase $|Y_{IJK}|$ at fixed $(g_{\mathrm{CY}}, H_I)$. Finally, comparing the exact $(g_{\mathrm{CY}}, H_I)$ and Fubini--Study $(g_{\mathrm{FS}}|_{TX}, H_I^{\mathrm{FS}})$ inner products on $T^*X \otimes U_I$ against the common volume form $\dd\mathrm{Vol}_{\mathrm{CY}}$ via the pointwise eigenvalue bounds
\begin{equation}
\begin{aligned}
  C_{\min,I} &:= \inf_{p \in X} \Big[e^{\beta_I(p)}\,\lambda_{\min}\!\big(g_{\mathrm{CY}}^{-1}(p) g_{\mathrm{FS}}|_{TX}(p)\big)\Big]\,, \\
  C_{\max,I} &:= \sup_{p \in X} \Big[e^{\beta_I(p)}\,\lambda_{\max}\!\big(g_{\mathrm{CY}}^{-1}(p) g_{\mathrm{FS}}|_{TX}(p)\big)\Big]\,,
\end{aligned}
  \label{eq:Cmin_Cmax}
\end{equation}
gives
\begin{equation}
  C_{\min,I}\,\cK_{I\bar{I}}^{\mathrm{harm}}(\mathrm{FS}) \le \cK_{I\bar{I}}^{\mathrm{harm}}(\mathrm{CY}, \mathrm{HYM}) \le C_{\max,I}\,\cK_{I\bar{I}}^{\mathrm{harm}}(\mathrm{FS})\,,
  \label{eq:sandwich_inequality}
\end{equation}
where $\cK_{I\bar{I}}^{\mathrm{harm}}(\mathrm{FS})$ denotes the minimum of Eq.~\eqref{eq:dimensionless_matter_metric} over all $\sigma_I$ with $\nu_I = \nu_I^{\mathrm{ref}} + \bar{\partial}_X \sigma_I$ in the Fubini--Study metrics. As a result, we can tune the Yukawa couplings using reference quantities first, and estimate in which direction and by how much the physical Yukawa couplings will change once we include neural networks to approximate the exact physical Yukawa couplings (for machine learning details see Appendix~\ref{app:ml_setup}).

\subsection{Moduli dependence of Yukawa couplings}
\label{subsec:moduli_dependence}

Physical Yukawa couplings in heterotic line bundle models depend on three classes of moduli: the K\"ahler moduli $t^i$, the complex structure (CS) moduli $C$, and the bundle moduli $S_{ab}$. Examining how each enters the Yukawa couplings~\eqref{eq:exact_physical_yukawa} clarifies which moduli can generate fermion mass hierarchies at tree level.

By the heterotic worldsheet non-renormalization theorem~\cite{Witten:1986,Distler:1987}, the holomorphic superpotential couplings $W_{IJK}(C)$ are independent of the K\"ahler moduli $t^i$ to all orders in $\alpha'$ perturbation theory, receiving only non-perturbative worldsheet instanton corrections of order $\cO(e^{-2\pi t^i})$. Hence, at tree level, the K\"ahler moduli enter~\eqref{eq:exact_physical_yukawa} exclusively through the dimensionless matter field K\"ahler metrics $\cK_{I\bar{J}}(t, C)$ evaluated along the zero-slope locus~\eqref{eq:duy_slope_zero}. Under an overall rescaling of the K\"ahler parameters $t^i \mapsto \lambda t^i$ ($\lambda > 0$) at fixed four-dimensional gauge coupling $g_{\mathrm{GUT}}$, the Ricci-flat metric scales as $g_{\mathrm{CY},a\bar{b}} \mapsto \lambda g_{\mathrm{CY},a\bar{b}}$, and the volume thus scales as $\cV_X \mapsto \lambda^3 \cV_X$. The HYM and harmonic $(0,1)$-form equations do not scale with $\lambda$, and hence $\beta_I$ and $\nu_I^{\mathrm{harm}}$ are unchanged. Consequently, the unit-volume metric $\hat{g}_{\mathrm{CY}}$, the dimensionless matter metrics $\cK_{I\bar{J}}$ in Eq.~\eqref{eq:dimensionless_matter_metric}, and the tree-level physical Yukawa couplings $Y_{IJK}$ in Eq.~\eqref{eq:exact_physical_yukawa} are all invariant under overall volume scalings. However, the relative K\"ahler moduli $[t^1 : \dots : t^{h^{1,1}}]$ do enter the tree-level masses, while large enough $\lambda$ suppresses $\alpha'$ and worldsheet-instanton corrections.

In contrast, the complex structure moduli $C$ enter both the holomorphic couplings $I_{IJK}(C)$ and the matter K\"ahler metrics $\cK_{I\bar{J}}(t, C)$ through the coefficients of the defining polynomials $P_1, \dots, P_K$. Evaluating $\cK_{I\bar{J}}(t, C)$ requires solving the coupled PDEs on $X$ for the Ricci-flat metric $g_{\mathrm{CY}}$, the HYM metrics $H_I$, and the exact shifts $\bar{\partial}_X \sigma_I^*$ in the harmonic $(0,1)$-forms $\nu_I^{\mathrm{harm}} = \nu_I^{\mathrm{ref}} + \bar{\partial}_X \sigma_I^*$. On the other hand, the zero-mode cohomology $H^1(X, L|_X)$, the analytic reference $(0,1)$-forms $\nu_I^{\mathrm{ref}}$, and the holomorphic couplings $I_{IJK}(C)$ can all be computed algebraically via the Koszul sequence, which expresses bundle-valued sections on the submanifold
\begin{equation}
    X = \{P_1 = \dots = P_K = 0\} \subset \cA\,
\end{equation}
in terms of ambient bundles on $\cA$ modulo multiples of $P_1, \dots, P_K$. For an ambient line bundle $L = \cO_{\cA}(q)$, this yields the exact sheaf sequence~\cite{Candelas:1987rx,Anderson:2008uw,Blesneag_2016,Blesneag_2017}
\begin{equation}
  0 \longrightarrow \bigwedge^K \cN^* \otimes L \xrightarrow{\;d_K\;} \bigwedge^{K-1} \cN^* \otimes L \xrightarrow{\;d_K\;} \cdots \xrightarrow{\;d_K\;} \cN^* \otimes L \xrightarrow{\;d_K\;} L \xrightarrow{\phantom{\;d_K\;}} L|_X \longrightarrow 0\,,
  \label{eq:koszul_resolution_general}
\end{equation}
where $\cN^* = \bigoplus_{\alpha=1}^K \cO_{\cA}(-d_\alpha)$ is the conormal bundle carrying the opposite multidegrees $-d_\alpha$ of the defining polynomials, and $d_K$ contracts exterior generators of $\cN^*$ with $(P_1, \dots, P_K)$. 
Computing $H^k(X, L|_X)$ from Eq.~\eqref{eq:koszul_resolution_general} proceeds iteratively through a spectral sequence: a sequence of two-dimensional grids $E_1, E_2, \dots$, called pages, where each page $E_{r+1}$ is the cohomology $\ker d_r / \im d_r$ of a linear map $d_r$ acting on the previous page. Its first page collects the ambient cohomology groups of each term in Eq.~\eqref{eq:koszul_resolution_general},
\begin{equation}
  E_1^{-p,q}(L) = H^q\!\left(\cA, \bigwedge^p \cN^* \otimes L\right)\,, \qquad 0 \le p \le K\,,
  \label{eq:koszul_E1_page}
\end{equation}
which follow in closed form from Bott's formula and K\"unneth's theorem on $\cA = \prod_{r=1}^m \PP^{n_r}$ and do not yet depend on $C$. The complex structure dependence enters when we pass to subsequent pages by taking cohomology successively with respect to the differentials
\begin{equation}
  d_r : E_r^{-p,q}(L) \to E_r^{-p+r, q-r+1}(L)\,,
\end{equation}
which are polynomial maps in the complex structure coefficients $C$ of $P_1, \dots, P_K$. Since $d_r = 0$ for all $r > K$, the sequence stabilizes at $K+1$, $E_\infty^{-p,q}(L) = E_{K+1}^{-p,q}(L)$, and gives
\begin{equation}
  H^k(X, L|_X) \cong \bigoplus_{q - p = k} E_{K+1}^{-p,q}(L)\,.
\end{equation}
This spectral sequence behaves differently for chiral families and vector-like Higgs fields. For chiral matter bundles with $\chi(\hat{X}, L_a) = -1$ and $h^0 = h^2 = h^3 = 0$, the dimension $h^1(\hat{X}, L_a) = 1$ is constant across all smooth points of the complex structure moduli space $\cM_{\mathrm{CS}}(\hat{X})$, although the cohomology representatives inside $E_r^{-p,p+1}(L_a)$ vary algebraically with $C$. Vector-like Higgs bundles $(H_u, H_d)$, by contrast, have $\chi(\hat{X}, L_H) = 0$; when $h^1(\hat{X}, L_H) = 0$ at generic $C$, massless Higgs doublets arise on complex structure cohomology jumping loci $\cJ_{\mathrm{Higgs}} \subset \cM_{\mathrm{CS}}(\hat{X})$ along which a Koszul differential $d_r(C)$ drops in rank~\cite{Anderson:2010,Buchbinder_2014,Gray:2019}.

Finally, at the Abelian bundle split locus $V = \bigoplus_{a=1}^5 L_a$, all off-diagonal bundle moduli $S_{ab} \in H^1(\hat{X}, L_a \otimes L_b^*)$ ($a \neq b$) have vanishing VEVs, $\langle S_{ab} \rangle = 0$. Moreover, the Atiyah map $H^1(X, TX) \to H^2(X, \End_0(V))$ vanishes identically because $\bigoplus_a (L_a \otimes L_a^*) \cong \cO_X^{\oplus 5}$ and $H^2(X, \cO_X) = 0$. Lastly, the tree-level holomorphic Chern--Simons superpotential vanishes in the absence of three-form flux~\cite{Anderson:2010,Buchbinder_2014}. Turning on VEVs $\langle S_{ab} \rangle \neq 0$ for the bundle moduli deforms $V$ into a non-Abelian extension or monad bundle~\cite{Anderson:2010,Buchbinder_2014,Ashmore:2018}, generating both higher-dimensional operators $(S_{ab}/M_{\mathrm{KK}})^n\, \Phi^I \Phi^J \Phi^K$ that fill Yukawa entries forbidden by $S(U(1)^5)$ at the split locus and an F-term superpotential that can stabilize complex structure moduli onto cohomology jumping loci. Whereas heterotic Froggatt--Nielsen models~\cite{Constantin:2024fn,Constantin:2025fn} assume generic $\cO(1)$ geometric couplings and generate hierarchies from multi-scale singlet VEVs $\langle S_{ab} \rangle$ and instantons $\langle e^{-2\pi t^i} \rangle$, our focus here is on the tree-level hierarchies generated directly by the Calabi--Yau geometry~$\hat{X}$.

\subsection{Hierarchy mechanisms}
\label{subsec:two_mechanisms}

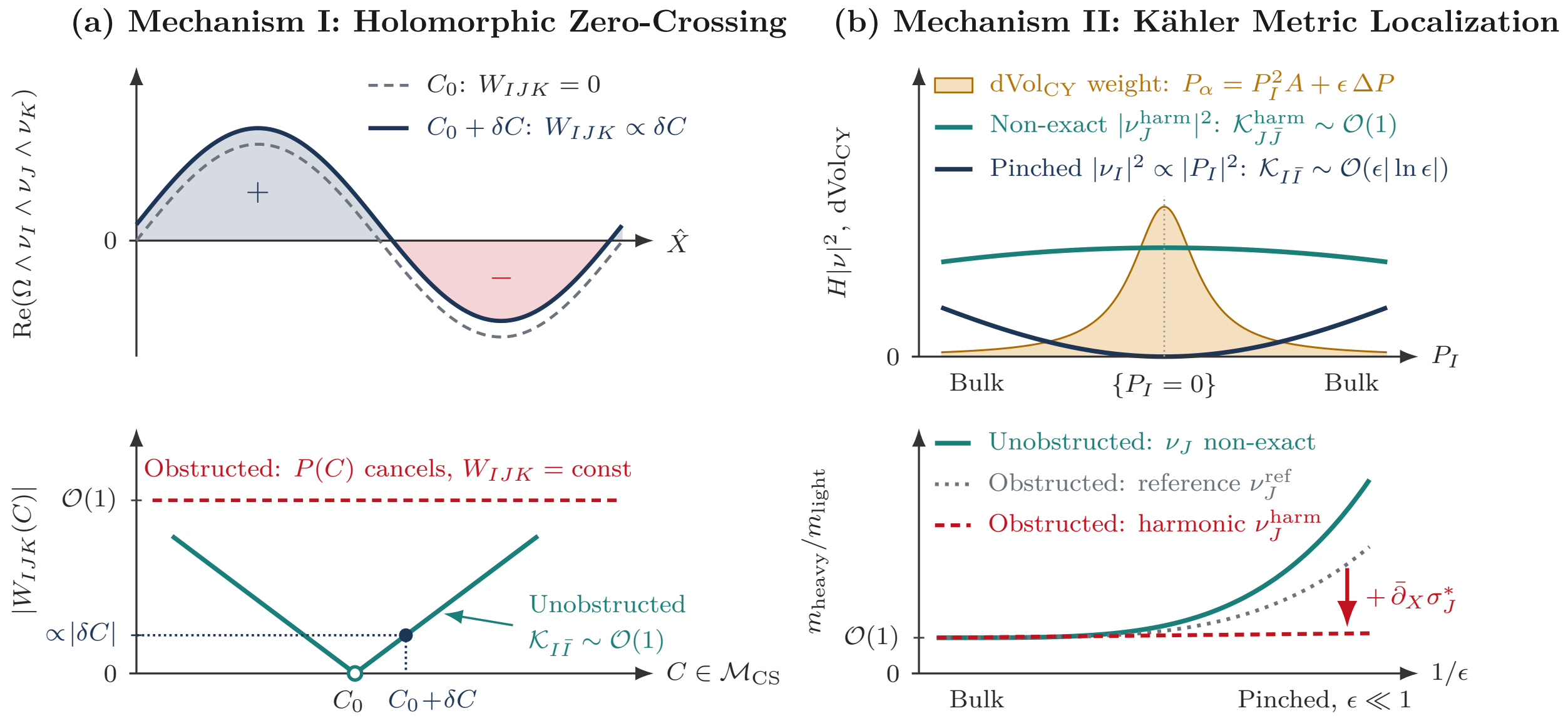
\begin{figure}[t]
\centering
\definecolor{mechNavy}{HTML}{1D3557}
\definecolor{mechCrimson}{HTML}{C1121F}
\definecolor{mechTeal}{HTML}{1B7F79}
\definecolor{mechAmber}{HTML}{D48806}
\definecolor{mechSlate}{HTML}{6C757D}
\resizebox{\textwidth}{!}{%
\begin{tikzpicture}[
    font=\footnotesize,
    >=Latex,
    axis/.style={->, line width=0.65pt, draw=black!80},
    lbl/.style={font=\scriptsize, text=black!85},
    paneltitle/.style={font=\small\bfseries, text=black!90}
]

\node[paneltitle, anchor=west] at (0.0, 7.78) {(a) Mechanism I: Holomorphic Zero-Crossing};

\begin{scope}[shift={(0.88, 5.35)}]
    \fill[mechNavy!18, domain=0:0.5265, smooth, variable=\s, samples=80]
        (0,0) -- plot ({5.45*\s}, {1.08*sin(360*\s) + 0.18}) -- ({5.45*0.5265}, 0) -- cycle;
    \fill[mechCrimson!18, domain=0.5265:0.9735, smooth, variable=\s, samples=80]
        ({5.45*0.5265}, 0) -- plot ({5.45*\s}, {1.08*sin(360*\s) + 0.18}) -- ({5.45*0.9735}, 0) -- cycle;
    \fill[mechNavy!18, domain=0.9735:1.0, smooth, variable=\s, samples=20]
        ({5.45*0.9735}, 0) -- plot ({5.45*\s}, {1.08*sin(360*\s) + 0.18}) -- (5.45, 0) -- cycle;

    \draw[axis] (0, -1.30) -- (0, 1.95);
    \draw[axis] (-0.08, 0) -- (5.80, 0) node[right, lbl] {$\hat{X}$};
    \node[lbl, anchor=east] at (-0.08, 0) {$0$};

    \draw[mechSlate, densely dashed, line width=0.95pt, domain=0:1.0, smooth, variable=\s, samples=100]
        plot ({5.45*\s}, {1.08*sin(360*\s)});
    \draw[mechNavy, line width=1.35pt, domain=0:1.0, smooth, variable=\s, samples=100]
        plot ({5.45*\s}, {1.08*sin(360*\s) + 0.18});

    \node[font=\small\bfseries, text=mechNavy] at (1.36, 0.54) {$+$};
    \node[font=\small\bfseries, text=mechCrimson] at (4.09, -0.42) {$-$};

    \node[lbl, rotate=90, anchor=south] at (-0.98, 0.30) {$\operatorname{Re}(\Omega \wedge \nu_I \wedge \nu_J \wedge \nu_K)$};

    \draw[mechSlate, densely dashed, line width=0.95pt] (2.60, 1.74) -- (3.05, 1.74);
    \node[lbl, anchor=west] at (3.12, 1.74) {$C_0$: $W_{IJK}=0$};
    \draw[mechNavy, line width=1.35pt] (2.60, 1.26) -- (3.05, 1.26);
    \node[lbl, anchor=west, text=mechNavy] at (3.12, 1.26) {$C_0+\delta C$: $W_{IJK} \propto \delta C$};
\end{scope}

\begin{scope}[shift={(0.88, 0.50)}]
    \draw[axis] (0, 0) -- (0, 2.75);
    \draw[axis] (-0.08, 0) -- (5.80, 0) node[right, lbl] {$C \in \cM_{\mathrm{CS}}$};
    \node[lbl, rotate=90, anchor=south] at (-0.98, 1.38) {$|W_{IJK}(C)|$};

    \node[lbl, anchor=east] at (-0.08, 0.0) {$0$};
    \draw[black!75, line width=0.55pt] (-0.06, 1.94) -- (0.06, 1.94);
    \node[lbl, anchor=east] at (-0.08, 1.94) {$\cO(1)$};

    \draw[mechCrimson, densely dashed, line width=1.15pt] (0.18, 1.94) -- (5.40, 1.94);
    \node[lbl, text=mechCrimson, anchor=south] at (2.82, 2.00) {Obstructed: $P(C)$ cancels, $W_{IJK}=\text{const}$};

    \draw[mechTeal, line width=1.45pt] (0.40, 1.54) -- (2.45, 0) -- (4.50, 1.54);

    \filldraw[fill=white, draw=mechTeal, line width=1.1pt] (2.45, 0) circle (2.1pt);
    \node[lbl, anchor=north] at (2.38, -0.06) {$C_0$};

    \draw[mechNavy, densely dotted, line width=0.75pt] (0, 0.428) -- (3.02, 0.428) -- (3.02, 0);
    \draw[mechNavy, line width=0.55pt] (-0.06, 0.428) -- (0.06, 0.428);
    \node[lbl, anchor=east, text=mechNavy] at (-0.08, 0.44) {$\propto\!|\delta C|$};
    \fill[mechNavy] (3.02, 0.428) circle (2.2pt);
    \node[lbl, anchor=north, text=mechNavy] at (3.32, -0.05) {$C_0\!+\!\delta C$};

    \draw[->, mechTeal, line width=0.7pt] (4.22, 0.54) -- (3.44, 0.67);
    \node[lbl, text=mechTeal, align=left, anchor=west] at (4.26, 0.54) {Unobstructed\\[2pt]$\cK_{I\bar{I}} \sim \cO(1)$};
\end{scope}

\node[paneltitle, anchor=west] at (8.55, 7.78) {(b) Mechanism II: K\"ahler Metric Localization};

\begin{scope}[shift={(9.65, 4.05)}]
    \fill[mechAmber!26, domain=-1.0:1.0, smooth, variable=\u, samples=140]
        ({2.75 - 2.50}, 0) -- plot ({2.75 + 2.50*\u}, {1.68*0.030/(\u*\u + 0.030)}) -- ({2.75 + 2.50}, 0) -- cycle;
    \draw[mechAmber!80!black, line width=0.6pt, domain=-1.0:1.0, smooth, variable=\u, samples=140]
        plot ({2.75 + 2.50*\u}, {1.68*0.030/(\u*\u + 0.030)});

    \draw[mechTeal, line width=1.5pt, domain=-1.0:1.0, smooth, variable=\u, samples=80]
        plot ({2.75 + 2.50*\u}, {1.22 - 0.16*\u*\u});

    \draw[mechNavy, line width=1.45pt, domain=-1.0:1.0, smooth, variable=\u, samples=100]
        plot ({2.75 + 2.50*\u}, {0.64*(\u*\u)/(1 + 0.16*\u*\u)});

    \draw[axis] (0, 0) -- (0, 3.25);
    \draw[axis] (-0.08, 0) -- (5.60, 0) node[right, lbl] {$P_I$};
    \draw[black!40, densely dotted, line width=0.55pt] (2.75, 0) -- (2.75, 1.80);

    \node[lbl, anchor=east] at (-0.08, 0) {$0$};
    \node[lbl, anchor=north] at (2.75, -0.05) {$\{P_I = 0\}$};
    \node[lbl, anchor=north] at (0.65, -0.05) {Bulk};
    \node[lbl, anchor=north] at (4.85, -0.05) {Bulk};

    \node[lbl, rotate=90, anchor=south] at (-0.62, 1.62) {$H|\nu|^2,\; \dd\mathrm{Vol}_{\mathrm{CY}}$};

    \filldraw[fill=mechAmber!26, draw=mechAmber!80!black, line width=0.55pt] (0.18, 2.96) rectangle (0.60, 3.12);
    \node[lbl, anchor=west, text=mechAmber!85!black] at (0.66, 3.04) {$\dd\mathrm{Vol}_{\mathrm{CY}}$ weight: $P_\alpha = P_I^2 A + \epsilon\,\Delta P$};

    \draw[mechTeal, line width=1.5pt] (0.18, 2.57) -- (0.60, 2.57);
    \node[lbl, anchor=west, text=mechTeal] at (0.66, 2.57) {Non-exact $|\nu_J^{\mathrm{harm}}|^2$: $\cK_{J\bar{J}}^{\mathrm{harm}} \sim \cO(1)$};

    \draw[mechNavy, line width=1.45pt] (0.18, 2.10) -- (0.60, 2.10);
    \node[lbl, anchor=west, text=mechNavy] at (0.66, 2.10) {Pinched $|\nu_I|^2 \propto |P_I|^2$: $\cK_{I\bar{I}} \sim \cO(\epsilon|\ln\epsilon|)$};
\end{scope}

\begin{scope}[shift={(9.65, 0.50)}]
    \draw[axis] (0, 0) -- (0, 2.75);
    \draw[axis] (-0.08, 0) -- (5.60, 0) node[right, lbl] {$1/\epsilon$};
    \node[lbl, rotate=90, anchor=south] at (-0.88, 1.38) {$m_{\mathrm{heavy}} / m_{\mathrm{light}}$};

    \node[lbl, anchor=east] at (-0.08, 0.0) {$0$};
    \draw[black!75, line width=0.55pt] (-0.06, 0.40) -- (0.06, 0.40);
    \node[lbl, anchor=east] at (-0.08, 0.40) {$\cO(1)$};

    \node[lbl, anchor=north] at (0.65, -0.05) {Bulk};
    \node[lbl, anchor=north] at (4.55, -0.05) {Pinched, $\epsilon \ll 1$};

    \draw[mechTeal, line width=1.5pt, domain=0.20:5.05, smooth, variable=\x, samples=90]
        plot (\x, {0.40 + 0.0032*(\x - 0.20)*(\x - 0.20)*(\x - 0.20)*(\x - 0.20)});

    \draw[mechSlate, dotted, line width=1.2pt, domain=0.20:5.05, smooth, variable=\x, samples=90]
        plot (\x, {0.40 + 0.00185*(\x - 0.20)*(\x - 0.20)*(\x - 0.20)*(\x - 0.20)});

    \draw[mechCrimson, densely dashed, line width=1.25pt] (0.20, 0.40) -- (5.05, 0.45);

    \draw[->, mechCrimson, line width=1.25pt] (4.80, 1.18) -- (4.80, 0.51);
    \node[lbl, text=mechCrimson, anchor=west] at (4.87, 0.85) {$+\,\bar{\partial}_X \sigma_J^*$};

    \draw[mechTeal, line width=1.5pt] (0.18, 2.58) -- (0.58, 2.58);
    \node[lbl, anchor=west, text=mechTeal] at (0.64, 2.58) {Unobstructed: $\nu_J$ non-exact};

    \draw[mechSlate, dotted, line width=1.2pt] (0.18, 2.12) -- (0.58, 2.12);
    \node[lbl, anchor=west, text=mechSlate!95!black] at (0.64, 2.12) {Obstructed: reference $\nu_J^{\mathrm{ref}}$};

    \draw[mechCrimson, densely dashed, line width=1.25pt] (0.18, 1.66) -- (0.58, 1.66);
    \node[lbl, anchor=west, text=mechCrimson] at (0.64, 1.66) {Obstructed: harmonic $\nu_J^{\mathrm{harm}}$};
\end{scope}

\end{tikzpicture}%
}
\caption{Schematic comparison of the two tree-level hierarchy mechanisms and their cohomological obstructions. \textbf{(a)}~In Mechanism~I, positive and negative lobes of $\operatorname{Re}(\Omega \wedge \nu_I \wedge \nu_J \wedge \nu_K)$ on $\hat{X}$ (top) cancel at a smooth zero-crossing locus $C_0 \in \cM_{\mathrm{CS}}$ (bottom), so displacing $C = C_0 + \delta C$ yields $|W_{IJK}| \propto |\delta C| \ll 1$ at $\cK_{I\bar{I}} \sim \cO(1)$; when $P(C)$ cancels out of the residue lift, $W_{IJK}$ is constant and the mechanism is obstructed. \textbf{(b)}~In Mechanism~II, approaching $P_\alpha = P_I^2 A + \epsilon\,\Delta P$ concentrates $\dd\mathrm{Vol}_{\mathrm{CY}}$ onto $\{P_I = 0\}$ (top), where $|\nu_I|^2 \propto |P_I|^2$ vanishes while a non-exact competing form $|\nu_J^{\mathrm{harm}}|^2$ stays $\cO(1)$. As pinching $1/\epsilon$ increases (bottom), $m_{\mathrm{heavy}}/m_{\mathrm{light}}$ grows in the unobstructed case, whereas if $\nu_J^{\mathrm{ref}}|_{\{P_I=0\}}$ is $\bar{\partial}_X$-exact, harmonic projection $+\bar{\partial}_X\sigma_J^*$ collapses the physical ratio back to $\cO(1)$.}
\label{fig:hierarchy_mechanisms}
\end{figure}

Let us discuss next how we can generate large mass hierarchies from tuning the complex structure moduli $C\in \cM_{\mathrm{CS}}(\hat{X})$. From Eq.~\eqref{eq:exact_physical_yukawa}, a large hierarchy at tree level can originate from either the holomorphic numerator $W_{IJK}(C)$ or the K\"ahler metric denominator $\cK_{I\bar{J}}(t, C)$. We refer to these two possibilities as Mechanism I and II, illustrated schematically in Figure~\ref{fig:hierarchy_mechanisms}.

\textbf{Mechanism I.} The first mechanism operates by identifying smooth zeros of the holomorphic superpotential couplings. Because the top-form integrand $\Omega \wedge \nu_I \wedge \nu_J \wedge \nu_K$ is complex-valued, $W_{IJK}(C)$ can vanish along a complex codimension-one locus $\{W_{IJK}(C) = 0\} \subset \cM_{\mathrm{CS}}(\hat{X})$ in the smooth interior of moduli space through phase or sign cancellation across the CY $\hat X$, even though none of the harmonic $(0,1)$-forms vanishes identically and all K\"ahler metrics $\cK_{I\bar{J}}(C)$ remain smooth and $\cO(1)$. Displacing the complex structure by $\delta C$ from smooth intersections of such zero-crossing loci yields parametric hierarchies $m_{\mathrm{light}} / m_{\mathrm{heavy}} \propto |\delta C|$ deep inside the supergravity regime, protected from perturbative $\alpha'$ corrections by worldsheet non-renormalization~\cite{Witten:1986,Distler:1987} and unaffected by metric corrections due to the topological nature of $W_{IJK}$.

However, for this holomorphic mechanism to work, $W_{IJK}(C)$ must depend non-trivially on the CS coefficients $C$ in addition to the overall dependence entering through $I_{\Omega}^{-1/2}$. The latter just rescales all Yukawa entries uniformly and hence cannot generate any hierarchies. Suppose the three participating fields originate at Koszul steps $p_I, p_J, p_K \in \{0, \dots, K\}$ in the spectral sequence~\eqref{eq:koszul_E1_page}. Descending from $E_1^{-p_I, p_I+1}(U_I)$ to $H^1(X, U_I)$ multiplies an ambient $(0, p_I+1)$-form by $p_I$ defining polynomials and inverts $p_I$ times the differential operator $\bar{\partial}_{\cA}$, whereas lifting $\int_X \Omega \wedge (\cdots)$ to $\cA$ via the Poincar\'e residue $\Omega = \Res_X\!\big(\Omega_{\cA} / \prod_{\alpha=1}^K P_\alpha\big)$ with $\Omega_{\cA} \in H^0(\mathcal{A}, \Omega^{D,0}_{\mathcal{A}} \otimes \bigwedge^K \mathcal{N})$ places $K$ defining polynomials in the denominator and integrates $\bar{\partial}_{\cA}$ by parts $K$ times~\cite{Blesneag_2016,Blesneag_2017}. Hence $W_{IJK}(C)$ vanishes identically unless $p_I + p_J + p_K \ge K$, and $I_{IJK}(C)$ can depend on $C$ in one of two ways:
\begin{enumerate}[label=(\roman*)]
  \item If $p_I + p_J + p_K > K$, uncanceled factors of $P_\alpha(C)$ survive in the ambient numerator.
  \item If at least one differential $d_r(C)$ in the spectral sequence of $U_I, U_J,$ or $U_K$ maps between two non-zero ambient cohomology spaces so that its kernel or cokernel representative varies with~$C$.
\end{enumerate}
Conversely, whenever $p_I + p_J + p_K = K$ with each $P_\alpha$ appearing once and all relevant Koszul differentials have trivial sources or targets, every $P_\alpha(C)$ cancels identically and $I_{IJK}$ reduces to a moduli-independent constant.

\textbf{Mechanism II.} The second mechanism generates hierarchies by tuning the denominator of $Y_{IJK}$. It localizes the K\"ahler metrics near loci where the defining polynomials factorize. Unlike $W_{IJK}(C)$, the integrand of $\cK_{I\bar{I}}$ in Eq.~\eqref{eq:dimensionless_matter_metric} is $H_I(p)\,|\nu_I(p)|_{g_{\mathrm{CY}}}^2 \ge 0$, so it cannot vanish by destructive interference and is $\cO(1)$ at a generic point in complex structure moduli space. To suppress the coupling pointwise, consider an ambient reference $(0,1)$-form $\nu_I^{\mathrm{ref}} = P_I\,\omega_I|_X$, where $\omega_I$ is a Fubini--Study harmonic $(0,1)$-form on a negative-degree $\PP^1$ factor and $P_I$ is a holomorphic polynomial on the remaining factors. Now suppose that a defining polynomial $P_\alpha$ approaches a completed-square factorization,
\begin{equation}
  P_\alpha(\epsilon) = P_I^2 A + \epsilon\,\Delta P\,,
\end{equation}
with $0 < \epsilon \ll 1$. As $\epsilon \to 0$, the CY volume measure
\begin{equation}
  \dd\mathrm{Vol}_{\mathrm{CY}} \propto \Omega \wedge \bar{\Omega} \propto \|\dd P_1 \wedge \cdots \wedge \dd P_K\|^{-2} \dd\mathrm{Vol}_{\mathrm{FS}}\,,
\end{equation}
concentrates onto the region where $|P_I|^2 = \cO(\epsilon)$, suppressing $\cK_{I\bar{I}}^{\mathrm{ref}} \sim \cO(\epsilon |\ln\epsilon|)$. By Eq.~\eqref{eq:pythagorean_monotonicity}, $\cK_{I\bar{I}}^{\mathrm{harm}} \le \cK_{I\bar{I}}^{\mathrm{ref}}$ then vanishes at least as fast as $\cO(\epsilon|\ln\epsilon|)$. For this suppression to produce a physical mass hierarchy relative to a different generation $\nu_J$, the competing harmonic norm $\cK_{J\bar{J}}^{\mathrm{harm}} = \min_{\sigma_J} \|\nu_J^{\mathrm{ref}} + \bar{\partial}_X \sigma_J\|^2$ must remain $\cO(1)$ rather than collapsing at the same rate, which requires the restriction of $\nu_J^{\mathrm{ref}}$ to the pinched locus $\{P_I = 0\}$ to remain cohomologically non-exact.

Whether this non-exactness holds depends on how the multidegrees of $P_I^2$ and $P_\alpha(\epsilon)$ overlap the ambient factors supporting $\nu_J$. Since $c_1(T\PP^1) = 2$ fixes the total polynomial degree on any $\PP^1$ factor to $2$, squaring a section $P_I$ of degree $1$ on a $\PP^1$ factor exhausts the Calabi--Yau degree budget on that factor. If $P_I^2$ saturates the degree of $P_\alpha(\epsilon)$ across all negative-degree $\PP^1$ factors carrying the anti-holomorphic form legs of $\nu_J$, this degree saturation can render $\nu_J^{\mathrm{ref}}|_{\{P_I = 0\}}$ $\bar{\partial}_X$-exact and collapse all $\cK_{I\bar{I}}^{\mathrm{harm}} \sim \cO(\epsilon |\ln\epsilon|)$ at the same time, as we show for the tetra-quadric in Section~\ref{subsec:tq_mech_ii}. In higher-codimension CICYs with $K \ge 2$, a defining polynomial $P_\alpha$ can instead carry degree $2$ on a single negative-degree $\PP^1$ factor of $U_J$ and degree $0$ on a second negative-degree $\PP^1$ factor of $U_J$, with its remaining degree absorbed by other ambient factors. As illustrated in Section~\ref{sec:cicy_5273}, this leaves the harmonic $(0,1)$-form leg of $\nu_J^{\mathrm{ref}}$ on the untouched $\PP^1$ factor unpinched and non-exact. Moreover, while taking $\epsilon \to 0$ in isolation drives the CICY toward high curvature, combining this mechanism with Mechanism I can generate multi-order-of-magnitude hierarchies well inside the smooth supergravity regime.

\subsection{Exact integration}
\label{subsec:exact_integration}

Evaluating the Yukawa couplings~\eqref{eq:exact_physical_yukawa} numerically across five orders of magnitude in the quark masses poses very different integration problems in the denominator and the numerator. The K\"ahler metrics $\cK_{I\bar{J}}$ and the volume integral $I_{\Omega}$ have smooth, positive integrands on $\hat X$, so Monte Carlo or spherical quadrature evaluates them reliably to sub-percent accuracy. In contrast, reproducing $m_c / m_t \sim 3 \times 10^{-3}$ and $m_u / m_t \sim 5 \times 10^{-6}$ via Mechanism~I requires the complex integrand of $I_{IJK}(C)$, which remains $\cO(1)$ pointwise, to cancel to three to five orders of magnitude when integrated over $\hat X$. Resolving such delicate cancellations via Monte Carlo sampling is not feasible, since statistical noise or undersampling even a small region spoils the cancellation. Similarly, cubature grids encounter branch-point singularities where quadratic roots coalesce, leading to numerically unstable results.

For the holomorphic integral $I_{IJK}(C)$, Eq.~\eqref{eq:stokes_holomorphic_invariance} allows us to avoid numerical integration altogether: since the smooth, exact correction terms $\bar{\partial}_X \sigma_I$ drop out identically, $I_{IJK}(C)$ depends only on $C$ and the cohomology classes $[\nu_I^{\mathrm{ref}}] \in H^1(\hat{X}_C, U_I)$. To perform the integration analytically, we extend the algebraic residue method of~\cite{Blesneag_2016,Blesneag_2017} to arbitrary Koszul steps $1 \le r_{\max} \le K$ on higher-codimension CICYs. We summarize the procedure here, while more technical details and exact closed-form expressions are given in Appendix~\ref{app:koszul}.

We first introduce the Koszul--Dolbeault double complex on $\cA$,
\begin{equation}
  K^{-r,q}(U) := \Omega^{0,q}\!\left(\cA, \bigwedge^r \cN^* \otimes U\right)\,,
  \label{eq:koszul_dolbeault_double_complex}
\end{equation}
equipped with commuting differentials
\begin{equation}
  \bar{\partial}_{\cA} : K^{-r,q} \to K^{-r,q+1} \qquad\text{and}\qquad d_K : K^{-r,q} \to K^{-r+1,q}\,.
\end{equation}
These can be combined into the total differential $\mathcal{D}$ with $\mathcal{D}^2 = 0$,
\begin{equation}
  \mathcal{D} = \bar{\partial}_{\cA} + \tilde{d}_K\qquad\text{with}\qquad \tilde{d}_K|_{K^{-r,q}} = (-1)^q d_K\,.
\end{equation}
By exactness of the Koszul sequence~\eqref{eq:koszul_resolution_general}, $H^1_{\mathrm{tot}}(K^{\bullet,\bullet}(U_I)) \cong H^1(X, U_I|_X)$, so each class
\begin{equation}
  [\nu_I^{\mathrm{ref}}] = [\Psi_I^{(0,1)}\big|_X]
\end{equation}
lifts to a $\mathcal{D}$-closed $1$-cocycle
\begin{equation}
  \Psi_I = \sum_{r=0}^{r_{\max}(I)} \Psi_I^{(-r, r+1)}\,, \qquad \Psi_I^{(-r, r+1)} \in K^{-r, r+1}(U_I)\,,
  \label{eq:total_cocycle_lift}
\end{equation}
obeying the staircase descent equations
\begin{equation}
  \bar{\partial}_{\cA} \Psi_I^{(-r, r+1)} = (-1)^{r+1} d_K \Psi_I^{(-r-1, r+2)}\,, \qquad
  \bar{\partial}_{\cA} \Psi_I^{(-r_{\max}, r_{\max}+1)} = 0\,.
  \label{eq:staircase_equations}
\end{equation}
We solve these equations algebraically by inverting $\bar{\partial}_{\cA}$ using the Fubini--Study Green's operators, which map $\bar{\partial}$-exact $(0,q)$-forms to co-exact $(0,q-1)$-forms, see Appendix~\ref{app:subsec:homotopies} for details and expressions in closed polynomial form.

To lift the computation of $I_{IJK}(C)$ to the ambient space $\cA$, we multiply $\Psi_I$, $\Psi_J$, and $\Psi_K$ via the graded wedge product,
\begin{equation}
  \Xi_{IJK} = \Psi_I \star \Psi_J \star \Psi_K\,,
\end{equation}
wedging both the differential-form legs and the $\bigwedge^\bullet \cN^*$ indices (with the graded signs given in Appendix~\ref{app:koszul}). Writing the holomorphic top form $\Omega$ as the Poincar\'e residue of $\Omega_{\cA} / (P_1 \cdots P_K)$ and integrating $\bar{\partial}_{\cA}$ by parts $K$ times via Eq.~\eqref{eq:staircase_equations}, each polynomial factor $P_\alpha$ in $d_K$ cancels the corresponding pole $1/P_\alpha$ up to a factor of $2\pi\ii$. The top-dimensional component $\Xi_{IJK}^{(-K,D)}$ with $D = \dim_{\CC}\cA = K + 3$ saturates all $K$ indices and defines a $(0,D)$-form $\xi_{IJK}^{(0,D)} \in \Omega^{0,D}(\cA, \bigwedge^K\mathcal{N}^*)$, which contributes to the integral
\begin{equation}
  \int_{X_C} \Omega \wedge \nu_I^{\mathrm{ref}} \wedge \nu_J^{\mathrm{ref}} \wedge \nu_K^{\mathrm{ref}}
  = \frac{1}{(2\pi\ii)^K} \int_{\cA} \Omega_{\cA} \wedge \xi_{IJK}^{(0,D)}\,.
  \label{eq:poincare_residue_lift_master}
\end{equation}
For a term in $\Psi_I \star \Psi_J \star \Psi_K$ to reach bidegree $(-K, D)$, the Koszul indices of the three factors must partition $\{1, \dots, K\}$ and their anti-holomorphic form degrees must sum to $n_r$ on each $\PP^{n_r}$ factor, so most intermediate staircase terms in Eq.~\eqref{eq:total_cocycle_lift} drop out by degree counting and need not be computed. Since the Fubini--Study Green's operators on $\cA = \prod_{r=1}^m \PP^{n_r}$ map monomials to monomials, the integral in~\eqref{eq:poincare_residue_lift_master} evaluates in closed form to a rational function of the polynomial coefficients $C$. Finally, when $\Psi_I$ and the analytic reference $(0,1)$-form $\nu_I^{\mathrm{ref}}$ entering $\cK_{I\bar{J}}$ are normalized independently, we match their cohomology classes via $[\nu_I^{\mathrm{ref}}] = s_I(C)\,[\Psi_I^{(0,1)}|_X]$ so that the numerator and denominator of Eq.~\eqref{eq:exact_physical_yukawa} are evaluated in the same basis.

\section{The Tetra-Quadric}
\label{sec:tetraquadric}
The tetra-quadric has been previously used to study quark masses from the heterotic string~\cite{Buchbinder_2014,Blesneag_2016,constantin2024computationquarkmassesstring}, but no scan or attempt to get realistic quark masses has been undertaken. We illustrate our techniques in this model and show analytically why both tree-level hierarchy mechanisms are obstructed at the split-bundle locus. We also confirm these obstructions numerically across the model's $20$-dimensional complex structure moduli space.

\subsection{Geometry and spectrum}
\label{subsec:tq_geometry}

The tetra-quadric CICY 7862 is a codimension-one hypersurface $X \subset \cA = (\PP^1)^4$ of multidegree $(2,2,2,2)$ given by the zero locus $P(x,y,u,v) = 0$, where $(x,y,u,v)$ are homogeneous coordinates on $(\PP^1)^4$. The manifold allows for a freely acting group action $\Gamma = \ZZ_2 \times \ZZ_2$ generated by simultaneous coordinate inversions and permutations~\cite{Buchbinder_2014,Blesneag_2016}. The upstairs geometry $X$ before the free quotient has Hodge numbers $(h^{1,1}, h^{2,1}) = (4, 68)$, while the quotient manifold $\hat{X} = X/\Gamma$ has Hodge numbers $(4, 20)$. The most general $\Gamma$-invariant defining polynomial $P(x,y,u,v) = \sum_{i=1}^{21} C_i \mathfrak{m}_i(x,y,u,v)$ is a linear combination of $21$ invariant monomials $\mathfrak{m}_i$; subtracting one overall rescaling leaves $20$ complex structure moduli.

Following~\cite{Buchbinder_2014,Blesneag_2016,constantin2024computationquarkmassesstring}, the visible-sector rank-five bundle $V = \bigoplus_{a=1}^5 L_a$ on $X$ is chosen as
\begin{equation}
\begin{aligned}
  L_1 &= \cO_X(-1,0,0,1)\,, \qquad
  L_2 = \cO_X(-1,-3,2,2)\,, \qquad
  L_3 = \cO_X(0,1,-1,0)\,, \\
  L_4 &= \cO_X(1,1,-1,-1)\,, \hspace{5mm}
  L_5 = \cO_X(1,1,0,-2)\,,\hspace{9.5mm}
  L_H = L_2^* \otimes L_5^*\,,
\end{aligned}
  \label{eq:tq_bundles}
\end{equation}
which obeys $c_1(V) = 0$ and $c_2(V) \le c_2(TX) = (24,24,24,24)$, while the slope-zero conditions~\eqref{eq:duy_slope_zero} restrict the K\"ahler moduli to the symmetric ray $t^1 = t^2 = t^3 = t^4 = t$.

Upon breaking $SU(5) \to G_{\mathrm{SM}}$ by a $\ZZ_2 \times \ZZ_2$ Wilson line, two chiral $\mathbf{10}$ families $(Q_2^a, U_2^a)$ ($a=1,2$) arise from $H^1(\hat{X}, L_2)$, the third family $(Q_5, U_5)$ arises from $H^1(\hat{X}, L_5)$, and the unique up-type Higgs doublet $H_u$ arises from $H^1(\hat{X}, L_H)$. Gauge invariance in the up-quark superpotential then permits only couplings of fields transforming in $(L_2, L_5, L_2^* \otimes L_5^*)$, which leads to the rank-two holomorphic up-type Yukawa matrix
\begin{equation}
  W^u(C) = \begin{pmatrix}
    0 & 0 & \lambda_1(C) \\
    0 & 0 & \lambda_2(C) \\
    \lambda_3(C) & \lambda_4(C) & 0
  \end{pmatrix}\,, \qquad
  \lambda_Q := \begin{pmatrix} \lambda_1 \\ \lambda_2 \end{pmatrix}\,, \qquad
  \lambda_U := \begin{pmatrix} \lambda_3 \\ \lambda_4 \end{pmatrix}\,.
  \label{eq:tq_bordered_yukawa}
\end{equation}
In the down-type sector, two chiral $\overline{\mathbf{5}}$ families $d^c_{1,2}$ arise from $H^1(\hat{X}, L_4 \otimes L_5)$, the third family $d^c_3$ arises from $H^1(\hat{X}, L_2 \otimes L_4)$, and the down-type Higgs doublet $H_d$ arises from $H^1(\hat{X}, L_2 \otimes L_5)$. Since with these charges none of the $Q\, d^c H_d$ couplings are gauge-invariant, all tree-level down-quark Yukawa couplings vanish identically at the split locus, giving $Y^d(C) = 0$ ($m_b = m_s = m_d = 0$) prior to turning on bundle-moduli singlet VEVs~\cite{Buchbinder_2014}.

At the split locus, the up-sector matter field K\"ahler metric takes the block-diagonal form $\cK = \operatorname{diag}(\cK_H,\allowbreak \cK_{Q_5},\allowbreak \cK_{U_5},\allowbreak \cK^Q,\allowbreak \cK^u)$, where $\cK^Q$ and $\cK^u$ are the $2 \times 2$ Hermitian blocks for $(Q_2^1, Q_2^2)$ and $(U_2^1, U_2^2)$, respectively. Canonically normalizing~\eqref{eq:tq_bordered_yukawa} gives $\rk Y^u = 2$ ($m_u = 0$ identically) and yields the tree-level top-to-charm mass ratio
\begin{equation}
  \frac{m_t(C)}{m_c(C)}
  = \sqrt{\frac{\cK_{U_5}(C)}{\cK_{Q_5}(C)}} \;
    \sqrt{\frac{\lambda_U(C)^\dagger \big(\cK^u(C)\big)^{-1} \lambda_U(C)}{\lambda_Q(C)^\dagger \big(\cK^Q(C)\big)^{-1} \lambda_Q(C)}}\,.
  \label{eq:tq_exact_ratio}
\end{equation}
Note that $\cK_H(C)$ and the overall K\"ahler modulus $t$ cancel in this expression.

\subsection{Holomorphic couplings and Hierarchy Mechanism I}
\label{subsec:tq_mech_i}
We first compute the holomorphic couplings $\lambda_1, \dots, \lambda_4$ in Eq.~\eqref{eq:tq_bordered_yukawa} to test Mechanism~I. For the tetra-quadric, as for any codimension-one hypersurface, the short exact Koszul sequence is simply
\begin{equation}
  0 \to \cN^* \otimes L \xrightarrow{P} L \to L|_X \to 0\,,
\end{equation}
with $\cN^* = \cO_{\cA}(-2,-2,-2,-2)$. In the associated long exact sequence in cohomology, the contributions to $H^1(X, L)$ on the CY can come either from $H^1$ ($p = 0$) or from $H^2$ ($p = 1$) in the ambient space~\cite{Blesneag_2016},
\begin{equation}
  H^1(X, L) \cong \coker\!\Big(H^1(\cA, \cN^* \otimes L) \xrightarrow{P} H^1(\cA, L)\Big)
  \oplus \ker\!\Big(H^2(\cA, \cN^* \otimes L) \xrightarrow{P} H^2(\cA, L)\Big)\,.
  \label{eq:tq_type1_type2}
\end{equation}
In~\eqref{eq:tq_bundles}, both $L_H = \cO_X(0,2,-2,0)$ and $L_5 = \cO_X(1,1,0,-2)$ satisfy
\begin{equation}
  H^1(\cA, \cN^* \otimes L) = H^2(\cA, \cN^* \otimes L) = 0\,,
\end{equation}
and therefore originate at Koszul step $p_H = p_5 = 0$, with $P$-independent ambient $(0,1)$-form representatives
\begin{equation}
  \hat{\nu}_H = \kappa_3^{-2}(y_0 y_1)\,\bar{\mu}_3\,, \qquad
  \hat{\nu}_{Q_5} = \kappa_4^{-2} P_{Q_5}(x,y)\,\bar{\mu}_4\,, \qquad
  \hat{\nu}_{U_5} = \kappa_4^{-2} P_{U_5}(x,y)\,\bar{\mu}_4\,,
  \label{eq:tq_type1_forms}
\end{equation}
where for $z_r \in \{x, y, u, v\}$ ($r = 1,2,3,4$) we write
\begin{equation}
  \kappa_r = |z_{r,0}|^2 + |z_{r,1}|^2\,, \qquad
  \mu_r = z_{r,0}\dd z_{r,1} - z_{r,1}\dd z_{r,0}\,,\qquad
  P_{Q_5, U_5}(x,y) = y_0 x_1 \pm y_1 x_0\,.
\end{equation}
Conversely, $L_2 = \cO_X(-1,-3,2,2)$ carries degree $-1$ on $\PP^1_1$, hence
\begin{equation}
  H^1(\cA, L_2) = H^2(\cA, L_2) = 0\,.
\end{equation}
The two families $(Q_2^a, U_2^a)$ therefore originate at Koszul step $p_2 = 1$, descending from the ambient cohomology group $H^2(\cA, \cO_{\cA}(-3,-5,0,0))$ via
\begin{equation}
  \bar{\partial}_{\cA} \hat{\nu}_2 = P\,\hat{\omega}_2\,,
\end{equation}
where $(\hat{\nu}_2, \hat{\omega}_2)$ stands collectively for $(\hat{\nu}_\Phi, \hat{\omega}_\Phi)$ with $\Phi \in \{Q_2^1, Q_2^2, U_2^1, U_2^2\}$. Since the ambient space cohomology group $H^2(\cA, L_2) = 0$ vanishes identically, the ambient $(0,2)$-forms
\begin{equation}
  \hat{\omega}_{Q_2^a, U_2^a} = \kappa_1^{-3} \kappa_2^{-5} \tilde{S}_{Q_2^a, U_2^a}(\bar{x}, \bar{y})\,\bar{\mu}_1 \wedge \bar{\mu}_2\,,
\end{equation}
with
\begin{equation}
  \tilde{S}_{Q_2^1, U_2^1} = \mp \bar{x}_0 \bar{y}_0^3 + \bar{x}_1 \bar{y}_1^3\,,\qquad
  \tilde{S}_{Q_2^2, U_2^2} = \mp \bar{x}_0 \bar{y}_0 \bar{y}_1^2 + \bar{x}_1 \bar{y}_1 \bar{y}_0^2\,,
  \label{eq:StildeDef}
\end{equation}
are completely independent of $P(C)$. Thus $p_H + p_5 + p_2 = 0 + 0 + 1 = 1 = K$ while all Koszul kernel and cokernel maps are trivial.
To see how $P(C)$ drops out as predicted in Section~\ref{subsec:two_mechanisms}, we denote restrictions to $X$ without hats ($\nu_I = \hat{\nu}_I|_{X}$) and write $\hat{\nu}_5 \in \{\hat{\nu}_{Q_5}, \hat{\nu}_{U_5}\}$ for the gauge-allowed pairs $(\nu_5, \nu_2) \in \{(\nu_{Q_5}, \nu_{U_2^a}),\, (\nu_{U_5}, \nu_{Q_2^a})\}$. Using the $(1,1)$-current $\delta_X$ of integration along $X = \{P = 0\}$,
\begin{equation}
  \bar{\partial}_{\cA}\left(\frac{\Omega_{\cA}}{2\pi\ii P}\right) = \Omega \wedge \delta_{X}\,,
\end{equation}
we lift the superpotential integral to $\cA = (\PP^1)^4$, which cancels $P(C)$ identically~\cite{Blesneag_2016},
\begin{equation}
  \int_{X_C} \Omega \wedge \nu_H \wedge \nu_5 \wedge \nu_2
  = -\frac{1}{2\pi\ii} \int_{\cA} \frac{\Omega_{\cA}}{P} \wedge \hat{\nu}_H \wedge \hat{\nu}_5 \wedge (P\,\hat{\omega}_2)
  = -\frac{1}{2\pi\ii} \int_{\cA} \Omega_{\cA} \wedge \hat{\nu}_H \wedge \hat{\nu}_5 \wedge \hat{\omega}_2\,,
  \label{eq:tq_poincare_lelong}
\end{equation}
and yields the moduli-independent holomorphic Yukawa matrix
\begin{equation}
  W^u \propto \frac{\ii \pi^3}{6} \begin{pmatrix}
    0 & 0 & 0 \\
    0 & 0 & 1 \\
    0 & 1 & 0
  \end{pmatrix}\,,
  \qquad
  \lambda_1(C) = \lambda_3(C) = 0\,, \qquad \lambda_2(C) = \lambda_4(C) \neq 0\,.
  \label{eq:tq_exact_topological_yukawa}
\end{equation}
Since the holomorphic Yukawa matrix does not depend on the complex structure moduli, Mechanism~I cannot work. Moreover, the ambient integral in~\eqref{eq:tq_poincare_lelong} inherits an accidental $SL(2,\CC)_1 \times SL(2,\CC)_2$ symmetry on the first two ambient $\PP^1$ factors, which is broken on the CY. Under the Cartan subgroup $U(1)_2 \subset SL(2,\CC)_2$ with $(y_0, y_1)$ of charges $(+1, -1)$, $\Omega_{\cA}$ and $\hat{\nu}_H$ have charge $0$, whereas the constituent monomials of $P_{Q_5, U_5}$ and $\tilde{S}_{Q_2^a, U_2^a}$ carry $U(1)_2$ charges
\begin{equation}
  P_{Q_5, U_5}: \{\pm 1\}\,,\qquad
  \tilde{S}_{Q_2^2, U_2^2}: \{\pm 1\}\,,\qquad
  \tilde{S}_{Q_2^1, U_2^1}: \{\pm 3\}\,.
\end{equation}
Even though $(Q_2^1, Q_2^2)$ share the same line bundle $L_2$ and satisfy $p_H + p_5 + p_2 = K$, which avoids the cohomological vanishing conditions of~\cite{Blesneag_2016,Gray:2019,Anderson:2021}, the couplings $\lambda_1$ and $\lambda_3$ of $(Q_2^1, U_2^1)$ still vanish by conservation of this $U(1)_2$ charge. Meanwhile, the element $\operatorname{diag}(\ii, -\ii) \in SL(2,\CC)_1$ acts as $(x_0, x_1) \mapsto (\ii x_0, -\ii x_1)$ and maps $P_{Q_5}\tilde{S}_{U_2^2} \mapsto +P_{U_5}\tilde{S}_{Q_2^2}$ while leaving $\Omega_{\cA}$ and the measure factors invariant, enforcing $\lambda_4(C) = \lambda_2(C)$. The ratio between the two non-vanishing up-type quark masses~\eqref{eq:tq_exact_ratio} thus simplifies across the entire CS moduli space $\cM_{\mathrm{CS}}(\hat{X})$ to a pure ratio of matter field K\"ahler metrics,
\begin{equation}
  \frac{m_t(C)}{m_c(C)} = \sqrt{\frac{\cK_{U_5}(C)}{\cK_{Q_5}(C)}\,\frac{\big((\cK^u(C))^{-1}\big)_{22}}{\big((\cK^Q(C))^{-1}\big)_{22}}}\,.
  \label{eq:tq_pure_kahler_ratio}
\end{equation}
With $W^u$ independent of $C$ and Mechanism~I ruled out, any tree-level hierarchy $m_t / m_c \gg 1$ in this model must come from Mechanism~II through the K\"ahler metrics in Eq.~\eqref{eq:tq_pure_kahler_ratio}.

\subsection{K\"ahler metric localization and Hierarchy Mechanism II}
\label{subsec:tq_mech_ii}

Under the coordinate reflection $\cP_x : (x_0, x_1) \mapsto (x_0, -x_1)$, the ambient polynomials transform as $ P_{Q_5} \leftrightarrow -P_{U_5}$ and $\tilde{S}_{Q_2^a} \leftrightarrow -\tilde{S}_{U_2^a}$. On any $\cP_x$-symmetric locus such as the one-parameter Dwork family studied in~\cite{constantin2024computationquarkmassesstring}, $\cK_{Q_5} = \cK_{U_5}$ and $\cK^Q = \cK^u$ hold identically, enforcing $m_t / m_c = 1$. To break $\cP_x$ and test Mechanism~II of Section~\ref{subsec:two_mechanisms}, consider a one-parameter family of defining polynomials 
\begin{equation}
    P(\epsilon) = P(0) + \epsilon\,\Delta P(x,y,u,v)\,,\qquad 0 < \epsilon \ll 1\,,
\end{equation}
designed to suppress $\cK_{Q_5}$ by concentrating the hypersurface near $\{P_{Q_5}(x,y) = y_0 x_1 + y_1 x_0 = 0\}$. If $P(0)$ vanished only to first order along $\{P_{Q_5} = 0\}$, so that $P(0) = P_{Q_5}\,B$ for some polynomial $B$ different from $P_{Q_5}$, the variety would be reducible and along the $\{B=0\}$ component one would have $|P_{Q_5}| = \cO(1)$ generically. This means that the integral $\cK_{Q_5}$ is suppressed only logarithmically from the divergence of $I_{\Omega}$ along $\{P_{Q_5} = B = 0\}$. Power-law suppression of $\cK_{Q_5}$ therefore requires $P(0)$ to vanish to second order along $\{P_{Q_5} = 0\}$, so that $P(0) = P_{Q_5}(x,y)^2 A$. Since $P(0)$ has multidegree $(2,2,2,2)$ and $P_{Q_5}^2$ already has multidegree $(2,2,0,0)$, the factor $A$ must have multidegree $(0,0,2,2)$ and hence depends only on $(u,v)$:
\begin{equation}
  P(\epsilon) = (y_0 x_1 + y_1 x_0)^2 A(u,v) + \epsilon\,\Delta P(x,y,u,v)\,, \qquad 0 < \epsilon \ll 1\,.
  \label{eq:tq_completed_square}
\end{equation}
As $\epsilon \to 0$, the CY volume form concentrates onto the region $|P_{Q_5}| = \cO(\sqrt\epsilon)$. If we evaluate the K\"ahler metrics using the reference $(0,1)$-forms $\nu_I^{\mathrm{ref}}$ (i.e., we set the NN corrections $\sigma_I = 0$), this suppresses $\cK_{Q_5}^{\mathrm{ref}} \sim \cO(\epsilon |\ln\epsilon|)$ while $\cK_{U_5}^{\mathrm{ref}}, \cK_{Q_2^2}^{\mathrm{ref}}, \cK_{U_2^2}^{\mathrm{ref}}$ remain $\cO(1)$, producing a large apparent hierarchy $(m_t / m_c)_{\mathrm{ref}} \propto (\epsilon|\ln\epsilon|)^{-1/2} \gg 1$. 
However, this hierarchy is an artifact of $\bar{\partial}_X$-exact terms in $\nu_{Q_2^2}^{\mathrm{ref}}$ that are removed once we pass to harmonic representatives. Solving
\begin{equation}
    \bar{\partial}_{\PP^1_1} \hat{\nu}_2 = P(0)\,\hat{\omega}_2
\end{equation}
via the Fubini--Study Green's operator on $\PP^1_1$ with $P(0) = P_{Q_5}^2 A(u,v)$, and pulling back to $\{P_{Q_5} = 0\}$ via $[x_0 : x_1] = [y_0 : -y_1]$ gives very different cohomology classes for $Q_2^2$ and $U_2^2$. For $Q_2^2$, the relative minus sign in $\tilde{S}_{Q_2^2}$ in~\eqref{eq:StildeDef} produces $|y_0|^2 - |y_1|^2$, so although the pulled-back reference form $\nu_{Q_2^2}^{\mathrm{ref}}$ is pointwise non-zero along $\{P_{Q_5}=0\}$, it is $\bar{\partial}$-exact on $\PP^1_2$:
\begin{equation}
  \nu_{Q_2^2}^{\mathrm{ref}}\Big|_{\{P_{Q_5}=0\}}
  = -\frac{1}{2} A(u,v)\,\frac{\bar{y}_0 \bar{y}_1(|y_0|^2 - |y_1|^2)}{\kappa_2^5}\,\bar{\mu}_2
  = -\bar{\partial}_{\PP^1_2}\!\left(\frac{A(u,v)\,\bar{y}_0^2 \bar{y}_1^2}{4\,\kappa_2^4}\right)\,.
  \label{eq:nu_Q22_exact_on_neck}
\end{equation}
For $U_2^2$, the relative plus sign in $\tilde{S}_{U_2^2}$ produces $|y_0|^2 + |y_1|^2 = \kappa_2$, leaving a non-exact harmonic $(0,1)$-form on $\PP^1_2$,
\begin{equation}
  \nu_{U_2^2}^{\mathrm{ref}}\Big|_{\{P_{Q_5}=0\}}
  = -\frac{1}{2} A(u,v)\,\frac{\bar{y}_0 \bar{y}_1}{\kappa_2^4}\,\bar{\mu}_2\,.
\end{equation}
Lifting one factor via $[x_0 : x_1] = [y_0 : -y_1]$ gives a global $\Gamma$-equivariant section of $L_2 = \cO_X(-1,-3,2,2)$,
\begin{equation}
  \sigma_{Q_2^2} = \frac{A(u,v)\,\bar{y}_0 \bar{y}_1\,(\bar{x}_0 \bar{y}_1 - \bar{x}_1 \bar{y}_0)}{8\,\kappa_1\,\kappa_2^3} \in \Omega^{0,0}(\hat{X}, L_2)\,,
\end{equation}
whose restriction to $\{P_{Q_5}=0\}$ reproduces~\eqref{eq:nu_Q22_exact_on_neck}. Inserting $\sigma_{Q_2^2}$ into Eq.~\eqref{eq:pythagorean_monotonicity} on $X_\epsilon$ cancels the leading contribution to $\nu_{Q_2^2}^{\mathrm{ref}}$ along $\{P_{Q_5}=0\}$ and forces $\cK_{Q_2^2}^{\mathrm{harm}} = \cO(\epsilon)$ alongside $\cK_{Q_5}^{\mathrm{harm}} = \cO(\epsilon)$, whereas $\cK_{U_5}^{\mathrm{harm}}$ and $\cK_{U_2^2}^{\mathrm{harm}}$ remain $\cO(1)$. Since $(Q_5, U_5)$ and $(Q_2^2, U_2^2)$ are supported on the same region, the metric factors in Eq.~\eqref{eq:sandwich_inequality} remain comparable, and the quark mass ratio~\eqref{eq:tq_pure_kahler_ratio} becomes
\begin{equation}
  \frac{m_t}{m_c} \sim \sqrt{\frac{\cK_{U_5}^{\mathrm{harm}}}{\cK_{Q_5}^{\mathrm{harm}}}\,\frac{\cK_{Q_2^2}^{\mathrm{harm}}}{\cK_{U_2^2}^{\mathrm{harm}}}}
  \sim \sqrt{\frac{\cO(1)}{\cO(\epsilon)}\,\frac{\cO(\epsilon)}{\cO(1)}} \sim \cO(1)\,.
  \label{eq:tq_ratio_collapse}
\end{equation}
Thus, along the completed-square locus~\eqref{eq:tq_completed_square}, degree saturation on $\PP^1_1 \times \PP^1_2$ forces $\cK_{Q_2^2}^{\mathrm{harm}}$ and $\cK_{Q_5}^{\mathrm{harm}}$ to vanish at the same rate, obstructing Mechanism~II as well. Combined with the moduli-independence of $W^u$ in Eq.~\eqref{eq:tq_exact_topological_yukawa}, which rules out Mechanism~I, neither tree-level mechanism can generate a parametric hierarchy $m_t / m_c \gg 1$ at the split-bundle locus of this tetra-quadric model.

\subsection{Numerical results}
\label{subsec:tq_numerical}

To verify the analytic results obtained in the previous subsection quantitatively, we solve the Monge--Amp\`ere, HYM, and harmonic $(0,1)$-form equations numerically across the $20$-dimensional complex structure moduli space of $\hat{X}$ using neural networks~\cite{larfors2022numericalmetricscompleteintersection,constantin2024computationquarkmassesstring} and Shiffman--Zelditch point sampling~\cite{Douglas:2006hz,larfors2022numericalmetricscompleteintersection}. The resulting K\"ahler norms and mass ratios are summarized in Table~\ref{tab:tq_summary}.

\begin{table}[t]
\centering
\footnotesize
\setlength{\tabcolsep}{4pt}
\begin{tabular}{@{}llcccc@{}}
\toprule
Regime in $\cM_{\mathrm{CS}}(\hat{X})$ & Configuration & $\cK_{Q_5}^{\mathrm{harm}}$ & $\cK_{Q_2^2}^{\mathrm{ref}} \to \cK_{Q_2^2}^{\mathrm{harm}}$ & $(m_t/m_c)_{\mathrm{ref}}$ & $(m_t/m_c)_{\mathrm{harm}}$ \\
\midrule
Dwork ($\cP_x$-symmetric) & $\psi = 0.5$ & $\cK_{Q_5} = \cK_{U_5}$ & $\cK_{Q_2^2} = \cK_{U_2^2}$ & $1.00$ & $1.00$ \\
Smooth interior & Generic $C\in\mathcal{M}_{\text{CS}}$ & $\cO(1)$ & $\cO(1) \to \cO(1)$ & $1.94$ & $1.78$ \\
Completed square & $\epsilon = 3.0 \times 10^{-3}$ & $6.12 \times 10^{-5}$ & $0.167 \to 5.84 \times 10^{-5}$ & $155.1$ & $2.13$ \\
Completed square & $\epsilon = 1.5 \times 10^{-3}$ & $3.19 \times 10^{-5}$ & $0.168 \to 3.01 \times 10^{-5}$ & $220.4$ & $2.09$ \\
\bottomrule
\end{tabular}
\caption{Numerical evaluations of the top-to-charm mass ratio on the tetra-quadric line bundle model (CICY 7862), comparing the reference basis  $\nu^{\mathrm{ref}}$ and harmonic solution $\nu^{\mathrm{harm}} = \nu^{\mathrm{ref}} + \bar{\partial}_X \sigma^*$ for $N = 10^6$ sampled points. Along the completed-square family~\eqref{eq:tq_completed_square}, the harmonic representative reduces $\cK_{Q_2^2}$ by more than three orders of magnitude, collapsing $(m_t/m_c)_{\mathrm{ref}} = 220.4$ to $2.09$ in agreement with~\eqref{eq:tq_ratio_collapse}.}
\label{tab:tq_summary}
\end{table}

On the $\cP_x$-symmetric Dwork family and at generic $\cO(1)$ polynomials, we find $m_t / m_c = 1.00$ and $1.78$, respectively, reproducing~\cite{constantin2024computationquarkmassesstring}. Along the completed-square family~\eqref{eq:tq_completed_square} at $\epsilon = 3.0 \times 10^{-3}$ and $\epsilon = 1.5 \times 10^{-3}$, the reference wavefunctions produce large apparent hierarchies $(m_t/m_c)_{\mathrm{ref}} = 155.1$ and $220.4$, which also persist ($142.8$ and $198.6$) when the learned Ricci-flat metric and the HYM connections are used. Once the harmonic $(0,1)$-form equation is solved at $\epsilon = 1.5 \times 10^{-3}$, however, $\cK_{Q_2^2}$ drops by a factor of $5{,}580$, tracking $\cK_{Q_5}^{\mathrm{harm}} = 3.19 \times 10^{-5}$, and collapses the physical ratio back to $m_t / m_c = 2.09$, directly confirming Eqs.~\eqref{eq:nu_Q22_exact_on_neck} and~\eqref{eq:tq_ratio_collapse}. A similar behavior is observed at $\epsilon=3\times10^{-3}$. Across smooth interior and near-pinched regimes alike, the physical top-to-charm ratio on the split tetra-quadric remains bounded by $m_t / m_c \lesssim 2\text{--}3$, more than two orders of magnitude below the GUT-scale target $m_t / m_c \sim 200\text{--}350$.

\section{Scan of CICY Line Bundle Models}
\label{sec:cicy_scan}

In Section~\ref{sec:tetraquadric}, we showed that the tetra-quadric model gives only $\rk Y^u = 2$ and cannot generate a large top-to-charm ratio at the split-bundle locus because both Mechanism~I and Mechanism~II are obstructed. To move beyond this example, we turn to the existing databases of three-generation CICY heterotic line bundle models constructed in~\cite{Anderson:2012yf,Anderson:2013xka}, which catalog tens of thousands of candidate bundles across favorable CICY threefolds. Rather than analyzing each geometry individually, we first derive general group-theoretic conditions required for $\rk Y^u = 3$ and $m_t \gg m_c \gg m_u$ in Section~\ref{subsec:rank_bounds}, then apply them together with cohomological and smoothness criteria to filter the databases in Section~\ref{subsec:database_scan}, and finally explain how the unique surviving model evades the tetra-quadric obstructions in Section~\ref{subsec:why_5273_singled_out}.

\subsection{Rank bounds}
\label{subsec:rank_bounds}

To filter the CICY databases before computing any Calabi--Yau cohomology, we first derive geometry-independent conditions from $S(U(1)^5)$ charge conservation for a split bundle $V = \bigoplus_{a=1}^5 L_a$ to yield three non-zero up-quark masses ($\rk Y^u = 3$) and a large hierarchy ($m_t \gg m_c \gg m_u$). Suppose first that the low-energy spectrum contains a single up-type Higgs doublet $H_u$ arising from $H^1(\hat{X}, L_c^* \otimes L_d^*)$ ($1 \le c < d \le 5$), with $S(U(1)^5)$ charge $-e_c - e_d$. Since the three chiral $\mathbf{10}$ families carry charges $e_{a_i}$ ($i=1,2,3$), gauge invariance permits the coupling $\mathbf{10}_{a_i} \mathbf{10}_{a_j} \mathbf{5}^H_{cd}$ only when $\{a_i, a_j\} = \{c, d\}$. Thus $Y^u$ is non-zero only in the off-diagonal blocks which pair generations in $L_c$ with generations in $L_d$. Let $n_c$ and $n_d$ be the number of generations arising from $L_c$ and $L_d$, with $n_c + n_d \le 3$. Thus $(n_c, n_d)$ is at most $(2,1)$ or $(1,1)$,\footnote{A $(2,1)$ split gives $2 \times 1$ and $1 \times 2$ off-diagonal blocks as in Eq.~\eqref{eq:tq_bordered_yukawa}, whereas $(1,1)$ leaves the third row and column identically zero.} so any single-Higgs split model obeys
\begin{equation}
  \rk Y^u \le 2\,,
  \label{eq:single_higgs_bound}
\end{equation}
everywhere in K\"ahler and complex structure moduli space. Achieving $\rk Y^u = 3$ at the split locus therefore requires placing the three chiral $\mathbf{10}$ families in three distinct line bundles $L_a, L_b, L_c$, with family counts across $(L_1, \dots, L_5)$ forming the partition $(1,1,1,0,0)$ up to permutation. In addition, all three up-type Higgs sectors $\mathbf{5}^H_{ab}, \mathbf{5}^H_{ac}, \mathbf{5}^H_{bc}$ must remain light so that the electroweak doublet $H_u$, which is formed from a linear combination of $H_{ab}$, $H_{ac}$, and $H_{bc}$, populates all off-diagonal entries of $Y^u$.

Even on a $(1,1,1,0,0)$ model with three light Higgs sectors, a further constraint arises if left- and right-handed quarks share the same internal wavefunctions. This is the case in unbroken $SU(5)$ GUT models on the ``upstairs'' geometry $X$, where $Q_a$ and $u^c_a$ share a single wavefunction in $H^1(X, L_a)$. The physical Yukawa matrix $Y^u$ is then complex symmetric ($Y^u_{ab} = Y^u_{ba}$)  with $Y^u_{aa} = 0$, forcing its singular values to obey $m_t \le m_c + m_u$. In any ``downstairs'' Standard Model on the quotient $\hat{X} = X/\Gamma$, however, the hypercharge Wilson line breaking $SU(5) \to G_{\mathrm{SM}}$ automatically assigns distinct Wilson-line twists $\cW_Q \neq \cW_{u^c}$ to $Q$ ($Y = 1/6$) and $u^c$ ($Y = -2/3$), so $Y^u_{ab} \neq Y^u_{ba}$ and $m_t \gg m_c \gg m_u$ is unobstructed.

\subsection{Database scan}
\label{subsec:database_scan}

\begin{table}[t]
\centering
\footnotesize
\setlength{\tabcolsep}{3.5pt}
\begin{tabular}{@{}lrrr@{}}
\toprule
Selection criterion & $SU(5)$ models~\cite{Anderson:2013xka} & Line bundle sums~\cite{Anderson:2012yf} & Standard Models~\cite{Anderson:2012yf} \\
\midrule
Total entries ($\sum_{a=1}^5 \chi(L_a) = -3|\Gamma|$) & $63{,}325$ & $202$ & $2{,}112$ \\
$\Gamma$-divisible ($\chi(L_a) \equiv 0 \pmod{|\Gamma|}$) & $40{,}480$ & $202$ & $2{,}112$ \\
Strictly chiral ($h^2(X, L_a) = 0$, no $\overline{\mathbf{10}}$s) & $34{,}988$ & $202$ & $2{,}112$ \\
\midrule
Chiral $(3,0,0,0,0)$ ($\rk Y^u = 0$ identically) & $7{,}142$ & $59$ & $592$ \\
Chiral $(2,1,0,0,0)$ ($\rk Y^u \le 2$, Eq.~\eqref{eq:single_higgs_bound}) & $16{,}894$ & $93$ & $976$ \\
Chiral $(1,1,1,0,0)$ ($3$ distinct matter bundles) & $10{,}952$ & $50$ & $544$ \\
\midrule
Generic moduli ($h^1(X, L_{H_{ab}}) \ge |\Gamma|$, $3$ sectors) & $0$ & $0$ & $0$ \\
Koszul step $E_2$ ($\dim E_2^{-p,p+1} \ge |\Gamma|$, $3$ sectors) & $36$ & $0$ & $0$ \\
Ambient degree selection rule ($3$ sectors) & $3$ & $0$ & $0$ \\
Smooth quotient on $d_2$-kernel locus ($\rk Y^u = 3$) & $1$ & $0$ & $0$ \\
\bottomrule
\end{tabular}
\caption{Classification of tree-level up-quark Yukawa ranks across all $63{,}325$ upstairs $SU(5)$ line bundle models of~\cite{Anderson:2013xka} and all $202$ unique line bundle sums ($2{,}112$ downstairs Standard Models) of~\cite{Anderson:2012yf}. Only a single smooth quotient, CICY 5273, Model~500, admits a tree-level rank-three up-quark Yukawa matrix.}
\label{tab:cicy_landscape_census}
\end{table}

Armed with these selection rules, we scanned the two existing databases of CICY heterotic line bundle models, corresponding to the columns of Table~\ref{tab:cicy_landscape_census}:
\begin{enumerate}[label=(\roman*)]
    \item All $63{,}325$ upstairs $SU(5)$ bundles constructed in~\cite{Anderson:2013xka} across $67$ favorable CICYs with $h^{1,1}(X) \le 6$ and $100$ $(X, |\Gamma|)$ pairs, for which we explicitly construct the $\Gamma$-quotients and Wilson-line projections to $\hat{X}$.
    \item The earlier catalog~\cite{Anderson:2012yf} of $202$ line bundle sums on $15$ CICYs, together with the $2{,}112$ downstairs Standard Models obtained from them.
\end{enumerate}
As a check on the tetra-quadric (CICY 7862), none of the $64$ equivariant models in~\cite{Anderson:2013xka} and none of the $22$ models in~\cite{Anderson:2012yf} have $\rk Y^u = 3$; five models in each database have $\rk Y^u = 2$ and are permutation-equivalent to~\eqref{eq:tq_bundles}, while the remainder have $\rk Y^u = 0$.

Across these databases requiring $\chi(X, L_a)$ to be divisible by $|\Gamma|$ and excluding $\overline{\mathbf{10}}$ mirror generations ($h^2(X, L_a) = 0$) leaves $34{,}988$ strictly chiral models in~\cite{Anderson:2013xka} (both conditions are already built into~\cite{Anderson:2012yf}), as listed in the upper rows of Table~\ref{tab:cicy_landscape_census}. Applying the rank bound~\eqref{eq:single_higgs_bound} then rules out the $(3,0,0,0,0)$ and $(2,1,0,0,0)$ partitions, isolating $10{,}952$ chiral $(1,1,1,0,0)$ models. We tested each of these against three further geometric conditions needed for $\rk Y^u = 3$:
\begin{enumerate}[label=(\roman*)]
  \item \emph{Three light Higgs sectors.} We label the three matter line bundles without loss of generality as $L_2, L_3, L_4$. The three Higgs bundles $L_{H_{ab}} = L_a^* \otimes L_b^*$ with $ab \in \{23, 24, 34\}$ support massless doublets on $\hat{X}$ if they satisfy $h^1(X, L_{H_{ab}}) \ge |\Gamma|$. Notably, not a single model in either database satisfies this condition at generic complex structure. Restricting to jumping loci of higher Koszul differentials ($r \ge 2$) at generic $d_1$ requires $\sum_{p=0}^K \dim E_2^{-p,p+1}(L_{H_{ab}}) \ge |\Gamma|$ at generic $C$ for all three Higgs sectors, selecting $36$ candidate models from~\cite{Anderson:2013xka}.
  \item \emph{Ambient degree selection rules.} Extending the vanishing conditions of~\cite{Blesneag_2016,Blesneag_2017} to codimension $K$ and ambient dimension $D = K + 3$, a trilinear coupling involving $L_a, L_b,$ and $L_{H_{ab}}$ can be non-zero only if the Koszul steps satisfy
  \begin{equation}
    p_a + p_b + p_{H_{ab}} \ge K\,,
  \end{equation}
  and if they can saturate the top anti-holomorphic form degree $n_r$ on every $\PP^{n_r}$ factor of $\cA$. Imposing this condition on all three Higgs sectors $\{23, 24, 34\}$ reduces the $36$ candidates to three models: CICY 6804, Model~128 ($|\Gamma| = 2, K = 2$), CICY 5273, Model~500 ($|\Gamma| = 2, K = 5$), and CICY 5302, Model~114 ($|\Gamma| = 4, K = 5$).
  \item \emph{Equivariance and quotient smoothness.} Finally, we checked $\Gamma$-equivariance of the individual line bundles and smoothness of $\hat{X} = X/\Gamma$ on the Higgs jumping locus across all freely acting symmetries classified in~\cite{Braun:2010}. On CICY 6804, Model~128, every restricted polynomial that gives $d_2$ a kernel for all three Higgs bundles forces $X$ to intersect the fixed-point locus of the free involution at eight points, rendering $\hat{X}$ singular. On CICY 5302, Model~114 ($|\Gamma| = 4$), every free symmetry group of order four permutes the $\PP^1$ factors carrying the distinct negative charges $(-2)$ of $L_2, L_3, L_4$, so $g L_a \neq L_a$ for some non-trivial $g \in \Gamma$, and the split line bundles fail to descend individually to $\hat{X}$.
\end{enumerate}
Only CICY 5273, Model~500 satisfies all topological, equivariant, and smoothness conditions. Checking gauge invariance of the down-type couplings $\mathbf{10}_a \overline{\mathbf{5}}_{bc} \overline{\mathbf{5}}^H_{de}$ across all $34{,}988$ chiral entries further shows that no split line bundle model in the database achieves simultaneous tree-level $\rk Y^u = 3$ and $\rk Y^d = 3$.

\subsection{CICY 5273, Model 500}
\label{subsec:why_5273_singled_out}

The surviving candidate CICY 5273 with Model~500 in the classification of~\cite{Candelas:1987kf,Anderson:2013xka} is a codimension $K = 5$ complete intersection in the $D = 8$ dimensional ambient space $\cA = \PP^1_1 \times \PP^1_2 \times \PP^1_3 \times \PP^1_4 \times \PP^2_5 \times \PP^2_6$, quotiented by a freely acting symmetry $\Gamma \cong \ZZ_2$. Its three chiral $\mathbf{10}$ families reside in $L_2, L_3, L_4$. Since the free $\ZZ_2$ generator $g_{\mathrm{free}}$ exchanges the two $\PP^2$ factors ($\PP^2_5 \leftrightarrow \PP^2_6$), on which all five line bundles carry equal charges $q_a^5 = q_a^6$, every $L_a$ is individually $\Gamma$-equivariant, while the $\ZZ_2$ hypercharge Wilson line projects $Q_a$ and $u^c_a$ ($a=2,3,4$) onto opposite $\Gamma$-eigenspaces. Simultaneously, Serre duality guarantees that the three conjugate down-Higgs bundles $L_{H_{d,ab}} = L_{H_{ab}}^* = L_a \otimes L_b$ acquire massless modes on the exact same jumping locus $\cJ_{\mathrm{Higgs}}$ as the up-Higgs bundles. These couple $Q_2, Q_3, Q_4$ to the bottom anti-quark $b^c \in H^1(\hat{X}, L_1 \otimes L_5)$, giving $\rk Y^d = 1$ and hence a massive bottom quark $b^c$, while the down and strange quarks $d^c, s^c \in H^1(\hat{X}, L_2 \otimes L_5)$ remain massless at the split locus because $H^1(\hat{X}, L_1 \otimes L_a) = 0$ for $a \in \{2,3,4\}$.

In this model, both mechanisms for generating Yukawa hierarchies that we described in Section~\ref{subsec:two_mechanisms} are unobstructed. Each chiral quark $(Q_a, u^c_a)$ originates in $E_1^{-1,2}$ ($p_a = 1$, $q_a = 2$) via multiplication by the fifth defining polynomial $P_5$, whereas each up-Higgs $H_{ab}$ originates in $E_2^{-4,5}$ ($p_H = 4$, $q_H = 5$) from $\ker d_2$ along the smooth $14$-dimensional Higgs locus $\cJ_{\mathrm{Higgs}} \subset \cM_{\mathrm{CS}}(\hat{X})$. Consequently, $p_a + p_b + p_H = 1 + 1 + 4 = 6 > K$ and $q_a + q_b + q_H = 9 > D$ for all three Higgs sectors, so uncanceled polynomial factors survive in the ambient lift, and the Higgs cocycles depend on the complex structure moduli $C$ through $\ker d_2(C)$. This allows us to use Mechanism~I to tune smooth holomorphic zero-crossings in individual entries of $W^u(C)$. The polynomial $P_5(x_2, z_5, z_6)$ in~\eqref{eq:cicy5273_config} has multidegree $(0,2,0,0,1,1)$, so it carries degree $2$ on $\PP^1_2$ and degree $0$ on the remaining $\PP^1$ factors $\PP^1_1, \PP^1_3, \PP^1_4$ that support the harmonic $(0,1)$-legs of $(Q_a, u^c_a)$. Asymmetry between the $\PP^1_2$ sheets can therefore split the K\"ahler norms of opposite $\Gamma$-parities without rendering the $(0,1)$-forms exact. We construct this model explicitly and compute its physical quark masses in Section~\ref{sec:cicy_5273}.

\section{Quark Masses on CICY 5273}
\label{sec:cicy_5273}

We now construct the MSSM-like heterotic line bundle model on CICY 5273, Model~500 explicitly, determine its slope-zero K\"ahler cone, analyze its $14$-dimensional Higgs jumping locus and vector-like exotics lifting, summarize our numerical setup, and present the resulting physical quark masses. Explicit formulas for the $13$ reference $(0,1)$-forms and algebraic cocycle contractions are collected in Appendix~\ref{app:koszul}. The defining polynomials, benchmark complex structure points, and renormalization group equations are recorded in Appendix~\ref{app:benchmarks}, and details of the neural network architectures and training protocol are given in Appendix~\ref{app:ml_setup}.

\subsection{Geometry and K\"ahler cone}
\label{subsec:5273_geometry}

The covering threefold $X \subset \cA = \PP^1_1 \times \PP^1_2 \times \PP^1_3 \times \PP^1_4 \times \PP^2_5 \times \PP^2_6$ is the common zero locus of five homogeneous polynomials $P_1 = \dots = P_5 = 0$ with configuration matrix
\begin{equation}
  [X] = \left[\begin{array}{c|ccccc}
    \PP^1_1 & 1 & 1 & 0 & 0 & 0 \\
    \PP^1_2 & 0 & 0 & 0 & 0 & 2 \\
    \PP^1_3 & 0 & 0 & 1 & 1 & 0 \\
    \PP^1_4 & 0 & 0 & 1 & 1 & 0 \\
    \PP^2_5 & 1 & 0 & 1 & 0 & 1 \\
    \PP^2_6 & 0 & 1 & 0 & 1 & 1
  \end{array}\right]^{6,30}_{-48}\,,
  \label{eq:cicy5273_config}
\end{equation}
where the five columns list the multidegrees of $P_1, \dots, P_5$, the superscripts and subscript record $(h^{1,1}(X), h^{2,1}(X)) = (6,30)$ and $\chi(X) = -48$, and $c_2(TX) = (24, 24, 24, 24, 36, 36)$. In homogeneous coordinates $x_r = [x_{r,0} : x_{r,1}] \in \PP^1_r$ ($r=1,2,3,4$) and $z_s = [z_{s,0} : z_{s,1} : z_{s,2}] \in \PP^2_s$ ($s=5,6$), the unique freely acting $\ZZ_2$ involution $g_{\mathrm{free}}$ acts as~\cite{Braun:2010}
\begin{equation}
  g_{\mathrm{free}} : \quad x_1 \mapsto [x_{1,0} : -x_{1,1}]\,, \qquad
  x_r \mapsto [x_{r,1} : x_{r,0}] \;\; (r=2,3,4)\,, \qquad
  z_5 \leftrightarrow z_6\,,
  \label{eq:g_free_action}
\end{equation}
pairing $P_2 = g_{\mathrm{free}}(P_1)$ and $P_4 = g_{\mathrm{free}}(P_3)$ while leaving $P_5 = g_{\mathrm{free}}(P_5)$ invariant. Whenever $X$ avoids the ambient fixed locus $\mathrm{Fix}(g_{\mathrm{free}})$, which consists of $16$ copies of $\PP^2$, the quotient $\hat{X} = X/\Gamma$ is smooth with Hodge numbers $(h^{1,1}(\hat{X}), h^{2,1}(\hat{X})) = (5, 17)$ and Euler characteristic $\chi(\hat{X}) = -24$.

\begin{table}[t]
\centering
\small
\setlength{\tabcolsep}{5.5pt}
\begin{tabular}{@{}llcccc@{}}
\toprule
Sector & Fields & Ambient multidegree $q$ & $\chi(\hat{X}, U)$ & Koszul origin & $g_{\mathrm{free}}$ parity\\
\midrule
$\mathbf{10}_2$ & $Q_2,\; u^c_2$ & $(-2, -1, 0, 0, 1, 1)$ & $-1$ & $E_1^{-1,2}$ & $-1,\; +1$ \\
$\mathbf{10}_3$ & $Q_3,\; u^c_3$ & $(0, -1, -2, 0, 1, 1)$ & $-1$ & $E_1^{-1,2}$ & $+1,\; -1$ \\
$\mathbf{10}_4$ & $Q_4,\; u^c_4$ & $(0, -1, 0, -2, 1, 1)$ & $-1$ & $E_1^{-1,2}$ & $+1,\; -1$ \\
\midrule
$\mathbf{5}^H_{23}$ & $H_{23}$ & $(2, 2, 2, 0, -2, -2)$ & $0$ & $E_2^{-4,5}$ & $+1$ \\
$\mathbf{5}^H_{24}$ & $H_{24}$ & $(2, 2, 0, 2, -2, -2)$ & $0$ & $E_2^{-4,5}$ & $+1$ \\
$\mathbf{5}^H_{34}$ & $H_{34}$ & $(0, 2, 2, 2, -2, -2)$ & $0$ & $E_2^{-4,5}$ & $-1$ \\
\midrule
$\overline{\mathbf{5}}_{15}$ & $b^c$ & $(2, 3, 2, 2, -3, -3)$ & $-1$ & $E_2^{-3,4}$ & $+1$ \\
$\overline{\mathbf{5}}^H_{23}$ & $H_{d,23}$ & $(-2, -2, -2, 0, 2, 2)$ & $0$ & $E_2^{-3,4}$ & $+1$ \\
$\overline{\mathbf{5}}^H_{24}$ & $H_{d,24}$ & $(-2, -2, 0, -2, 2, 2)$ & $0$ & $E_2^{-3,4}$ & $+1$ \\
$\overline{\mathbf{5}}^H_{34}$ & $H_{d,34}$ & $(0, -2, -2, -2, 2, 2)$ & $0$ & $E_2^{-3,4}$ & $-1$ \\
\bottomrule
\end{tabular}
\caption{Matter and Higgs line bundles entering the Yukawa sector on CICY 5273, Model~500, showing their multidegrees on $(\PP^1)^4 \times (\PP^2)^2$, quotient indices $\chi(\hat{X}, U)$, Koszul spectral-sequence origins on $X$, and $g_{\mathrm{free}}$ transformation on $\cJ_{\mathrm{Higgs}}$. Every gauge-allowed Yukawa triple transforms with combined phase $+1$.}
\label{tab:5273_spectrum}
\end{table}

On this geometry, Model~500 of~\cite{Anderson:2013xka} equips $X$ with the rank-five line bundle sum $V = \bigoplus_{a=1}^5 L_a$ defined by
\begin{equation}
\begin{aligned}
  L_1 &= \cO_X(2, 2, 1, 1, -2, -2)\,, &\qquad
  L_2 &= \cO_X(-2, -1, 0, 0, 1, 1)\,, \\
  L_3 &= \cO_X(0, -1, -2, 0, 1, 1)\,, &\qquad
  L_4 &= \cO_X(0, -1, 0, -2, 1, 1)\,, \\
  L_5 &= \cO_X(0, 1, 1, 1, -1, -1)\,, & &
\end{aligned}
  \label{eq:5273_bundles}
\end{equation}
satisfying $c_1(V) = 0$ and $c_2(V) = (16, 24, 18, 18, 30, 30) \le c_2(TX)$, so that heterotic anomaly cancellation is satisfied with trivial hidden bundle and five-branes wrapping the effective $\Gamma$-invariant curve class $[W] = c_2(TX) - c_2(V) = J_2 \cdot (J_5 + J_6) + 4[\PP^1_1\text{ curve}] = (8, 0, 6, 6, 6, 6)$, descending to $[\hat{W}] = (4, 0, 3, 3, 3, 3)$ on $\hat{X}$. Table~\ref{tab:5273_spectrum} lists the $13$ matter and Higgs bundles entering the quark Yukawa couplings together with their $\ZZ_2$ transformation $g_{\mathrm{free}} \nu_I = \pm \nu_I$. Under~\eqref{eq:g_free_action}, the holomorphic $(3,0)$-form satisfies $g_{\mathrm{free}} \Omega = +\Omega$, so every gauge-allowed up-type triple $H_{ab} Q_a u^c_b$ and down-type triple $Q_a b^c H_{d,jk}$ ($\{a,j,k\} = \{2,3,4\}$) has $g_\mathrm{free}(\nu_I \wedge \nu_J \wedge \nu_K) = +\nu_I \wedge \nu_J \wedge \nu_K$ and is $\Gamma$-invariant.

To preserve supersymmetry at the split locus, the K\"ahler moduli must satisfy the slope-zero conditions~\eqref{eq:duy_slope_zero}. Since $g_{\mathrm{free}}$ swaps $\PP^2_5 \leftrightarrow \PP^2_6$, invariance on $\hat{X}$ sets $t_5 = t_6 > 0$ (matching $h^{1,1}(\hat{X}) = 5$). Inside the $\Gamma$-invariant K\"ahler cone ($t_5 = t_6 > 0$, $t_1, t_2, t_3, t_4 > 0$), the five slope-zero equations $\mu_t(L_a) = 0$ in~\eqref{eq:duy_slope_zero} enforce $t_3 = t_4$ and select a two-dimensional cone of rays parameterized by the ratio
\begin{equation}
  \xi := \frac{t_3}{t_5} \in \left(\frac{\sqrt{13} - 2}{3},\; \frac{\sqrt{17} - 1}{2}\right)\,,
  \label{eq:xi_kahler_window}
\end{equation}
along which $t_1 / t_5$ and $t_2 / t_5$ are positive algebraic functions of $\xi$. At the symmetric ray $\xi = 1$ normalized to $\cV_X = 62$, one finds 
\begin{equation}
    t_{\mathrm{HYM}}(\xi=1) \approx (0.7259, 0.3629, 1.1987, 1.1987, 1.1987, 1.1987)\,,
\end{equation}
where $t_2 = 0.3629$ gives a moderate worldsheet-instanton factor $e^{-2\pi t_2} \approx 0.102$. To suppress worldsheet instantons uniformly across all K\"ahler moduli, maximizing $\min_i t_i / \cV_X^{1/3}$ over~\eqref{eq:xi_kahler_window} selects $\xi_\star \approx 0.85$, where $t_1(\xi_\star) = t_2(\xi_\star)$. Scaling the overall volume so that $\min_i t_i = 1.00$ defines the supergravity point
\begin{equation}
  t_{\mathrm{SUGRA}} = (1.0000,\; 1.0000,\; 2.0433,\; 2.0433,\; 2.4062,\; 2.4062)\,,
  \label{eq:t_sugra_point}
\end{equation}
with $\cV_X = 428.60$, $\cV_{\hat{X}} = 214.30$, and uniform instanton suppression $\max_i e^{-2\pi t_i} = 1.87 \times 10^{-3}$. At $t_{\mathrm{SUGRA}}$, the one-loop heterotic gauge threshold correction~\cite{Lukas:1998tt,Blumenhagen:2005ga}
\begin{equation}
    \Delta \alpha_{\mathrm{GUT}}^{-1} = \frac{1}{4\pi |\Gamma|}\big(c_2(V) - \frac{1}{2}c_2(TX) + \sum_n (1 - z_n)^2 [W_n]\big) \cdot t_{\mathrm{SUGRA}}
\end{equation}
with $[W]=\sum_n[W_n]$ is between $3.91$ and $6.35$, depending on the five-brane interval positions $z_n \in [0, 1]$. Note that in either case this remains small compared to $\alpha_{\mathrm{GUT}}^{-1} = 24.30$.

\subsection{Higgs locus and vector-like mass lifting}
\label{subsec:5273_higgs_locus}

\begin{table}[t]
\centering
\footnotesize
\setlength{\tabcolsep}{3.8pt}
\renewcommand{\arraystretch}{1.16}
\begin{tabular}{@{}lcccc@{}}
\toprule
Bundle sector & \thead{Upstairs $X$\\($SU(5)$, generic)} & \thead{Downstairs $\hat{X}$\\(generic $\cM_{\mathrm{CS}}$)} & \thead{Downstairs $\hat{X}$\\(on $\cJ_{\mathrm{Higgs}}$)} & \thead{Downstairs $\hat{X}$\\($\cJ_{\mathrm{Higgs}}$ with $g_{\mathrm{int}}$)} \\
\midrule
$L_2$ & $2\,\mathbf{10}_2$ & $(Q_2, u^c_2, e^c_2)$ & $(Q_2, u^c_2, e^c_2)$ & $(Q_2, u^c_2, e^c_2)$ \\
$L_3$ & $2\,\mathbf{10}_3$ & $(Q_3, u^c_3, e^c_3)$ & $(Q_3, u^c_3, e^c_3)$ & $(Q_3, u^c_3, e^c_3)$ \\
$L_4$ & $2\,\mathbf{10}_4$ & $(Q_4, u^c_4, e^c_4)$ & $(Q_4, u^c_4, e^c_4)$ & $(Q_4, u^c_4, e^c_4)$ \\
$L_1 \oplus L_1^*$ & $0$ & $0$ & $0$ & $\mathbf{10}_1 \oplus \overline{\mathbf{10}}_1$ \\
\midrule
$L_2 \otimes L_5$ & $4\,\overline{\mathbf{5}}_{25}$ & $(d^c, L_1) \oplus (s^c, L_2)$ & $(d^c, L_1) \oplus (s^c, L_2)$ & $(d^c, L_1) \oplus (s^c, L_2)$ \\
$(L_1 \otimes L_5) \oplus \text{dual}$ & $2\,\overline{\mathbf{5}}_{15}$ & $(b^c, L_3)$ & $(b^c, L_3)$ & $(b^c, L_3) \oplus 2(\mathbf{5}_{15} \oplus \overline{\mathbf{5}}_{15})$ \\
$(L_2^* \otimes L_3^*) \oplus \text{dual}$ & $0$ & $0$ & $H_{23} \oplus H_{d,23}$ & $(H_{23} \oplus H_{d,23}) \oplus (T_{23} \oplus \bar{T}_{23})$ \\
$(L_2^* \otimes L_4^*) \oplus \text{dual}$ & $0$ & $0$ & $H_{24} \oplus H_{d,24}$ & $(H_{24} \oplus H_{d,24}) \oplus (T_{24} \oplus \bar{T}_{24})$ \\
$(L_3^* \otimes L_4^*) \oplus \text{dual}$ & $0$ & $0$ & $H_{34} \oplus H_{d,34}$ & $(H_{34} \oplus H_{d,34}) \oplus 2(T_{34} \oplus \bar{T}_{34})$ \\
\midrule
Total singlets ($\mathbf{1}$) & $36$ & $19$ & $19$ & $29$ \\
\bottomrule
\end{tabular}
\caption{Massless spectrum of CICY 5273, Model~500 across different complex structure regimes. Each $\mathbf{10}\oplus\overline{\mathbf{5}}$ host a chiral SM generation $(Q, u^c, e^c, d^c, L)$, $\mathbf{10}_1 \oplus \overline{\mathbf{10}}_1$ and $\mathbf{5}_{15} \oplus \overline{\mathbf{5}}_{15}$ denote complete vector-like multiplets, and $(T_{ab}, \bar{T}_{ab})$ denote vector-like $(\mathbf{3},\mathbf{1})_{-1/3} \oplus (\overline{\mathbf{3}},\mathbf{1})_{+1/3}$ color triplets.}
\label{tab:5273_full_spectrum_summary}
\end{table}

Having fixed the K\"ahler moduli along the slope-zero rays, we now turn to the complex structure moduli space $\cM_{\mathrm{CS}}(\hat{X})$. We first identify the locus supporting massless electroweak Higgs doublets and then explain how extra vector-like states are lifted. Whereas all chiral matter bundles have $\chi(\hat{X}, U) < 0$, the six vector-like Higgs bundles $H_{ab}$ and $H_{d,ab}$ for $ab \in \{23, 24, 34\}$ have $\chi(\hat{X}, U) = 0$. On $\hat{X}$, $H^1(\hat{X}, L_{H_{ab}})$ and its Serre dual $H^1(\hat{X}, L_{H_{d,ab}})$ are governed by the kernel of the $\Gamma$-equivariant Koszul differential block $D_{H_{ab}}(C)$. Keeping all three vector-like pairs $(H_{ab}, H_{d,ab})$ massless imposes the three holomorphic determinantal equations
\begin{equation}
  \det D_{H_{ab}}(C) = 0\,, \qquad (a,b) \in \{(2,3), (2,4), (3,4)\}\,,
  \label{eq:5273_determinantal_locus}
\end{equation}
which cut out a smooth $14$-dimensional jumping locus $\cJ_{\mathrm{Higgs}} \subset \cM_{\mathrm{CS}}(\hat{X})$ by imposing $3$ conditions in the $17$-dimensional moduli space. At generic points of $\cJ_{\mathrm{Higgs}}$ on the split-bundle locus, all color-triplet partners remain massive and the massless charged spectrum is MSSM-like, containing three chiral Standard Model families together with three vector-like pairs of electroweak Higgs doublets $(H_{ab}, H_{d,ab})$, as summarized in Table~\ref{tab:5273_full_spectrum_summary}. Electroweak symmetry breaking then proceeds through one light up-type combination $H_u = \sum_{ab} \hat{v}^u_{ab} H_{ab}$ and one light down-type combination $H_d = \sum_{ab} \hat{v}^d_{ab} H_{d,ab}$ ($\|\hat{v}^u\| = \|\hat{v}^d\| = 1$), with $\langle H_u \rangle = v \sin\beta$, $\langle H_d \rangle = v \cos\beta$, and $v \approx 174\text{ GeV}$.

To construct explicit points on $\cJ_{\mathrm{Higgs}}$, we start from a symmetric sub-locus invariant under an accidental $\ZZ_2$ involution $g_{\mathrm{int}}$, defined in Appendix~\ref{app:benchmarks}. On this symmetric slice, parity selection rules satisfy Eq.~\eqref{eq:5273_determinantal_locus} automatically and set $W_{Q_a b^c H_d} = 0$. However, the extra symmetry also causes additional blocks of $d_1(C)$ and $d_2(C)$ to drop rank, so accidental vector-like color triplets, mirror matter, and bundle singlets become massless on $\hat{X}$. Because all of these extra multiplets are vector-like under $G_{\mathrm{SM}} \times S(U(1)^5)$, lifting them does not require leaving the split-bundle locus. Turning on the nine $g_{\mathrm{int}}$-odd polynomial deformations while maintaining Eq.~\eqref{eq:5273_determinantal_locus} on the three Higgs-doublet blocks restores full generic rank to all other Koszul differentials. As we verified explicitly at Benchmarks~1--4, every accidental $g_{\mathrm{int}}$ exotic acquires a mass near the compactification scale.

Moving slightly off the $g_{\mathrm{int}}$-symmetric slice along $\cJ_{\mathrm{Higgs}}$ also activates both hierarchy mechanisms of Section~\ref{subsec:two_mechanisms}. Breaking $g_{\mathrm{int}}$ mixes a small $g_{\mathrm{int}}$-odd component into $b^c \in H^1(\hat{X}, L_1 \otimes L_5)$, generating a realistic bottom Yukawa coupling. Simultaneously, smooth zero-crossings in the off-diagonal entries of $W^u(C)$ produce $m_u \ll m_c \ll m_t$ via Mechanism~I. Finally, moderately localizing the $\PP^1_2$ roots of $P_5(x_2, z_5, z_6) = 0$ toward $x_{2,0} = \pm x_{2,1}$ implements Mechanism~II at finite $\epsilon$: it suppresses $\cK_{Q_3}$ and $\cK_{u^c_4}$ and enhances the top Yukawa coupling while keeping $\min_{X} \dd P \sim 0.01\text{--}0.035$ well away from conifold singularities.

Whereas the accidental $g_{\mathrm{int}}$ exotics lift within $\cJ_{\mathrm{Higgs}}$ at the split locus, lifting two of the three remaining Higgs doublet pairs $(H_{ab}, H_{d,ab})$ requires moving slightly off both loci. Recovering the minimal one-pair MSSM from the three remaining Higgs doublet pairs $(H_{ab}, H_{d,ab})$ requires generating a $3 \times 3$ bilinear mass matrix $W_{\mathrm{Higgs}} = \sum_{ab,cd} M^{\mathrm{Higgs}}_{ab,cd}(C, S)\, H_{ab} H_{d,cd}$ of rank two, with left and right null vectors $(\hat{v}^u, \hat{v}^d)$. Purely complex-structure displacements $\delta C_\perp$ normal to $\cJ_{\mathrm{Higgs}}$ at $\langle S_{ab} \rangle = 0$ preserve $S(U(1)^5)$ and only turn on diagonal masses $\mu_{ab}(\delta C_\perp)\, H_{ab} H_{d,ab}$; setting two $\mu_{ab} \neq 0$ and one $\mu_{ab} = 0$ would align $H_u$ along a single $H_{ab}$ and reduce $\rk Y^u \le 2$ by~\eqref{eq:single_higgs_bound}. Populating the off-diagonal entries requires VEVs of the generation-mixing bundle singlets $S_{ab} \in H^1(\hat{X}, L_a \otimes L_b^*)$ with $a \neq b \in \{2,3,4\}$, which couple linearly via $S_{cd} H_{ac} H_{d,ad}$. All six such singlet sectors are present everywhere on $\cM_{\mathrm{CS}}(\hat{X})$. Because they come in conjugate pairs $(S_{ab}, S_{ba})$ of opposite $S(U(1)^5)$ charge, they admit $D$-flat directions along the slope-zero K\"ahler cone. Enforcing the single determinantal condition $\det M^{\mathrm{Higgs}}(C, \langle S \rangle) = 0$ in place of~\eqref{eq:5273_determinantal_locus} then lifts two heavy combinations at $M_H \sim \varepsilon_S M_{\mathrm{KK}}$, with $\varepsilon_S \sim \max(|\delta C_\perp|, |\langle S_{ab} \rangle| / M_{\mathrm{KK}})$, while leaving one light pair $(H_u, H_d)$ along misaligned unit vectors $(\hat{v}^u, \hat{v}^d)$.

Since these same singlets $S_{ab}$ ($a,b \in \{2,3,4\}$) also permit higher-dimensional up-quark couplings $(S_{ac}/M_{\mathrm{KK}})\,\mathbf{10}_c \mathbf{10}_b H_{ab}$ (including diagonal entries for $c = b$), the choice of decoupling scale $M_H$ involves a trade-off between Yukawa stability at fixed $C$ and gauge unification:
\begin{itemize}[nosep]
  \item \emph{GUT-scale decoupling ($M_H \sim M_{\mathrm{GUT}}$):} For $\varepsilon_S \sim 10^{-2}\text{--}10^{-1}$, the two heavy doublet pairs decouple near $M_{\mathrm{GUT}}$, preserving standard one-pair MSSM gauge coupling unification ($\alpha_{\mathrm{GUT}}^{-1} \approx 24.3\text{--}25.5$) and the two-loop RG evolution used in Section~\ref{subsec:5273_results}. Although $\cO(\varepsilon_S)$ singlet insertions shift the entries of $Y^u$ by $\Delta Y^u \sim \cO(\varepsilon_S)$ at fixed $C$, exceeding the small split-locus eigenvalue $Y_u \simeq 3.2 \times 10^{-6}$, Mechanism~I relies on smooth codimension-one holomorphic zero-crossing hypersurfaces, specifically $\{\det W^u(C, \langle S \rangle) = 0\}$. Under a smooth bundle deformation $\varepsilon_S \ll 1$, such codimension-one zero loci do not disappear, but merely shift smoothly in $\cM_{\mathrm{CS}}(\hat{X})$ by $\Delta C \sim \cO(\varepsilon_S)$, so $Y_u \sim 3 \times 10^{-6}$ can be maintained by tracking the shifted locus.
  \item \emph{Intermediate-scale decoupling ($M_H \ll M_{\mathrm{GUT}}$):} Conversely, taking $\varepsilon_S \lesssim 10^{-6}$ leaves $Y_u$ unperturbed at fixed $C$, but keeps two extra electroweak doublet pairs light down to $M_H \sim \varepsilon_S M_{\mathrm{KK}}$. Their color-triplet partners $(T_{ab}, \bar{T}_{ab})$ must remain near the compactification scale to satisfy dimension-five proton decay bounds. Between $M_H$ and $M_{\mathrm{GUT}}$, these two incomplete $SU(5)$ doublet pairs modify the running of the gauge and Yukawa couplings.\footnote{In $SU(5)$ normalization, they shift the one-loop $G_{\mathrm{SM}}$ beta-function coefficients by $(\Delta b_1, \Delta b_2, \Delta b_3) = (\frac{6}{5}, 2, 0)$, changing $\alpha_i^{-1}(M_{\mathrm{GUT}})$ by $-\frac{\Delta b_i}{2\pi}\ln(M_{\mathrm{GUT}}/M_H)$ and adding positive Higgs wave-function anomalous dimensions to the Yukawa RGEs.} This spoils precision one-step unification unless $M_H$ lies close to $M_{\mathrm{GUT}}$ or threshold corrections compensate.
\end{itemize}
Constructing the deformed non-Abelian bundle metrics and solving for the stabilized point $(C + \Delta C, \langle S_{ab} \rangle)$ explicitly is beyond the scope of the present split-locus analysis. In what follows, we therefore work directly on $\cJ_{\mathrm{Higgs}}$ at $\langle S_{ab} \rangle = 0$ and treat the light Higgs directions $\hat{v}^u$ and $\hat{v}^d$ as effective parameters. When evolving couplings to low energies, we adopt GUT-scale decoupling of the two heavy Higgs pairs.

\subsection{Moduli search and numerical setup}
\label{subsec:5273_numerical}

\begin{figure}[t]
\centering
\begin{tikzpicture}[
    font=\footnotesize,
    >=Latex,
    box/.style={
        rounded corners=3pt,
        line width=0.65pt,
        draw=black!65,
        fill=black!3,
        align=center,
        inner xsep=4pt,
        inner ysep=5pt
    },
    algbox/.style={
        box,
        draw=blue!50!black,
        fill=blue!4,
        minimum height=1.18cm
    },
    pdebox/.style={
        box,
        draw=teal!60!black,
        fill=teal!4,
        minimum height=1.18cm
    },
    arrow/.style={->, line width=0.75pt, draw=black!75},
    lineonly/.style={line width=0.75pt, draw=black!75},
    lbl/.style={font=\scriptsize, text=black!80, inner sep=2pt}
]

\node[box, text width=2.95cm, minimum height=1.55cm, anchor=west] (search) at (0, 0) {
    \textbf{Moduli Search}\\[2pt]
    Optimize $C \in \cJ_{\mathrm{Higgs}}$\\
    for smooth $X,\, \hat{X}$
};

\coordinate (split) at (3.68, 0);

\node[algbox, text width=6.54cm, anchor=west] (alg) at (4.15, 0.92) {
    \textbf{Holomorphic Couplings (Algebraic)}\\[2pt]
    Koszul residue $I_{IJK}(C)$ and volume $I_{\Omega}$
};

\node[pdebox, text width=1.74cm, anchor=west] (pde1) at (4.15, -0.92) {
    \textbf{Ricci-Flat}\\[2pt]
    $g_{\mathrm{CY}}$
};

\node[pdebox, text width=1.74cm, anchor=west] (pde2) at (6.55, -0.92) {
    \textbf{HYM}\\[2pt]
    $H_a$
};

\node[pdebox, text width=1.74cm, anchor=west] (pde3) at (8.95, -0.92) {
    \textbf{Harmonic}\\[2pt]
    $\nu_I^{\mathrm{NN}}$
};

\node[lbl, anchor=north] at (7.56, -1.62) {Matter field K\"ahler metrics $\cK_{I\bar{I}}$ (neural PDEs)};

\coordinate (topmerge) at (12.42, 0.92);
\coordinate (botmerge) at (12.42, -0.92);
\coordinate (merge) at (12.42, 0);

\node[box, text width=3.05cm, minimum height=1.55cm, anchor=west] (phys) at (12.87, 0) {
    \textbf{Physical Yukawas}\\[2pt]
    $Y_{IJK}$ and 2-loop RG\\
    to $m_q,\, V_{\mathrm{CKM}}$
};

\draw[lineonly] (search.east) -- (split);
\draw[arrow] (split) |- (alg.west);
\draw[arrow] (split) |- (pde1.west);

\draw[arrow] (pde1.east) -- (pde2.west);
\draw[arrow] (pde2.east) -- (pde3.west);

\draw[lineonly] (alg.east) -- node[above=2pt, lbl] {$I_{IJK},\, I_{\Omega}$} (topmerge) -- (merge);
\draw[lineonly] (pde3.east) -- node[above=2pt, lbl] {$\cK_{I\bar{I}}$} (botmerge) -- (merge);
\draw[arrow] (merge) -- (phys.west);

\end{tikzpicture}
\caption{Workflow for computing physical quark masses on CICY 5273, Model~500. Smooth candidate geometries $\hat{X}$ are found on the Higgs locus $\cJ_{\mathrm{Higgs}}$ using reference-metric estimates and global smoothness checks. At each selected geometry, the holomorphic couplings $I_{IJK}(C)$ are computed algebraically via Eq.~\eqref{eq:poincare_residue_lift_master}, while the matter field K\"ahler metrics $\cK_{I\bar{I}}$ are computed by solving the Ricci-flat, HYM, and harmonic $(0,1)$-form equations sequentially with neural networks, before evolving $Y_{IJK}$ to low energies via two-loop renormalization group equations.}
\label{fig:5273_pipeline_schematic}
\end{figure}
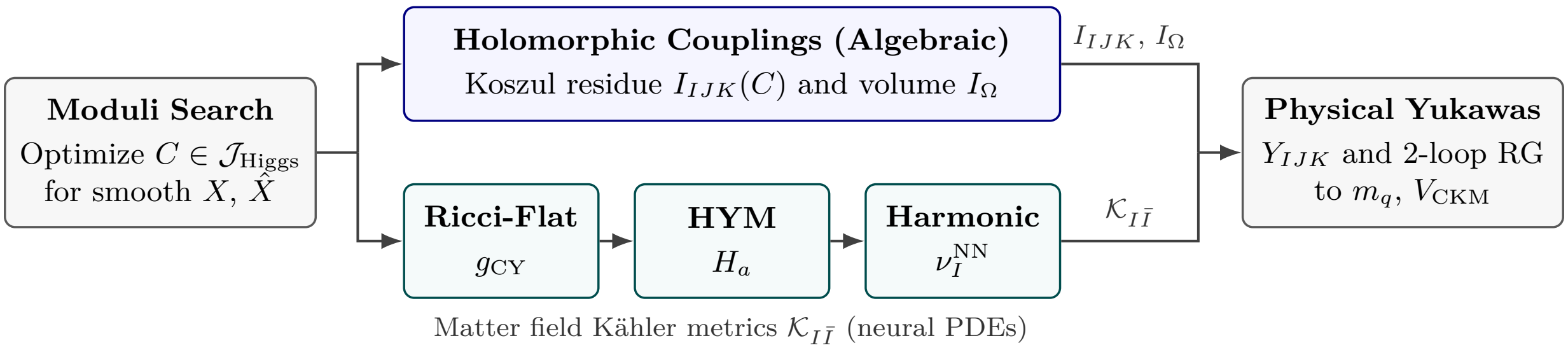

Searching the 14-dimensional sub-locus $\cJ_{\mathrm{Higgs}}$ directly with full numerical PDE solutions is impractical, since solving the Ricci-flat, HYM, and harmonic $(0,1)$-form equations on $N = 10^6$ points takes $\cO(1\text{--}2)$ GPU-hours per complex structure point. Following Section~\ref{subsec:yukawa_couplings} and~\cite{Chervonyi:2026lehmer}, we therefore split the search and Yukawa evaluation into three steps, cf.\ Figure~\ref{fig:5273_pipeline_schematic}:
\begin{enumerate}
  \item \emph{Moduli search on $\cJ_{\mathrm{Higgs}}$.} We parameterize the defining polynomials $P_1, \dots, P_5$ along $\cJ_{\mathrm{Higgs}}$ and optimize the polynomial coefficients $C$ across parallel workers and Higgs directions $(\hat{v}^u, \hat{v}^d)$ to obtain realistic quark mass ratios and CKM mixings. At each step, the six up-type couplings $W_{H_{ab} Q_a u^c_b}$ and three down-type couplings $W_{Q_a b^c H_{d,jk}}$ are computed algebraically via Eq.~\eqref{eq:poincare_residue_lift_master}. The volume integral $I_{\Omega}$ is evaluated using numerical quadrature integration procedures. The matter field K\"ahler metrics $\cK_{I\bar{I}}$ are approximated using the reference forms $\nu_I^{\mathrm{ref}}$.
  \item \emph{Smoothness checks.} Searching for large hierarchies often drives $\hat X$ toward singular loci. Since checking smoothness over a finite point sample can miss singularities, we verify globally over $\hat X$ that the Jacobian $\dd P$ retains full rank to ensure that the complex structure search does not stabilize close to or at conifold points, or tune the complex structure such that the generically free action $g_{\mathrm{free}}$ develops singularities.
  \item \emph{Numerical PDEs and RG evolution.} At smooth candidate points, we solve the geometric PDEs on $N = 10^6$ points~\cite{Douglas:2006hz,larfors2022numericalmetricscompleteintersection} ($90\%$ training, $10\%$ validation) across three independent random seeds on NVIDIA B200 GPUs, evaluate $\cK_{I\bar{I}}$ on an independent test sample of $9 \times 10^5$ points, and evolve the resulting GUT-scale Yukawa couplings~\eqref{eq:exact_physical_yukawa} to low energies via two-loop MSSM renormalization group equations, see Appendix~\ref{app:benchmarks}.
\end{enumerate}

To compute $\cK_{I\bar{I}}$ in the third step, we solve the three differential equations sequentially with neural networks~\cite{larfors2022numericalmetricscompleteintersection,constantin2024computationquarkmassesstring,Lust:2026warped}, see Appendix~\ref{app:ml_setup} for details. First, one network solves the Monge--Amp\`ere equation for the K\"ahler potential deformation $\phi$ in 
\begin{equation}
    g_{\mathrm{CY}} = g_{\mathrm{FS}} + \partial\bar{\partial}\phi\,.
\end{equation}
Second, at fixed $g_{\mathrm{CY}}$, three networks solve the zero-slope HYM equations for $\beta_{L_2}, \beta_{L_3}, \beta_{L_4}$ in 
\begin{equation}
    H_a = H_a^{\mathrm{FS}} e^{\beta_a}\,,
\end{equation} 
from which the remaining composite bundle metrics follow algebraically via $L_1 \otimes L_5 = (L_2 \otimes L_3 \otimes L_4)^*$. Normalizing each $H_I$ on $X$ sets $c_{IJK} = 1$ in Eq.~\eqref{eq:exact_physical_yukawa}. Third, at fixed $(g_{\mathrm{CY}}, H_I)$, $13$ networks solve the harmonic $(0,1)$-form equations 
\begin{align}
\bar{\partial}_{X,H_I}^\dagger(\nu_I^{\mathrm{ref}} + \bar{\partial}_X \sigma_I) = 0    
\end{align}
for the sections $\sigma_I \in \Omega^{0,0}(\hat{X}, U_I)$, giving $17$ networks per seed in total. When computing $Y_{IJK}$ in~\eqref{eq:exact_physical_yukawa}, we evaluate $\cK_{I\bar{I}}$ with the trained harmonic forms $\nu_I^{\mathrm{NN}} = \nu_I^{\mathrm{ref}} + \bar{\partial}_X \sigma_I^{\mathrm{NN}}$ while keeping the algebraic values~\eqref{eq:poincare_residue_lift_master} for $I_{IJK}(C)$: by Eq.~\eqref{eq:stokes_holomorphic_invariance}, the exact shifts $\bar{\partial}_X \sigma_I$ drop out of $I_{IJK}(C)$, whereas Monte Carlo integration of $\nu_I^{\mathrm{NN}}$ introduces $\cO(10^{-3})$ to $\cO(10^{-4})$ sampling noise that would drown out the light up-quark couplings $|W^u| \sim 10^{-8}$.

\subsection{Quark mass results}
\label{subsec:5273_results}

We present results for the quark masses for two choices of electroweak Higgs mixing: first with equal ``democratic'' Higgs VEVs at $t_{\mathrm{SUGRA}}$,  compared against GUT-scale target windows to verify that the Calabi--Yau geometry alone generates the hierarchy without Higgs tuning, and second for aligned Higgs VEVs evolved to low energies under the two-loop MSSM renormalization group to compare directly with experiment.

\begin{table}[t]
\centering
\footnotesize
\setlength{\tabcolsep}{4.2pt}
\begin{tabular}{@{}llccccc@{}}
\toprule
Benchmark & Target regime & $\min_{X} \dd P$ & $Y_t v$ & $Y_c v$ & $Y_u v$ & $Y_b v$ \\
 & & & [GeV] & [MeV] & [MeV] & [GeV] \\
\midrule
Target window & MSSM ($\tan\beta = 6$) & -- & $[55, 100]$ & $[160, 320]$ & $[0.30, 0.85]$ & $[4.8, 7.5]$ \\
Benchmark 1 & MSSM ($\tan\beta = 6$) & $0.03459$ & $50.43 \pm 0.14$ & $161.0 \pm 0.5$ & $0.969 \pm 0.034$ & $5.58 \pm 0.06$ \\
\midrule
Target window & MSSM ($\tan\beta = 10$) & -- & $[55, 100]$ & $[160, 320]$ & $[0.30, 0.85]$ & $[7.5, 12.0]$ \\
Benchmark 2 & MSSM ($\tan\beta = 10$) & $0.01054$ & $54.83 \pm 0.22$ & $258.5 \pm 1.3$ & $0.615 \pm 0.013$ & $6.97 \pm 0.05$ \\
\midrule
Target window & SM GUT ($\tan\beta \simeq 1$) & -- & $[55, 95]$ & $[170, 330]$ & $[0.30, 0.85]$ & $[0.85, 1.35]$ \\
Benchmark 3 & SM GUT ($\tan\beta \simeq 1$) & $0.02831$ & $53.30 \pm 0.50$ & $243.5 \pm 2.3$ & $0.688 \pm 0.033$ & $1.170 \pm 0.030$ \\
\bottomrule
\end{tabular}
\caption{Democratic-VEV benchmarks ($\hat{v}^u = \hat{v}^d = (1,1,1)/\sqrt{3}$) on $\cJ_{\mathrm{Higgs}}$, evaluated across three random seeds ($N = 10^6$ points) at the supergravity K\"ahler point $t_{\mathrm{SUGRA}}$ in~\eqref{eq:t_sugra_point} with $\alpha_{\mathrm{GUT}}^{-1} = 24.30$. Uncertainties show the standard deviation across the three seeds. For direct comparison across regimes, we list $Y_f v$, where $Y_f$ are the singular values of $Y^{u,d}$ ($m_{t,c,u}(M_{\mathrm{GUT}}) = Y_{t,c,u} v \sin\beta$ and $m_b(M_{\mathrm{GUT}}) = Y_b v \cos\beta$, giving $m_b(M_{\mathrm{GUT}}) = 0.918 \pm 0.010\text{ GeV}$ on Benchmark~1 at $\tan\beta = 6$ and $0.693 \pm 0.005\text{ GeV}$ on Benchmark~2 at $\tan\beta = 10$).}
\label{tab:5273_democratic_sugra}
\end{table}

Fixing democratic Higgs mixing $\hat{v}^u = \hat{v}^d = (1,1,1)/\sqrt{3}$ ensures that no hierarchy is introduced through VEV alignment. To construct Benchmarks~1--3 in Table~\ref{tab:5273_democratic_sugra}, we first optimize the polynomial coefficients on $\cJ_{\mathrm{Higgs}}$ using reference K\"ahler metrics at $t_{\mathrm{HYM}}(1)$ while keeping $\hat{X}$ smooth and $g_{\mathrm{free}}$ fixed-point-free, and then solve the Monge--Amp\`ere, HYM, and harmonic $(0,1)$-form equations directly at $t_{\mathrm{SUGRA}}$. These correspond to GUT-scale target windows for the minimal one-Higgs-pair MSSM at $\tan\beta = 6$ and $\tan\beta = 10$ and for non-supersymmetric SM running~\cite{Ross:2007az,Xing:2007fb,Antusch:2013jca}, assuming that two of the three Higgs pairs decouple near $M_{\mathrm{GUT}}$ as discussed in Section~\ref{subsec:5273_higgs_locus}. Across all three benchmarks, the numerical K\"ahler volumes $\cV_X^{\mathrm{NN}}$ agree with the intersection volume $\cV_X = 428.60$ to within $0.08\%$, and the relative variations across seeds are small: $0.28\%\text{--}0.95\%$ for $(Y_t v, Y_c v)$, $2.1\%\text{--}4.8\%$ for $Y_u v$, and $0.66\%\text{--}2.55\%$ for $Y_b v$. Even for democratic Higgs VEVs and well away from conifold points, the geometry alone produces the five-order-of-magnitude hierarchy across $(m_t, m_c, m_u, m_b)$. Moving from $t_{\mathrm{HYM}}(1)$ to $t_{\mathrm{SUGRA}}$ lowers $Y_t v$ by $10\%\text{--}20\%$, leaving $Y_t v \simeq 50.4\text{--}54.8\text{ GeV}$ just at the lower bound of the $[55, 100]\text{ GeV}$ target window when the dominant Higgs channel is diluted by $1/\sqrt{3}$. Mild $\cO(1)$ Higgs alignment brings all four masses inside their target windows.

\begin{table}[t]
\centering
\footnotesize
\setlength{\tabcolsep}{3.5pt}
\begin{tabular}{@{}lcccc@{}}
\toprule
Observable & \thead{GUT ($M_{\mathrm{GUT}}$)\\($\alpha_{\mathrm{GUT}}^{-1} = 24.30$)} & \thead{Low energy\\($\alpha_{\mathrm{GUT}}^{-1} = 24.30$)} & \thead{Low energy\\($\alpha_{\mathrm{GUT}}^{-1} = 25.53$)} & \thead{Experiment\\\cite{PDG:2024,Antusch:2013jca}} \\
\midrule
$m_t$ ($Y_t v$ / $m_t^{\mathrm{pole}}$) & $85.45 \pm 0.02\text{ GeV}$ & $170.85 \pm 0.02\text{ GeV}$ & $169.47 \pm 0.02\text{ GeV}$ & $172.57 \pm 0.29\text{ GeV}$ \\
$m_c$ ($Y_c v$ / $m_c(m_c)$) & $252.9 \pm 2.1\text{ MeV}$ & $1.300 \pm 0.008\text{ GeV}$ & $1.284 \pm 0.008\text{ GeV}$ & $1.273 \pm 0.005\text{ GeV}$ \\
$m_u$ ($Y_u v$ / $m_u(2\text{ GeV})$) & $0.563 \pm 0.011\text{ MeV}$ & $2.51 \pm 0.05\text{ MeV}$ & $2.47 \pm 0.05\text{ MeV}$ & $2.16 \pm 0.07\text{ MeV}$ \\
$m_b$ ($m_b(M_{\mathrm{GUT}})$ / $m_b(m_b)$) & $1.046 \pm 0.018\text{ GeV}$ & $4.56 \pm 0.07\text{ GeV}$ & $4.47 \pm 0.06\text{ GeV}$ & $4.18 \pm 0.03\text{ GeV}$ \\
$m_s,\; m_d$ (split locus) & $0$ & $0$ & $0$ & $93.5,\; 4.70\text{ MeV}$ \\
\midrule
$|V_{tb}|(M_Z)$ & -- & $0.99916$ & $0.99916$ & $0.99912 \pm 0.00004$ \\
$|V_{cb}|(M_Z)$ & -- & $0.04088$ & $0.04088$ & $0.0408 \pm 0.0014$ \\
$|V_{ub}|(M_Z)$ & -- & $0.00333$ & $0.00333$ & $0.00382 \pm 0.00020$ \\
\bottomrule
\end{tabular}
\caption{Physical quark masses and heavy-quark CKM mixing at Benchmark~4 on $\cJ_{\mathrm{Higgs}}$, evaluated across three random seeds on an independent test sample of $9 \times 10^5$ points at $t_{\mathrm{HYM}}(1)$ with $\tan\beta = 6.0$, $M_{\mathrm{SUSY}} = 3\text{ TeV}$, $\hat{v}^u = (0.8148, 0.4099, 0.4099)$, and $\hat{v}^d = (0.9936, 0.1122, 0.0123)$. Uncertainties show the standard deviation across the three seeds; at $M_{\mathrm{GUT}}$, the up-type rows list $Y_f v = m_f(M_{\mathrm{GUT}}) / \sin\beta$, while for the bottom quark $Y_b v = m_b(M_{\mathrm{GUT}}) / \cos\beta = 6.36 \pm 0.11\text{ GeV}$ at $\alpha_{\mathrm{GUT}}^{-1} = 24.30$.}
\label{tab:5273_aligned_benchmark}
\end{table}

\begin{figure}[t]
  \centering
  \includegraphics[width=\textwidth]{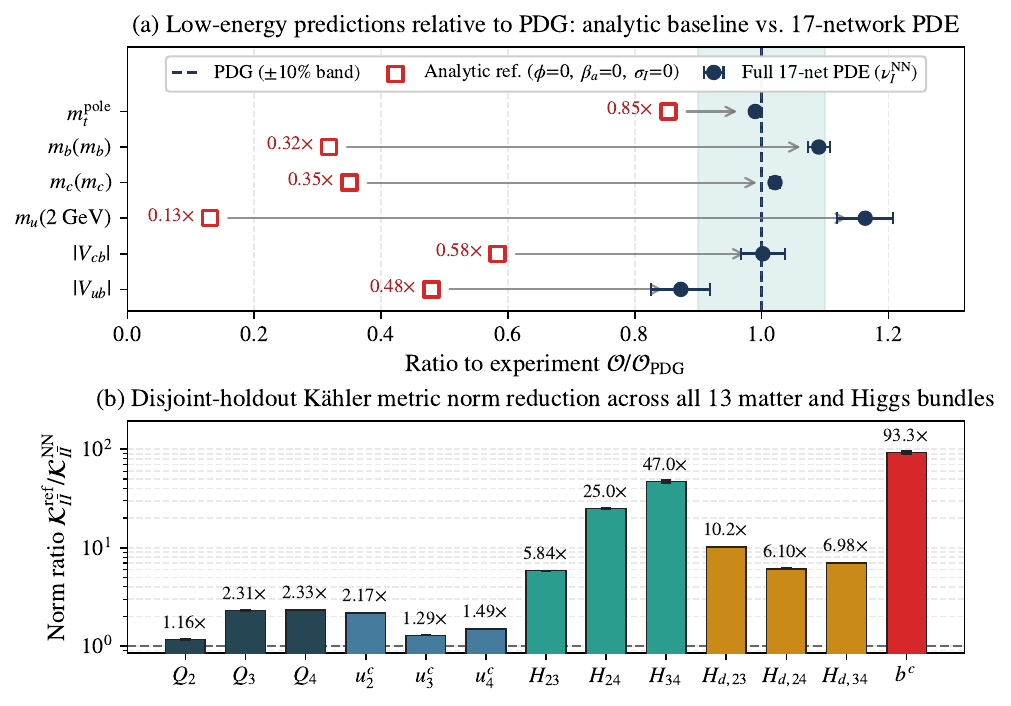}
  \caption{Effect of the numerical metric and harmonic-form solutions on physical observables and matter field K\"ahler metrics at Benchmark~4 ($\tan\beta = 6.0$, $M_{\mathrm{SUSY}} = 3\text{ TeV}$, $\alpha_{\mathrm{GUT}}^{-1} = 24.30$; three seeds evaluated on an independent $9 \times 10^5$-point test sample). \textbf{(a)} Ratio of predicted low-energy observables to PDG central values ($\pm 10\%$ shaded band), comparing the analytic reference basis ($\phi = 0, \beta_a = 0, \sigma_I = 0$, open squares) with the full numerical PDE solution (filled circles). \textbf{(b)} K\"ahler metric ratios $\cK_{I\bar{I}}^{\mathrm{ref}} / \cK_{I\bar{I}}^{\mathrm{NN}}$ across all $13$ matter and Higgs bundles, showing $\cK_{I\bar{I}}^{\mathrm{NN}} < \cK_{I\bar{I}}^{\mathrm{ref}}$ and large exact-form subtractions on the multi-step Koszul bundles ($H_{ab}, H_{d,ab}, b^c$).}
  \label{fig:5273_pde_impact}
\end{figure}

To compare directly with low-energy data, allowing the light electroweak Higgs doublets to align along 
\begin{equation}
    \hat{v}^u \approx (0.81,\allowbreak 0.41,\allowbreak 0.41)\,,\qquad
    \hat{v}^d \approx (0.99,\allowbreak 0.11,\allowbreak 0.01)
\end{equation}
at $\tan\beta = 6.0$ and $M_{\mathrm{SUSY}} = 3\text{ TeV}$ yields Benchmark~4 shown in Table~\ref{tab:5273_aligned_benchmark} and Figure~\ref{fig:5273_pde_impact}. Although the PDEs for Benchmark~4 are solved along the $\xi = 1$ ray at $t_{\mathrm{HYM}}(1)$ rather than at the tilted ray $t_{\mathrm{SUGRA}}$, tree-level physical Yukawa couplings and the HYM and harmonic-form equations are invariant under overall K\"ahler rescalings $t^i \mapsto \lambda t^i$, as shown in Section~\ref{subsec:moduli_dependence}. The resulting masses therefore are identical at any rescaled supergravity point $\lambda\, t_{\mathrm{HYM}}(1)$ with $\lambda \ge 2.76$, where $\min_i t_i \ge 1$. Evaluated across three independent random seeds on a common test sample of $9 \times 10^5$ points, the numerical volume $\cV_X^{\mathrm{NN}} = 62.237 \pm 0.006$ reproduces the intersection volume $\cV_X = 62$, and every bundle and seed obeys $0 < \cK_{I\bar{I}}^{\mathrm{NN}} < \cK_{I\bar{I}}^{\mathrm{ref}}$ as expected from~\eqref{eq:pythagorean_monotonicity}. Note that $\cK_{I\bar{I}}$ are reduced by factors of up to $93.3$ for $b^c$ and $47.0$ for $H_{34}$. The relative variation across seeds lies between $0.02\%$ and $1.95\%$ on $(m_t, m_c, m_u, m_b)$, with a separate Monte Carlo sampling uncertainty of $0.3\%\text{--}4\%$ across independent $9 \times 10^5$-point samples. Running the GUT-scale Yukawa matrices down to low energies at two loops\footnote{This is evaluated in Table~\ref{tab:5273_aligned_benchmark} both at the canonical MSSM unification benchmark $\alpha_{\mathrm{GUT}}^{-1} = 24.30$ and at the two-loop RGE-matched value $\alpha_{\mathrm{GUT}}^{-1} = 25.53$ at $M_{\mathrm{GUT}} = 2 \times 10^{16}\text{ GeV}$ for $M_{\mathrm{SUSY}} = 3\text{ TeV}$.} reproduces $m_t^{\mathrm{pole}}$, $m_c(m_c)$, $m_u(2\text{ GeV})$, and $m_b(m_b)$ as well as $|V_{tb}|$, $|V_{cb}|$, and $|V_{ub}|$ in close agreement with experiment~\cite{PDG:2024}. In contrast, evaluating the same holomorphic couplings in the analytic reference basis with $\phi = 0$, $\beta_a = 0$, and $\sigma_I = 0$ underestimates $(m_b, m_c, m_u)$ by factors of $3\text{--}8$ because of exact-form contributions to $\cK_{I\bar{I}}^{\mathrm{ref}}$. The point on $\cJ_{\mathrm{Higgs}}$, the Higgs directions $(\hat{v}^u, \hat{v}^d)$, and $(\tan\beta, M_{\mathrm{SUSY}})$ are chosen phenomenologically to show that smooth realistic points exist within the split-bundle moduli space. We have not studied whether dynamical moduli stabilization mechanisms would drive the complex structure to such points. Note also that our RG evolution assumes standard one-pair MSSM (and single-Higgs SM below $M_{\mathrm{SUSY}}$) beta functions, treating the two orthogonal Higgs pairs as decoupled at $M_{\mathrm{GUT}}$ rather than running a three-Higgs-doublet model, and omits model-dependent one-loop supersymmetric threshold corrections of a few percent.

\section{Conclusions}
\label{sec:conclusions}

In this paper, we studied whether realistic quark mass hierarchies can arise at tree level from smooth points in the complex structure moduli space of Calabi--Yau compactifications in $E_8 \times E_8$ heterotic line bundle models. Combining Koszul spectral-sequence methods, an algebraic Koszul--Dolbeault residue integration scheme for the holomorphic superpotential couplings, and numerical solutions of the Ricci-flat, Hermitian Yang--Mills, and harmonic $(0,1)$-form equations, we obtained both sharp no-go results on previously studied tetra-quadric models and an explicit realization of five-order-of-magnitude quark mass hierarchies in the smooth supergravity regime.

We showed that tree-level hierarchies across complex structure moduli space can arise from two distinct mechanisms: smooth zero-crossings of moduli-dependent holomorphic superpotential couplings (Mechanism~I) and K\"ahler metric localization near factorized polynomial loci (Mechanism~II). Both are governed by cohomological selection rules. Mechanism~I requires uncanceled defining polynomials or moduli-dependent Koszul kernels so that the holomorphic Yukawa coupling integral does not reduce to a moduli-independent constant. And Mechanism~II requires the competing harmonic $(0,1)$-forms to remain cohomologically non-exact on the pinched locus to generate a hierarchy from the matter field K\"ahler metrics.

On the $\ZZ_2 \times \ZZ_2$ quotient of the tetra-quadric (CICY 7862) studied in~\cite{Buchbinder_2014,Blesneag_2016,constantin2024computationquarkmassesstring}, both mechanisms are obstructed at tree level at the split-bundle locus. First, integration by parts cancels the defining polynomial out of the holomorphic couplings and exposes an accidental continuous ambient $U(1)_2$ symmetry, ruling out Mechanism~I. Second, along completed-square degenerations where the reference basis produces $(m_t/m_c)_{\mathrm{ref}} \approx 220$, degree-$(2,2)$ saturation on $\PP^1_1 \times \PP^1_2$ renders the competing Koszul $(0,1)$-form $\bar{\partial}_X$-exact and thereby collapses the physical quark mass ratio back to $m_t / m_c \simeq 2.1$, obstructing Mechanism~II. Across the smooth interior of its $20$-dimensional complex structure moduli space, $m_t / m_c$ remains bounded by $2\text{--}3$, with $m_u = 0$ and $Y^d = 0$.

Scanning all $63{,}325$ $SU(5)$ and $2{,}112$ Standard Model line bundle configurations in the CICY databases~\cite{Anderson:2012yf,Anderson:2013xka}, we found that achieving $\rk Y^u = 3$ at the split locus requires each family to arise from a separate line bundle and three light Higgs. The latter requirement was not realized in any of the known models, so we extended the search to computing spectra along cohomology jumping loci where the number of massless particles can increase. We also imposed opposite hypercharge Wilson-line parities on $Q_a$ and $u^c_a$ to avoid the bound $m_t \le m_c + m_u$. CICY 5273, Model~500 on $[(\PP^1)^4 \times (\PP^2)^2] / \ZZ_2$ is the unique smooth quotient in the database satisfying all of these requirements. It has a full-rank up-type Yukawa matrix, $\rk Y^u = 3$, and a rank-one down-type Yukawa matrix, $\rk Y^d = 1$. We emphasize that our mechanism makes it possible in principle to obtain realistic mass hierarchies for all Standard Model quarks. The fact that $Y^d$ has rank one is a limitation of the currently existing Standard Model databases and not a fundamental obstruction towards obtaining fully realistic quark masses with the mechanisms laid out in this paper.

Since the ambient space of the model contains $\PP^2 \times \PP^2$ factors and the CICY is at codimension $K = 5$, the model supports moduli-dependent Koszul kernels while leaving harmonic forms on some of the $\PP^1$ factors unpinched. This allows us to employ both hierarchy mechanisms inside the smooth interior of the model's $14$-dimensional complex structure locus $\cJ_{\mathrm{Higgs}}$ where the number of massless Higgs fields jumps from zero to three. Combining this exact algebraic structure with a three-stage search and a neural PDE solver pipeline in Google Antigravity~\cite{antigravity2026} with Gemini~\cite{geminiteam2023gemini,geminiteam2024gemini15,geminiteam2025gemini25,gemini4argon2026,gemini4carbon2026}, we evaluated the ten-dimensional $E_8 \times E_8$ supergravity couplings numerically at smooth points far from conifold singularities and at a point $t_{\mathrm{SUGRA}}$ in K\"ahler moduli space where all K\"ahler moduli are larger than 1 and worldsheet instanton corrections are small. We reproduced realistic hierarchies for all four tree-level massive quarks $(t, c, u, b)$ as well as heavy-quark CKM mixing, both with democratic Higgs VEVs and after two-loop MSSM renormalization group evolution.

Several directions remain open in extending this split-locus benchmark toward a complete heterotic Standard Model vacuum. Since $\rk Y^d \le 1$ across all split line bundle models with $\rk Y^u = 3$ in the existing databases, generating non-zero strange and down quark masses ($m_s, m_d > 0$) and the Cabibbo angle within these models requires deforming $V = \bigoplus_{a=1}^5 L_a$ into a non-Abelian extension bundle via non-vanishing VEVs $\langle S_{ab} \rangle \neq 0$ of $S(U(1)^5)$-charged bundle moduli singlets~\cite{Anderson:2010,Buchbinder_2014,Constantin:2024fn}. Computing the resulting higher-order couplings and non-Abelian HYM metrics numerically on CICY 5273, Model~500 is a natural next step. Alternatively, we could use an AI-guided search for new line bundle models that can satisfy all constraints at the split locus. In addition, we selected the complex structure point on $\cJ_{\mathrm{Higgs}}$, the K\"ahler moduli, and the light Higgs directions $(\hat{v}^u, \hat{v}^d)$ by hand to illustrate the existence of points in the moduli space that allow for phenomenologically viable quark masses. Embedding them into a dynamical moduli-stabilization framework via holomorphic Chern--Simons superpotentials and hidden-sector gaugino condensation~\cite{Anderson:2011,Cicoli:2013,Ashmore:2018}, and extending our analysis to the charged-lepton and neutrino sectors, are interesting avenues for future work.

\section*{Acknowledgments}
Throughout this work, we used internal versions of Gemini~\cite{geminiteam2023gemini,geminiteam2024gemini15,geminiteam2025gemini25}, including Gemini~4 Argon~\cite{gemini4argon2026} and Carbon~\cite{gemini4carbon2026}, paired with Google Antigravity~\cite{antigravity2026} and compute resources on Google Cloud. We thank Alex Davies, who motivated Yuri Chervonyi to write a proposal for this project and connected him with Fabian Ruehle and Nenad Tomasev. We thank Quoc Le and Thang Luong for support during this work. We thank Johannes Bausch, Ray Jiang, Sergei Gukov, Pushmeet Kohli, and Demis Hassabis for useful discussions and comments at the early stages of the project.

\appendix
\section{Koszul Forms and Residue Contractions}
\label{app:koszul}

In this appendix, we summarize the analytic construction of $\bar{\partial}_X$-closed reference $(0,1)$-forms $\nu_I^{\mathrm{ref}}$ for the $13$ matter and Higgs line bundles on CICY 5273, Model~500 and the algebraic total-cocycle contractions used to evaluate $I_{IJK}(C)$ in closed form.

\subsection{Fubini--Study Green's operators}
\label{app:subsec:homotopies}
To solve the Koszul--Dolbeault descent equations~\eqref{eq:staircase_equations} starting from an ambient $(0, r_{\max}+1)$-form ($0 \le r_{\max} \le K$), we invert $\bar{\partial}_{\cA}$ step by step down to $\nu^{\mathrm{ref}} = \Psi^{(0,1)}|_X$ using the Green's operator 
\begin{equation}
    G_r = \bar{\partial}_{\PP^{n_r}}^\dagger \Delta_{\bar{\partial}}^{-1}
\end{equation}
on each projective factor $\PP^{n_r}$. Since $G_r$ maps any $\bar{\partial}$-exact bundle-valued $(0,q)$-form $\beta$ to its unique co-exact preimage $\alpha = G_r \beta$ ($\bar{\partial}_{\PP^{n_r}} \alpha = \beta$, $\bar{\partial}_{\PP^{n_r}}^\dagger \alpha = 0$), each step of the descent reduces to algebraic polynomial identities~\cite{Blesneag_2016,Blesneag_2017,constantin2024computationquarkmassesstring}. 

On $\PP^1$ with homogeneous coordinates $w = [w_0 : w_1]$, $\kappa = |w_0|^2 + |w_1|^2$, and Fubini--Study harmonic $(0,1)$-form 
\begin{equation}
    \mu_{\PP^1} = \frac{-\bar{w}_1 \dd\bar{w}_0 + \bar{w}_0 \dd\bar{w}_1}{\kappa^{2}} \quad\in \Omega^{0,1}(\PP^1, \cO(-2))\,,
\end{equation}
any polynomial $R(w, \bar{w})$ homogeneous of degree $k - 1 \ge 1$ in $\bar{w}$ obeys
\begin{equation}
  \bar{\partial}_{\PP^1}\!\left(\frac{R(w, \bar{w})}{\kappa^{k-1}}\right)
  = \frac{w_0 \partial_{\bar{w}_1} R - w_1 \partial_{\bar{w}_0} R}{\kappa^{k-2}}\,\mu_{\PP^1}\,.
  \label{eq:app_p1_master}
\end{equation}
We use~\eqref{eq:app_p1_master} in three different ways for CICY 5273. 

\textbf{i.)} For $k = 2$ and a linear form $P(w) = b_0 w_0 + b_1 w_1$, setting $R(w, \bar{w}) = (-b_1 \bar{w}_0 + b_0 \bar{w}_1)$ gives
\begin{equation}
\bar{\partial}_{\PP^1} f_P(w,\bar{w}) =:\bar{\partial}_{\PP^1} \left(\frac{R(w, \bar{w})}{\kappa}\right) = P(w)\,\mu_{\PP^1}\,.
\end{equation}

\textbf{ii.)} For $k = 3$ on $\PP^1_2$, let
\begin{equation}
  P_5(x_2, z_5, z_6) = (x_{2,0}^2 + x_{2,1}^2) P_+ + (x_{2,0}^2 - x_{2,1}^2) P_- + x_{2,0} x_{2,1} P_\times
\end{equation}
with $P_\pm = z_5^{\mathsf{T}} S_\pm z_6$ and $P_\times = z_5^{\mathsf{T}} S_{\mathrm{cross}} z_6$ (viewing $z_s = (z_{s,0}, z_{s,1}, z_{s,2})^{\mathsf{T}}$ as column vectors) for the $3 \times 3$ matrices $S_\pm, S_{\mathrm{cross}}$ in~\eqref{eq:app_S_matrices}. Defining
\begin{equation}
  u_\pm = x_{2,0} \pm x_{2,1}\,,\qquad
  d_{\bar{x}} = \bar{x}_{2,0}^2 - \bar{x}_{2,1}^2\,,\qquad
  e_{\bar{x}} = \bar{x}_{2,0}^2 + \bar{x}_{2,1}^2\,,
\end{equation}
we have
\begin{equation}
  R_{\pm}(x_2, \bar{x}_2, z_5, z_6)
  := \pm\frac{1}{2} P_- u_\pm \bar{u}_\pm^2 + P_+\!\left(u_\mp \bar{x}_{2,0}\bar{x}_{2,1} \mp \frac{1}{2} u_\pm d_{\bar{x}}\right) - \frac{1}{4} P_\times(u_\pm d_{\bar{x}} + u_\mp e_{\bar{x}})\,,
  \label{eq:app_R_r1}
\end{equation}
so that 
\begin{equation}
    \bar{\partial}_{\PP^1_2}\left(\frac{R_\pm}{\kappa_2^2}\right) = \frac{P_5\bar{u}_\pm}{\kappa_2}\,\mu_{\PP_2^1}
\end{equation}
identically.

\textbf{iii.)} For $k = 4$ on $\PP^1_2$, defining
\begin{equation}
\begin{aligned}
\Psi_2 &= \frac{1}{3\kappa_2^3}\Big[P_-\big(x_{2,0}\bar{x}_{2,1}(3\bar{x}_{2,0}^2 + \bar{x}_{2,1}^2) + x_{2,1}\bar{x}_{2,0}(\bar{x}_{2,0}^2 + 3\bar{x}_{2,1}^2)\big) \\
&\qquad\quad - P_\times (x_{2,0}\bar{x}_{2,0}^3 - x_{2,1}\bar{x}_{2,1}^3) + P_+ N_+\Big]\,,
\end{aligned}
\label{eq:app_Psi1}
\end{equation}
with $N_+ = x_{2,0}\bar{x}_{2,0}^3 - x_{2,1}\bar{x}_{2,1}^3 - 3(x_{2,0}\bar{x}_{2,0}\bar{x}_{2,1}^2 - x_{2,1}\bar{x}_{2,0}^2\bar{x}_{2,1})$, Eq.~\eqref{eq:app_p1_master} gives 
\begin{equation}
    \bar{\partial}_{\PP^1_2}\Psi_2 = \frac{P_5 (\bar{x}_{2,0}^2 + \bar{x}_{2,1}^2)}{\kappa_2^{2}}\mu_{\PP_2^1}
\end{equation}
up to a harmonic remainder proportional to $P_+$ on $\PP^1_2$ that cancels at the second Koszul step on $(\PP^1)^3$ along $\cJ_{\mathrm{Higgs}}$.

Similarly, on the two $\PP^2_s$ ambient space factors labeled by $s \in \{5,6\}$ with homogeneous coordinates $z_s = [z_{s,0} : z_{s,1} : z_{s,2}]$, $\kappa_s = |z_{s,0}|^2 + |z_{s,1}|^2 + |z_{s,2}|^2$, and Fubini--Study $(0,2)$-form 
\begin{equation}
    \mu_{\PP^2_s} = \frac{1}{2\kappa_s^{3}} \epsilon_{ijk} \bar{z}_{s,i}\,\dd\bar{z}_{s,j} \wedge \dd\bar{z}_{s,k}\,,
\end{equation}
any vector $\vec{R}(z_s, \bar{z}_s)$ homogeneous of degree $k_s - 2 \ge 0$ in $\bar{z}_s$ satisfies
\begin{equation}
  \bar{\partial}_{\PP^2_s}\!\left(\frac{(\vec{R} \times \bar{z}_s) \cdot \dd\bar{z}_s}{k_s\,\kappa_s^{k_s}}\right)
  = \left[(z_s \cdot \vec{R}) - \frac{\kappa_s}{k_s}(\nabla_{\bar{z}_s} \cdot \vec{R})\right] \frac{\mu_{\PP^2_s}}{\kappa_s^{k_s - 2}}\,.
  \label{eq:app_p2_master}
\end{equation}
For $k_s = 2$ and $P_A = \vec{L}_A \cdot z_s$, the co-exact $(0,1)$-form $\alpha^{(s)}(\vec{L}_A) := \frac{1}{2}\kappa_s^{-2}(\vec{L}_A \times \bar{z}_s) \cdot \dd\bar{z}_s$ obeys 
\begin{equation}
    \bar{\partial}_{\PP^2_s} \alpha^{(s)}(\vec{L}_A) = P_A\,\mu_{\PP^2_s}\,,
\end{equation}
while at the second Koszul step on $\PP^2_s$ the vector triple-product identity gives the $0$-form preimage
\begin{equation}
  F^{(s)}(\vec{v}_A, \vec{v}_B) := \frac{(\vec{v}_A \times \vec{v}_B) \cdot \bar{z}_s}{2\kappa_s}\,, \qquad
  \bar{\partial}_{\PP^2_s} F^{(s)}(\vec{v}_A, \vec{v}_B) = (\vec{v}_B \cdot z_s)\,\alpha^{(s)}(\vec{v}_A) - (\vec{v}_A \cdot z_s)\,\alpha^{(s)}(\vec{v}_B)\,.
  \label{eq:app_F_s}
\end{equation}

\subsection{Explicit reference (0,1)-forms}
\label{app:subsec:explicit_forms}

For the six chiral quark multiplets $(Q_a, u^c_a)$ in $L_a$, $a \in \{2,3,4\}$, let $r_2 = 1$, $r_3 = 3$, and $r_4 = 4$ denote the $\PP^1_{r_a}$ factor on which $L_a$ carries degree $-2$ (while all three carry degree $-1$ on $\PP^1_2$). Their one-step Koszul $(0,1)$-forms take the compact form
\begin{equation}
  \nu_\Phi^{\mathrm{ref}} = \frac{R_{\epsilon(\Phi)}(x_2, \bar{x}_2, z_5, z_6)}{\kappa_2^2}\,\mu_{\PP^1_{r_a}}\,, \qquad
  \Phi \in \{Q_2, u^c_2, Q_3, u^c_3, Q_4, u^c_4\}\,,
  \label{eq:app_nu_matter}
\end{equation}
where $\epsilon(\Phi)= \pm1$ selects the $g_{\mathrm{free}}$ sign in Table~\ref{tab:5273_spectrum}: since $g_{\mathrm{free}} R_\epsilon = -\epsilon\,R_\epsilon$ and $g_{\mathrm{free}} \mu_{\PP^1_{1,3,4}} = -\mu_{\PP^1_{1,3,4}}$, one has $g_{\mathrm{free}} \nu_\Phi^{\mathrm{ref}} = +\epsilon(\Phi)\,\nu_\Phi^{\mathrm{ref}}$, with $\epsilon = +1$ for $u^c_2, Q_3, Q_4$ and $\epsilon = -1$ for $Q_2, u^c_3, u^c_4$. Because $R_\epsilon$ is holomorphic in $(z_5, z_6)$ and independent of $\bar{x}_{r \neq 2}$ while $\bar{\partial}_{\PP^1_{r_a}} \mu_{\PP^1_{r_a}} = 0$, only $\bar{\partial}_{\PP^1_2}$ acts non-trivially, so $\bar{\partial}_{\cA} \nu_\Phi^{\mathrm{ref}} = P_5\,\kappa_2^{-1}(\bar{x}_{2,0} + \epsilon\,\bar{x}_{2,1})\,\mu_{\PP^1_2} \wedge \mu_{\PP^1_{r_a}}$ vanishes identically upon restriction to $X = \{P_1 = \dots = P_5 = 0\}$ for all $C$.

For the up-Higgs bundles $H_{23}, H_{24}, H_{34}$ and the bottom anti-quark $b^c$, $P_5(x_2, z_5, z_6)$ couples $\PP^2_5$ and $\PP^2_6$, so a factorized product $F^{(6)} \alpha^{(5)}$ is not $\bar{\partial}_X$-closed. Instead, we perform the three-step Koszul descent on $\PP^2_5 \times \PP^2_6$. Denoting by $G_{s}$ and $\Pi_{s}$ the Fubini--Study Green's operator and harmonic projector on $\PP^2_s$ in the harmonic basis 
\begin{equation}
    \omega_{\bm{\alpha}_s}^{(s)} = \left(\frac{\bar{z}_s}{\kappa_s}\right)^{\bm{\alpha}_s} \mu_{\PP^2_s}\,, \qquad |\bm{\alpha}_s| = k_s - 3
\end{equation}
of $\cH^{0,2}(\PP^2_s, \cO(-k_s))$, the product Green's operator
\begin{equation}
  G_{56} = G_{5} \otimes \Pi_{6} + \mathrm{id}_{5} \otimes G_{6}
  \label{eq:app_K45}
\end{equation}
inverts $\bar{\partial}_{\PP^2_5 \times \PP^2_6}$ on any cochain with vanishing harmonic projection $\Pi_{5} \otimes \Pi_{6}$. Eliminating the degree-$0$ $\PP^1$ coordinate between $(P_1, P_2)$ for $H_{34}$ and $(P_3, P_4)$ for $H_{23}$ and $H_{24}$ defines four-equation resultant subvarieties $Y_{ab} \supset X$ on which the $\Gamma$-equivariant Koszul differential block $D_{H_{ab}}(C)$  of size $45 \times 45$ for $H_{23}$ and $H_{24}$, and size $72 \times 72$ for $H_{34}$ acquires a one-dimensional kernel along $\cJ_{\mathrm{Higgs}}$, while $b^c$ descends directly from the unique $\Gamma$-invariant class in $\ker d_2 \subset E_2^{-3,4}(X, L_1 \otimes L_5)$. Applying $G_{56}$ three times yields closed analytic $(0,1)$-forms
\begin{equation}
  \nu_k^{\mathrm{ref}} = \frac{\big(\vec{C}_{5,k}(x,z,\bar{z}) \times \bar{z}_5\big) \cdot \dd\bar{z}_5}{2\kappa_5^2} + \frac{\big(\vec{C}_{6,k}(x,z,\bar{z}) \times \bar{z}_6\big) \cdot \dd\bar{z}_6}{2\kappa_6^2}\,, \qquad k \in \{H_{23}, H_{24}, H_{34}, b^c\}\,,
  \label{eq:app_nu_kunneth}
\end{equation}
where $\vec{C}_{5,k}$ and $\vec{C}_{6,k}$ are explicit polynomials in $(x, z)$ and the Fubini--Study ratios $\bar{z}_s / \kappa_s$, determined by~\eqref{eq:app_p2_master}--\eqref{eq:app_F_s}.

Finally, each conjugate down-Higgs bundle $H_{d,ab} = H_{ab}^*$ carries degree $-2$ on three of the four $\PP^1$ factors and degree $(2,2)$ on $\PP^2_5 \times \PP^2_6$, concentrating its spectral sequence on $Y_{ab}$ in row $q = 3$. Along $\cJ_{\mathrm{Higgs}}$, Serre duality guarantees a one-dimensional $\Gamma$-equivariant kernel of $d_2$, and applying the two-step Fubini--Study Green's operator on $(\PP^1)^3$ produces $\bar{\partial}_X$-closed $(0,1)$-forms $\nu_{H_{d,ab}}^{\mathrm{ref}}$, simplifying to $\nu_{H_{d,23/24}}^{\mathrm{ref}} = f_{P_1} f_{P_2} \Psi_2\,\mu_{\PP^1_{3/4}}$ when the $\PP^1_2$ harmonic remainder of $\Psi_2$ vanishes. Numerically on $X$, all $13$ forms satisfy $\|\bar{\partial}_X \nu^{\mathrm{ref}}\|_{L^2} / \|\nu^{\mathrm{ref}}\|_{L^2} \lesssim 10^{-6}$, or down to $6 \times 10^{-10}$ in double precision.

\subsection{Residue contractions}
\label{app:subsec:residue_contraction}

\begin{table}[t]
\centering
\footnotesize
\setlength{\tabcolsep}{4.5pt}
\begin{tabular}{@{}llll@{}}
\toprule
Sector & Spectral origin & Constructed cocycle levels & Reason remaining levels drop out \\
\midrule
$Q_a,\; u^c_a$ & $E_1^{-1,2}$ ($S = \{5\}$) & $\Psi_{Q,U}^{(0,1)} + \Psi_{Q,U}^{(-1,2)}$ & Complete $r_{\max} = 1$ cocycle \\
$H_{23}, H_{24}, H_{34}$ & $E_2^{-4,5}$ on $X$ & $\Psi_H^{(-4,5)}$ ($S = \{1,2,3,4\}$) & $S_Q \cup S_U \subseteq \{5\} \implies S_H = \{1,2,3,4\}$ \\
$b^c$ & $E_2^{-3,4}$ & $\Psi_{b^c}^{(-3,4)}$ ($|S| = 3$, $q_{(\PP^2)^2} = 4$) & $Q_a, H_d$ have $q_{(\PP^2)^2} = 0 \implies q_{b^c} \ge 4$ \\
$H_{d,23}, H_{d,24}, H_{d,34}$ & $E_2^{-3,4}$ & $\sum_{r=0}^3 \Psi_{H_d}^{(-r,r+1)}$ & Exact $\PP^1$ Green's staircase \\
\bottomrule
\end{tabular}
\caption{Total cocycle levels in $K^{\bullet,\bullet}(U_I)$ contributing to the top bidegree $(-5,8)$ product on $\cA = (\PP^1)^4 \times (\PP^2)^2$. Here $S \subseteq \{1,\dots,5\}$ labels the subset of conormal summands $\cO_{\cA}(-d_\alpha)$ in $\bigwedge^{|S|} \cN^*$ ($d_K(e_\alpha) = P_\alpha$), and $q_{(\PP^2)^2}$ is the anti-holomorphic form degree on $\PP^2_5 \times \PP^2_6$.}
\label{tab:app_cocycle_truncation}
\end{table}

Writing homogeneous cochains in~\eqref{eq:koszul_dolbeault_double_complex} as $\omega^{(0,q)} \otimes e_S$, where $\omega^{(0,q)}$ has anti-holomorphic form degree $q$ and $e_S$ is an exterior basis element of $\bigwedge^{|S|} \cN^*$, the graded wedge product of Section~\ref{subsec:exact_integration} is defined by
\begin{equation}
  (\omega_1^{(0,q_1)} \otimes e_{S_1}) \star (\omega_2^{(0,q_2)} \otimes e_{S_2})
  = (-1)^{|S_1|\,q_2} (\omega_1^{(0,q_1)} \wedge \omega_2^{(0,q_2)}) \otimes (e_{S_1} \wedge e_{S_2})\,.
  \label{eq:app_binary_star}
\end{equation}
To evaluate $I_{IJK}(C)$ via~\eqref{eq:poincare_residue_lift_master} on $\cA = (\PP^1)^4 \times (\PP^2)^2$ ($K = 5$, $D = 8$), we extract the $(-5,8)$ component of $\Xi_{IJK} = \Psi_I \star \Psi_J \star \Psi_K$, which requires disjoint Koszul index sets $S_I \sqcup S_J \sqcup S_K = \{1,2,3,4,5\}$ and disjoint $(\PP^1)^4$ form-leg sets $T_I \sqcup T_J \sqcup T_K = \{1,2,3,4\}$ together with form degree $(2,2)$ on $\PP^2_5 \times \PP^2_6$. As summarized in Table~\ref{tab:app_cocycle_truncation}, degree counting truncates both the up-type and down-type products to two bidegree channels,
\begin{align}
  \Xi^{\mathrm{up}}_{(-5,8)} &= \Psi_H^{(-4,5)} \star \Psi_Q^{(0,1)} \star \Psi_U^{(-1,2)} + \Psi_H^{(-4,5)} \star \Psi_Q^{(-1,2)} \star \Psi_U^{(0,1)}\,, \label{eq:app_xi_up} \\
  \Xi^{\mathrm{down}}_{(-5,8)} &= \Psi_Q^{(0,1)} \star \Psi_{b^c}^{(-3,4)} \star \Psi_{H_d}^{(-2,3)} + \Psi_Q^{(-1,2)} \star \Psi_{b^c}^{(-3,4)} \star \Psi_{H_d}^{(-1,2)}\,. \label{eq:app_xi_down}
\end{align}
Since the $(2,2)$-form on $\PP^2_5 \times \PP^2_6$ has even total degree $4$ (so $q \equiv |T| \pmod 2$), each term in~\eqref{eq:app_xi_up}--\eqref{eq:app_xi_down} carries by~\eqref{eq:app_binary_star} the graded sign
\begin{equation}
  \sigma_{\mathrm{tot}} = (-1)^{|S_I|\,|T_J| + (|S_I|+|S_J|)|T_K|}\,\operatorname{sgn}(S_I, S_J, S_K)\,\operatorname{sgn}(T_I, T_J, T_K)\,,
  \label{eq:app_koszul_sign}
\end{equation}
where $\operatorname{sgn}(S_I, S_J, S_K)$ and $\operatorname{sgn}(T_I, T_J, T_K)$ denote the permutation signs ordering the disjoint unions into $(1,2,3,4,5)$ and $(1,2,3,4)$, respectively.

On $\PP^2_5 \times \PP^2_6$, classes in $H^2(\PP^2_s, \cO(-3-n))$ are represented in the Serre-dual monomial basis so that multiplication by $z_{s,0}^{m_0} z_{s,1}^{m_1} z_{s,2}^{m_2}$ subtracts $(m_0, m_1, m_2)$ from the dual exponent tuple and evaluates to unity on $(0,0,0) \in H^2(\PP^2_s, \cO(-3)) \cong \CC$. On each $\PP^1_r$ factor ($r=1,2,3,4$), every $(0,1)$-factor has the bihomogeneous Fubini--Study form 
\begin{equation}
    \alpha_{a_0,a_1,b_0,b_1} = \frac{x_0^{a_0} x_1^{a_1} \bar{x}_0^{b_0} \bar{x}_1^{b_1}}{\kappa^{b_0+b_1+2}}\,\bar{\theta}\,,\qquad a_0 + a_1 = b_0 + b_1\,,\qquad \theta = x_0 \dd x_1 - x_1 \dd x_0\,,
\end{equation}
whose integral reduces in polar coordinates $x = [1 : \rho e^{\ii\phi}]$ to an elementary Beta function,
\begin{equation}
\begin{aligned}
  -\frac{1}{2\pi\ii} \int_{\PP^1} \theta \wedge \alpha_{a_0,a_1,b_0,b_1}
  &= \frac{1}{\pi} \int_0^\infty \frac{\rho^{a_1 + b_1 + 1}}{(1 + \rho^2)^{b_0 + b_1 + 2}}\,\dd\rho \int_0^{2\pi} e^{\ii(a_1 - b_1)\phi}\,\dd\phi \\
  &= \delta_{a_0,b_0}\,\delta_{a_1,b_1}\,\frac{a_0!\,a_1!}{(a_0 + a_1 + 1)!}\,.
\end{aligned}
  \label{eq:app_p1_beta}
\end{equation}
Combining the $(2\pi\ii)^5$ residue factors with the coordinate normalization~\eqref{eq:raw_integral_conventions} absorbs all powers of $\pi$ into a single overall prefactor,
\begin{equation}
  I_{IJK}(C) = 6\pi^3\,C_{\mathrm{rat}}\!\big(\Xi^{(-5,8)}_{IJK}(C)\big)\,,
  \label{eq:app_6pi3_rational}
\end{equation}
where $C_{\mathrm{rat}}$ is a finite rational contraction of the $E_2$ kernel coefficients and the factorials~\eqref{eq:app_p1_beta}. Projecting both the unit-norm cocycles $\Psi_I$ and the analytic reference forms~\eqref{eq:app_nu_matter}--\eqref{eq:app_nu_kunneth} orthogonally modulo $\im(d_1)$ onto $E_2$ then fixes the relative normalization factors $s_I(C)$ (with collinearity verified to $10^{-14}$), allowing us to rescale $I_{IJK}(C) \mapsto s_I(C) s_J(C) s_K(C) I_{IJK}(C)$ such that it uses the same basis as $\cK_{I\bar{I}}$.

\section{Benchmark Points and RG Matching}
\label{app:benchmarks}

In this appendix, we record the explicit $\Gamma$-equivariant parameterization of the defining polynomials $P_1, \dots, P_5$ on CICY 5273, the complex structure coefficients of Benchmarks~1--4 in Section~\ref{sec:cicy_5273}, and our two-loop renormalization group matching conventions.

\subsection{Defining polynomials}
\label{app:subsec:defining_polys}

Under the free $\ZZ_2$ involution $g_{\mathrm{free}}$ in~\eqref{eq:g_free_action}, the defining polynomials satisfy $P_2 = g_{\mathrm{free}}(P_1)$, $P_4 = g_{\mathrm{free}}(P_3)$, and $P_5 = g_{\mathrm{free}}(P_5)$, and are parameterized by
\begin{align}
  P_1(x_1, z_5) &= x_{1,0}(C_1 z_{5,0} + C_2 z_{5,1}) + x_{1,1}(C_{24} z_{5,0} + C_{25} z_{5,1} + C_3 z_{5,2})\,, \\
  P_2(x_1, z_6) &= x_{1,0}(C_1 z_{6,0} + C_2 z_{6,1}) - x_{1,1}(C_{24} z_{6,0} + C_{25} z_{6,1} + C_3 z_{6,2})\,, \\
  P_3(x_3, x_4, z_5) &= x_{3,0} x_{4,0}(C_4 z_{5,0} + C_5 z_{5,1}) + x_{3,1} x_{4,1}(C_6 z_{5,0} + C_7 z_{5,1}) \nonumber \\
  &\quad + x_{3,0} x_{4,1}(C_{26} z_{5,0} + C_{27} z_{5,1} + C_8 z_{5,2}) \\
  &\quad + x_{3,1} x_{4,0}(C_{28} z_{5,0} + C_{29} z_{5,1} + C_9 z_{5,2})\,, \nonumber\\
  P_4(x_3, x_4, z_6) &= x_{3,1} x_{4,1}(C_4 z_{6,0} + C_5 z_{6,1}) + x_{3,0} x_{4,0}(C_6 z_{6,0} + C_7 z_{6,1}) \nonumber \\
  &\quad + x_{3,1} x_{4,0}(C_{26} z_{6,0} + C_{27} z_{6,1} + C_8 z_{6,2}) \\ 
  &\quad+ x_{3,0} x_{4,1}(C_{28} z_{6,0} + C_{29} z_{6,1} + C_9 z_{6,2})\,, \nonumber \\
  P_5(x_2, z_5, z_6) &= z_5^{\mathsf{T}}(S_+ + S_-) z_6\,x_{2,0}^2 + (z_5^{\mathsf{T}} S_{\mathrm{cross}} z_6)\,x_{2,0} x_{2,1} \nonumber \\
  &\quad + z_5^{\mathsf{T}}(S_+ - S_-) z_6\,x_{2,1}^2\,,
\end{align}
where we view $z_s = (z_{s,0}, z_{s,1}, z_{s,2})^{\mathsf{T}}$ ($s = 5, 6$) as column vectors with transpose $^{\mathsf{T}}$, and invariance under $g_{\mathrm{free}}$ ($x_{2,0} \leftrightarrow x_{2,1}$, $z_5 \leftrightarrow z_6$) requires $S_+$ and $S_{\mathrm{cross}}$ to be symmetric and $S_-$ to be antisymmetric:
\begin{equation}
\begin{aligned}
  S_+ &= \begin{pmatrix}
    2C_{10} & C_{11} + C_{12} & C_{30} \\
    C_{11} + C_{12} & 2C_{13} & 0 \\
    C_{30} & 0 & C_{18}
  \end{pmatrix}\,, \qquad
  S_- = \begin{pmatrix}
    0 & C_{22} & C_{20} \\
    -C_{22} & 0 & C_{21} \\
    -C_{20} & -C_{21} & 0
  \end{pmatrix}\,, \\[4pt]
  S_{\mathrm{cross}} &= \begin{pmatrix}
    C_{23} + C_{32} & C_{33} & C_{31} \\
    C_{33} & C_{23} - C_{32} & 0 \\
    C_{31} & 0 & C_{19}
  \end{pmatrix}\,.
\end{aligned}
  \label{eq:app_S_matrices}
\end{equation}
Here our indexing $\{C_1, \dots, C_{33}\}$ follows an ambient covering-space parameterization in which four $g_{\mathrm{free}}$-breaking coefficients $C_{14} = \dots = C_{17} = 0$ in $P_5$ vanish on $\hat{X}$ (and are omitted above), while setting $C_{11} = C_{12}$ fixes the redundant split in $C_{11} + C_{12}$.

\subsection{Benchmark points}
\label{app:subsec:dictionaries}

\begin{table}[t]
\centering
\footnotesize
\setlength{\tabcolsep}{7pt}
\begin{tabular}{@{}lrrrr@{}}
\toprule
Coefficient & Benchmark 1 & Benchmark 2 & Benchmark 3 & Benchmark 4 \\
\midrule
$C_1,\; C_4$ & $+1.000000000000$ & $+1.000000000000$ & $+1.000000000000$ & $+1.000000000000$ \\
$C_2$ & $+0.761741029838$ & $+0.758847562256$ & $+0.778027533560$ & $+0.649238343369$ \\
$C_3$ & $+1.939351353708$ & $+1.881987175793$ & $+1.992518690257$ & $+1.929617871661$ \\
$C_5$ & $-0.562245455876$ & $-0.520223331052$ & $-0.498050691075$ & $-0.514893213118$ \\
$C_6$ & $+0.412307332912$ & $+0.400162180108$ & $+0.300811928579$ & $+0.386867922884$ \\
$C_7$ & $+0.963214773418$ & $+0.986021690774$ & $+1.115247092947$ & $+1.107082599262$ \\
$C_8$ & $+1.734278127683$ & $+1.811430623260$ & $+1.816034764071$ & $+1.746832797347$ \\
$C_9$ & $-0.245437748058$ & $-0.184656606073$ & $-0.211312772147$ & $-0.183838522987$ \\
$C_{10}$ & $-0.283943172203$ & $-0.189586935161$ & $-0.303720714426$ & $-0.184522246296$ \\
$C_{11} = C_{12}$ & $+0.111101688074$ & $+0.097931442302$ & $+0.095511340053$ & $+0.111413758282$ \\
$C_{13}$ & $-0.057946384203$ & $-0.017685357459$ & $-0.048691047958$ & $-0.023867910389$ \\
$C_{18}$ & $+0.040182518923$ & $+0.049595554657$ & $-0.060333723869$ & $+0.029468956021$ \\
$C_{19}$ & $+1.903918053316$ & $+1.788166494604$ & $+1.872148038341$ & $+1.894964886735$ \\
$C_{20}$ & $+0.865668460081$ & $+0.778968743625$ & $+0.855584651830$ & $+0.855980168407$ \\
$C_{21}$ & $-0.286399599691$ & $-0.321357567234$ & $-0.227962075870$ & $-0.263889339256$ \\
$C_{22}$ & $-0.014378433384$ & $-0.010811936153$ & $+0.008217805618$ & $-0.073664310620$ \\
$C_{23}$ & $+0.604416739784$ & $+0.373197017710$ & $+0.730204929357$ & $+0.462367203855$ \\
$C_{24}$ & $-0.047191262619$ & $-0.077260571081$ & $-0.032335204705$ & $+0.071740758323$ \\
$C_{25}$ & $+0.051769509456$ & $+0.018954360283$ & $+0.004138784410$ & $-0.035519134612$ \\
$C_{26}$ & $+0.066896913656$ & $+0.037185692440$ & $+0.093005561554$ & $-0.058712615072$ \\
$C_{27}$ & $+0.009569279873$ & $-0.091196120276$ & $+0.015248260415$ & $-0.039715113386$ \\
$C_{28}$ & $-0.029904711116$ & $-0.054955676777$ & $+0.000930506202$ & $+0.047740816977$ \\
$C_{29}$ & $-0.146434025772$ & $-0.005360734711$ & $-0.036139425822$ & $+0.051993525541$ \\
$C_{30}$ & $+0.051405533662$ & $+0.058697388248$ & $+0.046788836151$ & $-0.004729525344$ \\
$C_{31}$ & $+0.069257265750$ & $-0.029605384990$ & $+0.011774665654$ & $+0.098659092418$ \\
$C_{32}$ & $-0.032334972025$ & $-0.105693273720$ & $-0.006763275321$ & $-0.243802487505$ \\
$C_{33}$ & $+0.177386824378$ & $+0.142888270033$ & $+0.067541302769$ & $+0.165200312798$ \\
\midrule
$\min_{X} (\dd P)$ & $0.034593$ & $0.010541$ & $0.028314$ & $0.012417$ \\
$\min_{\mathrm{Fix}(g_{\mathrm{free}})} \|P\|_2$ & $0.058635$ & $0.042920$ & $0.027575$ & $0.024880$ \\
\bottomrule
\end{tabular}
\caption{Non-zero complex structure coefficients $C_i$ (with $C_{14} = C_{15} = C_{16} = C_{17} = 0$) and global smoothness diagnostics for Benchmarks~1--3 (Table~\ref{tab:5273_democratic_sugra}) and Benchmark~4 (Table~\ref{tab:5273_aligned_benchmark}) on $\cJ_{\mathrm{Higgs}}$.}
\label{tab:app_benchmark_coeffs}
\end{table}

Table~\ref{tab:app_benchmark_coeffs} lists the complex structure parameters $C_i$ and global smoothness diagnostics for Benchmarks~1--4 analyzed in Section~\ref{subsec:5273_results}. To evaluate smoothness, we check that $P=\dd P=0$ does not have a solution. More precisely, we compute the smallest singular value of the $5 \times 14$ ambient Jacobian $\dd P = (\partial P_A / \partial Z_M)$ and check that it is bounded away from zero, $\min_{X} (\dd P) > 0$, and that $X$ avoids all $16$ $\PP^2$ components of the ambient fixed locus $\mathrm{Fix}(g_{\mathrm{free}})$ via $\min_{\mathrm{Fix}(g_{\mathrm{free}})} \|P\|_2 > 0$, where $\|P\|_2 = (\sum_{A=1}^5 |P_A|^2)^{1/2}$.

All four points are chosen to lie on the Higgs locus $\cJ_{\mathrm{Higgs}}$~\eqref{eq:5273_determinantal_locus} where three Higgs multiplets become massless. On the symmetric sub-locus invariant under the accidental $\ZZ_2$ involution
\begin{equation}
  g_{\mathrm{int}} : \quad x_r \mapsto [x_{r,0} : -x_{r,1}]\,, \qquad
  x_2 \mapsto [x_{2,1} : x_{2,0}]\,, \qquad
  z_s \mapsto [z_{s,0} : z_{s,1} : -z_{s,2}]\,,
\end{equation}
with $r=1,3,4$ and $s=5,6$, Eq.~\eqref{eq:5273_determinantal_locus} holds automatically while parity forces $W^d = 0$. Turning on the nine $g_{\mathrm{int}}$-odd coefficients $\{C_{22}, C_{24}, \dots, C_{31}\}$ breaks $g_{\mathrm{int}}$, mixes a $g_{\mathrm{int}}$-odd component into the $b^c$ representative in $\ker d_2$, and generates a non-zero tree-level bottom-quark Yukawa coupling. We choose these coefficients so that the three determinantal equations~\eqref{eq:5273_determinantal_locus} continue to hold for Benchmarks~1--4.

\subsection{Two-loop RG matching}
\label{app:subsec:rg_matching}

Comparing string predictions with experiment requires bridging a large gap in energy: Eq.~\eqref{eq:exact_physical_yukawa} gives the Yukawa couplings at $M_{\mathrm{GUT}} = 2 \times 10^{16}\text{ GeV}$, while quark masses and CKM mixings are measured between $2\text{ GeV}$ and $173\text{ GeV}$~\cite{PDG:2024}. Over these fourteen orders of magnitude in energy, quantum loop corrections, dominated by strong QCD interactions as $\alpha_s(\mu)$ grows toward lower energies, increase the effective quark Yukawa couplings by factors of roughly two to five~\cite{Ross:2007az,Xing:2007fb,Antusch:2013jca}. To compare Benchmark~4 directly with experimental data in Table~\ref{tab:5273_aligned_benchmark}, we evolve the couplings down in energy scale $\mu$ across three standard effective-field-theory regimes, assuming that two linear combinations of the three Higgs pairs $(H_{ab}, H_{d,ab})$ acquire masses near $M_{\mathrm{GUT}}$ so that only a single light MSSM Higgs pair $(H_u, H_d)$ remains below $M_{\mathrm{GUT}}$, with Yukawa matrices $Y^u = \sum_{ab} \hat{v}^u_{ab} Y^u_{ab}$ and $Y^d = \sum_{ab} \hat{v}^d_{ab} Y^d_{ab}$.

First, between $\mu = M_{\mathrm{GUT}}$ and an effective supersymmetry scale $M_{\mathrm{SUSY}} = 3\text{ TeV}$, all MSSM superpartners are active. We evolve $(Y^u, Y^d)$ downward using the two-loop MSSM renormalization group equations~\cite{Martin:1993zk,Antusch:2013jca},
\begin{equation}
  \frac{\dd Y^{u,d}}{\dd \ln \mu} = \frac{1}{16\pi^2} \beta_1^{u,d} + \frac{1}{(16\pi^2)^2} \beta_2^{u,d}\,,
  \label{eq:app_2loop_rge}
\end{equation}
where $\beta_1^{u,d}$ and $\beta_2^{u,d}$ are the one- and two-loop MSSM beta functions, which depend on $Y^{u,d}$ and on the three gauge couplings $(g_1, g_2, g_3)$ in $SU(5)$ normalization ($\alpha_1 = \frac{5}{3}\alpha_Y$). These equations are formulated in the $\overline{\mathrm{DR}}$ (modified dimensional reduction) renormalization scheme, which preserves supersymmetry in loop diagrams. The gauge couplings are fixed by matching to the measured values at the $Z$-boson mass $M_Z = 91.1876\text{ GeV}$~\cite{PDG:2024}. Running these couplings up to $M_{\mathrm{GUT}}$ with $M_{\mathrm{SUSY}} = 3\text{ TeV}$ yields the unified coupling $\alpha_{\mathrm{GUT}}^{-1} = 25.53$ shown alongside the reference value $\alpha_{\mathrm{GUT}}^{-1} = 24.30$ in Table~\ref{tab:5273_aligned_benchmark}.

Second, at $\mu = M_{\mathrm{SUSY}}$, heavy superpartners and the heavy Higgs states decouple, leaving the ordinary Standard Model with a single light Higgs doublet. Since $\langle H_u \rangle = v \sin\beta$ and $\langle H_d \rangle = v \cos\beta$, matching onto the Standard Model Yukawa matrices sets $y^u_{\mathrm{SM}} = Y^u \sin\beta$ and $y^d_{\mathrm{SM}} = Y^d \cos\beta$, together with a small finite one-loop conversion~\cite{Martin:1993zk} from the supersymmetric $\overline{\mathrm{DR}}$ scheme to the Standard Model $\overline{\mathrm{MS}}$ (modified minimal subtraction) scheme used in experimental analyses. We then evolve $y^{u,d}_{\mathrm{SM}}$ at two loops from $M_{\mathrm{SUSY}}$ down to $M_Z$. At $\mu = M_Z$, diagonalizing $y^{u,d}_{\mathrm{SM}} y^{u,d\,\dagger}_{\mathrm{SM}}$ via unitary left-handed quark rotations $U_L^{u,d}$ gives the CKM mixing matrix $V_{\mathrm{CKM}}(M_Z) = U_L^{u\dagger} U_L^d$.

Finally, below $M_Z$, we convert the running Yukawa couplings into the standard quark mass observables reported by the Particle Data Group~\cite{PDG:2024}. Because the top quark decays before hadronizing, its mass is quoted as the kinematic pole mass $m_t^{\mathrm{pole}}$, which we compute from its running $\overline{\mathrm{MS}}$ mass via the three-loop QCD relation~\cite{Chetyrkin:1997un}. For the lighter quarks, as $\mu$ drops below each heavy-quark threshold, that quark is integrated out of the QCD beta function (stepping from five active flavors down to four), using four-loop $\overline{\mathrm{MS}}$ QCD running and two-loop threshold matching~\cite{Chetyrkin:1997un}. Following PDG conventions, the bottom and charm running masses are evaluated at their own scales, $m_b(\mu = m_b)$ and $m_c(\mu = m_c)$, whereas the light up-quark mass $m_u(2\text{ GeV})$ is evaluated at $\mu = 2\text{ GeV}$ (with four active flavors $u, d, s, c$) so as to remain safely in the perturbative QCD regime.

\section{Neural PDE Architecture and Training Details}
\label{app:ml_setup}

In this appendix, we summarize the sampling scheme, loss functionals, training protocol and neural network architectures, used to compute the Ricci-flat K\"ahler metric $g_{\mathrm{CY}}$, the slope-zero HYM bundle metrics $H_I$, and the harmonic bundle-valued $(0,1)$-forms $\nu_I^{\mathrm{NN}}$ on CICY 7862 (Section~\ref{sec:tetraquadric}) and CICY 5273, Model~500 (Section~\ref{sec:cicy_5273}).

\subsection{Sampling and sequential PDE loss functionals}
\label{app:subsec:ml_losses}

For a smooth complete intersection $X = \{P_1 = \dots = P_K = 0\} \subset \cA = \prod_{r=1}^m \PP^{n_r}$, we generate $N = 10^6$ points on $X$ using the sampling method of~\cite{Douglas:2006hz,larfors2022numericalmetricscompleteintersection}, split $90\%$--$10\%$ into training and validation sets, and use an additional independent test sample of $9 \times 10^5$ points to evaluate physical quark masses. At each point $p \in X$, the ambient Fubini--Study metric $g_{\mathrm{FS}}(t) = \sum_{r=1}^m t^r g_{\mathrm{FS},r}$ and the implicit-function-theorem pullback to $T_p X$ define an exact holomorphic tangent frame along with importance weights $w(p) = |\Omega(p)|^2 / \dd\mu_{\mathrm{sample}}(p)$ such that weighted Monte Carlo expectations
\begin{equation}
  \langle f \rangle_w := \frac{\sum_p w(p)\,f(p)}{\sum_p w(p)}
\end{equation}
approximate continuum integrals against the CY volume form $\Omega \wedge \bar{\Omega}$.

Since the HYM and harmonic $(0,1)$-form equations depend on the Monge--Amp\`ere solution, we train separate networks sequentially across three stages:
\begin{enumerate}
  \item \emph{Stage 1: Ricci-flat Calabi--Yau metric.} We parameterize the K\"ahler metric in the fixed K\"ahler class $[J] = [J_{\mathrm{FS}}(t)]$ via a globally defined smooth $\Gamma$-invariant potential $\phi_{\mathrm{sym}} \in C^\infty(\hat{X})$,
  \begin{equation}
    g_{\mathrm{CY}}(p) = g_{\mathrm{FS}}(p) + \partial\bar{\partial}\phi_{\mathrm{sym}}(p)\,,
    \label{eq:app_cy_ansatz}
  \end{equation}
  evaluated by exact nested forward/reverse-mode automatic differentiation in JAX and pulled back to $TX$. Because~\eqref{eq:app_cy_ansatz} is globally defined on $\cA$, K\"ahlerity ($\dd J = 0$), K\"ahler class preservation, and coordinate-patch transition compatibility hold identically. We train $\phi_{\mathrm{sym}}$ by minimizing the positive-branch Monge--Amp\`ere $L^1$ residual,
  \begin{equation}
    \cL_{\mathrm{MA}}(\phi) = \left\langle \left| 1 - \frac{\det g_{\mathrm{pos}}(p)}{\kappa\,|\Omega(p)|^2} \right| \right\rangle_w\,, \qquad
    \det g_{\mathrm{pos}}(p) := \prod_{a=1}^3 \varepsilon\,\mathrm{softplus}\!\left(\frac{\lambda_a(g_{\mathrm{CY}}(p))}{\varepsilon}\right)\,,
    \label{eq:app_ma_loss}
  \end{equation}
  where $\lambda_a(g_{\mathrm{CY}})$ are the eigenvalues of the $3 \times 3$ Hermitian metric $g_{\mathrm{CY}}$, $\varepsilon = 10^{-7}$ smoothly prevents signature flips during early training steps, and $\kappa = \langle \det g_{\mathrm{FS}}|_{TX} / |\Omega|^2 \rangle_w$ fixes the K\"ahler volume normalization.

  \item \emph{Stage 2: Slope-zero Hermitian Yang--Mills metrics.} Freezing $g_{\mathrm{CY}}$, we parameterize the Hermitian fiber metric on each independent line bundle $L_a = \cO_X(q_a)$ along the zero-slope K\"ahler locus~\eqref{eq:duy_slope_zero} as $H_a(p) = H_a^{\mathrm{FS}}(p)\,e^{\beta_a(p)}$, where 
  \begin{equation}
      H_a^{\mathrm{FS}}(p) = \prod_{r=1}^m (\kappa_r)^{-q_a^r}
  \end{equation}
  and $\beta_a \in C^\infty(\hat{X})$. Each network minimizes the pulled-back curvature trace
  \begin{equation}
    \cL_{\mathrm{HYM}}(\beta_a) = \big\langle |\mu_{\mathrm{HYM}}(H_a)| \big\rangle_w\,,\qquad \mu_{\mathrm{HYM}}(H_a) := g_{\mathrm{CY}}^{i\bar{\jmath}}\,\partial_i \partial_{\bar{\jmath}}(\ln H_a^{\mathrm{FS}} + \beta_a)
    \label{eq:app_hym_loss}
  \end{equation}
  and we monitor convergence relative to the Fubini--Study curvature trace $\mu_{\mathrm{FS}} := \mu_{\mathrm{HYM}}(H_a^{\mathrm{FS}})$ at $\beta_a = 0$. On CICY 5273, Model~500, we train three networks for the basis bundles $L_2, L_3, L_4$ and construct the remaining seven composite matter and Higgs bundle metrics ($b^c, H_{ab}, H_{d,ab}$) algebraically from $(\beta_{L_2}, \beta_{L_3}, \beta_{L_4})$ via $L_1 \otimes L_5 = (L_2 \otimes L_3 \otimes L_4)^*$.

  \item \emph{Stage 3: Harmonic bundle-valued $(0,1)$-forms.} Freezing $(g_{\mathrm{CY}}, H_I)$, we represent the harmonic $(0,1)$-form in the Dolbeault class of the analytic reference form $\nu_I^{\mathrm{ref}}$ as
  \begin{equation}
    \nu_I^{\mathrm{NN}}(p) = \nu_I^{\mathrm{ref}}(p) + \bar{\partial}_X \sigma_I^{\mathrm{NN}}(p)\,,
    \label{eq:app_harm_ansatz}
  \end{equation}
  which is $\bar{\partial}_X$-closed and stays in the cohomology class $[\nu_I^{\mathrm{ref}}] \in H^1(\hat{X}, U_I)$ throughout training. Precomputing the static tangent frames and reference curvature contractions once prior to optimization, we train a dedicated network for each bundle sector to minimize the co-closedness PDE residual plus a zero-mean gauge-fixing term on $\sigma_I^{\mathrm{NN}}$,
  \begin{equation}
    \cL_{\mathrm{harm}}(\sigma_I) = \Big\langle \big| g_{\mathrm{CY}}^{a\bar{b}}(p)\,\partial_a\big[H_I(p)\,\big(\nu_{I,\bar{b}}^{\mathrm{ref}}(p) + \partial_{\bar{b}}\sigma_I^{\mathrm{NN}}(p)\big)\big] \big| \Big\rangle_w + \big|\langle \sigma_I^{\mathrm{NN}} \rangle_w\big|^2\,.
    \label{eq:app_harm_loss}
  \end{equation}
\end{enumerate}

\subsection{Bihomogeneous spectral networks and equivariant section ansatz}
\label{app:subsec:ml_arch}

To guarantee smoothness and exact $(\CC^\times)^m$ projective invariance across all coordinate charts without transition penalties, every network takes as input the real degree-zero bihomogeneous spectral invariants~\cite{Berglund:2022gvm,constantin2024computationquarkmassesstring,Lust:2026warped} of each ambient projective factor $\PP^{n_r}$,
\begin{equation}
  \bm{\xi}_r(z_r, \bar{z}_r) = \left\{ \frac{\operatorname{Re}(z_{r,i}\bar{z}_{r,j})}{\kappa_r} \right\}_{0 \le i \le j \le n_r} \;\cup\;
  \left\{ -\frac{\operatorname{Im}(z_{r,i}\bar{z}_{r,j})}{\kappa_r} \right\}_{0 \le i < j \le n_r} \in \RR^{(n_r + 1)^2}\,,
  \label{eq:app_spectral_inputs}
\end{equation}
giving input dimension $d_{\mathrm{in}} = \sum_{r=1}^m (n_r + 1)^2 = 4(2^2) = 16$ on the tetra-quadric and $d_{\mathrm{in}} = 4(2^2) + 2(3^2) = 34$ on CICY 5273. Discrete $\Gamma$-equivariance is enforced directly in the forward pass: for the CY metric and the HYM connection networks,
\begin{equation}
    \phi_{\mathrm{sym}}(p) = \frac{1}{|\Gamma|} \sum_{\gamma \in \Gamma} \phi_{\mathrm{NN}}(\gamma \cdot p)\,,\qquad
    \beta_a(p) = \frac{1}{|\Gamma|} \sum_{\gamma \in \Gamma} \beta_{\mathrm{NN},a}(\gamma \cdot p)\,.
\end{equation}

For the harmonic form networks of Stage~3, a smooth section $\sigma_I \in \Omega^{0,0}(\hat{X}, U_I)$ of $U_I = \cO_X(q_I^1, \dots, q_I^m)$ transforms on $X$ with multidegree $q_I$ and Wilson-line pullback parity $\eta_I \in \{\pm 1\}$. We expand $\sigma_I^{\mathrm{NN}}$ globally over the ambient Fubini--Study monomial frame $\{s_{I,\alpha}(z, \bar{z})\}_{\alpha=1}^{M_I}$ of $U_I$,
\begin{equation}
  \sigma_I^{\mathrm{NN}}(p) = \sum_{\alpha=1}^{M_I} c_{I,\alpha}^{\mathrm{sym}}(p)\,s_{I,\alpha}(z, \bar{z})\,, \qquad
  s_{I,\alpha}(z, \bar{z}) = \prod_{r=1}^m \kappa_r^{\min(q_I^r,0)}\mathfrak{m}_{r,\alpha_r}(z_r, \bar{z}_r)\,,
  \label{eq:app_global_section_basis}
\end{equation}
where $\mathfrak{m}_{r,\alpha_r}$ runs over all holomorphic monomials of degree $q_I^r$ when $q_I^r \ge 0$ and all anti-holomorphic monomials of degree $|q_I^r|$ when $q_I^r < 0$, so that $s_{I,\alpha}$ carries holomorphic multidegree $q_I$ and unit Fubini--Study scaling across $\cA$. The neural network outputs the $2M_I$ real and imaginary parts of the degree-zero coefficient functions $c_{I,\alpha}(p)$, which are projected onto the $\Gamma$-eigenspace of parity $\eta_I$ via 
\begin{equation}
    c_{I,\alpha}^{\mathrm{sym}}(p) = \frac{1}{2}[c_{I,\alpha}(p) + \eta_I \epsilon_\alpha c_{I,\pi(\alpha)}(g_{\mathrm{free}}(p))]\,,
\end{equation}
where $g_{\mathrm{free}} s_{I,\alpha} = \epsilon_\alpha s_{I,\pi(\alpha)}$. For Model 500 on CICY 5273, the resulting basis dimensions $M_I = \prod_{r=1}^6 \binom{|q_I^r| + n_r}{n_r}$ are $M_I = 54$ for the six quark multiplets $Q_{2,3,4}$ and $u^c_{2,3,4}$, $M_I = 972$ for the six Higgs multiplets $H_{23,24,34}$ and $H_{d,23,24,34}$, and $M_I = 10{,}800$ for $b^c$.

\subsection{Training protocol and learning curves}
\label{app:subsec:ml_training}

\begin{table}[t]
\centering
\footnotesize
\setlength{\tabcolsep}{3.4pt}
\begin{tabular}{@{}llll@{}}
\toprule
Component & Stage 1 ($g_{\mathrm{CY}}$) & Stage 2 ($H_{L_2, L_3, L_4}$) & Stage 3 ($\nu_I^{\mathrm{NN}}$, $13$ bundles) \\
\midrule
Networks / seed & $1$ & $3$ (parallel) & $13$ (parallel) \\
MLP architecture & $34 \to [128 \times 4] \to 1$ & $34 \to [128 \times 3] \to 1$ & $34 \to [128 \times 3] \to 2M_I$ \\
Hidden / output & GELU / linear & GELU / bias-free linear & GELU / bias-free linear \\
$\Gamma$-equivariance & $\phi_{\mathrm{sym}}$ ($\eta = +1$) & $\beta_a$ ($\eta = +1$) & $c_{I,\alpha}^{\mathrm{sym}}$ ($\eta_I \in \{\pm 1\}$) \\
Batch / epochs / clip & $4{,}096$ / $300$ / none & $4{,}096$ / $300$ / $5.0$ & $4{,}096$ / $300$ / $5.0$ \\
Val residual (3-seed) & $\sigma_{\mathrm{MA}}^{\mathrm{val}} = 0.0215$ & $\|\mu_{\mathrm{HYM}}\|/\|\mu_{\mathrm{FS}}\| = 0.055\text{--}0.072$ & $\|\bar{\partial}^\dagger \nu^{\mathrm{NN}}\|/\|\bar{\partial}^\dagger \nu^{\mathrm{ref}}\| = 0.005\text{--}0.048$ \\
\midrule
Train / val / test & \multicolumn{3}{c}{$9 \times 10^5$ / $10^5$ / $9 \times 10^5$} \\
Optimizer \& schedule & \multicolumn{3}{c}{AdamW ($\mathrm{lr}_{\max} = 10^{-3}$, weight decay $10^{-4}$, $5\%$ warmup + cosine decay to $10^{-6}$)} \\
\bottomrule
\end{tabular}
\caption{Architecture and optimization hyperparameters for the $17$-network neural PDE pipeline on CICY 5273, Model~500 ($M_I = 54$ for $Q_a, u^c_a$; $M_I = 972$ for $H_{ab}, H_{d,ab}$; $M_I = 10{,}800$ for $b^c$). Final validation residuals list three-seed means at Benchmark~4, with $\sigma_{\mathrm{MA}}^{\mathrm{val}}$ denoting the validation Monge--Amp\`ere residual~\eqref{eq:app_ma_loss}.}
\label{tab:app_ml_hyperparams}
\end{table}

\begin{figure}[t]
\centering
\includegraphics[width=\textwidth]{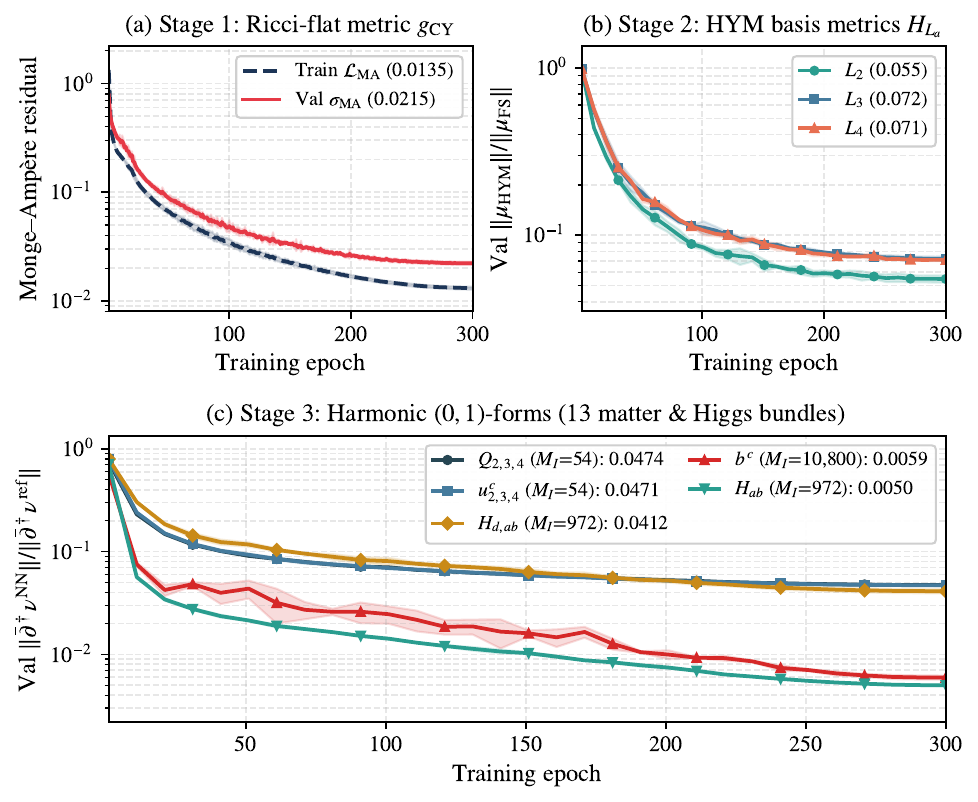}
\caption{Three-seed training and validation convergence for CICY 5273, Model~500 at Benchmark~4 ($N = 10^6$ sampled points; solid curves show validation means, shaded bands show $\pm 1$ standard deviation). (a)~Stage~1 Monge--Amp\`ere training loss and validation residual. (b)~Stage~2 validation HYM curvature residuals for $L_2, L_3, L_4$. (c)~Stage~3 validation co-closedness residuals across all $13$ matter and Higgs line bundles, grouped by sector.}
\label{fig:cicy5273_learning_curves}
\end{figure}

Table~\ref{tab:app_ml_hyperparams} summarizes the network architectures and optimization hyperparameters for CICY 5273, Model~500. All $17$ models per seed are implemented in JAX/Equinox and trained in single precision for $300$ epochs at batch size $4{,}096$ using AdamW $\epsilon_{\mathrm{Adam}} = 10^{-7}$ with weight decay $10^{-4}$, a $15$-epoch linear warmup from $10^{-5}$ to $10^{-3}$, and cosine decay to $10^{-6}$. Exploiting the dependency graph of the equations, Stage~2 trains the three HYM networks in parallel once Stage~1 completes, and Stage~3 trains all $13$ harmonic $(0,1)$-form networks in parallel on NVIDIA B200 GPUs, completing a full $17$-network seed in roughly $1.5$ hours of wall-clock time. Per-epoch wall-clock times on a single B200 GPU are $\simeq 5.0\text{ s}$ for Stage~1, $0.8\text{ s}$ for Stage~2, $1.3\text{--}2.8\text{ s}$ for $Q_a, u^c_a, H_{ab}, H_{d,ab}$, and $11.7\text{ s}$ for the $M_I = 10{,}800$ network of $b^c$.

Figure~\ref{fig:cicy5273_learning_curves} shows the resulting three-seed learning curves at Benchmark point~4. Across the three independent random initializations, the validation Monge--Amp\`ere error decreases smoothly from $\sigma_{\mathrm{MA}} = 0.781$ to $0.0215 \pm 0.0001$, the normalized HYM curvature residuals decrease by more than an order of magnitude to $0.055\text{--}0.072$, and the normalized co-closedness residuals drop to $4.1\times 10^{-2}\text{--}4.8\times 10^{-2}$ for $(Q_a, u^c_a, H_{d,ab})$ and $5.0\times 10^{-3}\text{--}5.9\times 10^{-3}$ for $(H_{ab}, b^c)$, accompanied by monotonic reductions of the matter field K\"ahler norms $\cK_{I\bar{I}}$ across all $39$ bundle-seed pairs (up to $93\times$ on $b^c$ and $47\times$ on $H_{34}$).

\clearpage
\bibliographystyle{JHEP}
\bibliography{references}

\end{document}